\documentclass[11pt, a4paper]{article}
\usepackage{jheppub, amsmath, amssymb, bbm, color, graphicx, hyperref, pgf, tikz}
\usepackage{braket,mathtools, subcaption}
\renewcommand\[{\begin{equation}}
\renewcommand\]{\end{equation}}
\newcommand{\ba}{\begin{eqnarray}}
\newcommand{\ea}{\end{eqnarray}}

\renewcommand{\eqref}[1]{Eq.\,(\ref{#1})}

\newcommand{\vm}{\vec m}
\newcommand{\vn}{\vec n}

\newcommand{\N}{\mathbb N}

\newcommand{\R}{\mathbb R}

\newcommand{\SU}{\mathrm{SU}(2)}
\newcommand{\Spin}{\mathrm{Spin}(4)}
\newcommand{\defeq}{\vcentcolon=}

\newcommand{\Tr}{\mathrm{Tr}}

\newcommand{\std}{D}				
\def\Ds{D_{\textsc s}}
\newcommand{\bi}{{\gamma_{\textsc{bi}}}}
\newcommand{\sfs}{\gamma}
\newcommand{\sfsc}{\gamma_\mathrm{cons}}

\newcommand{\jmin}{{j_{\text{min}}}}
\newcommand{\jmax}{{j_{\text{max}}}}

\def\d{\mathrm{d}}
\def\e{\textrm e}

\newcommand{\expec}[1]{\langle #1 \rangle}

\newcommand{\COTa}{Calcagni:2013ku}

\begin{document}

\title{Curvature effects in the spectral dimension \\ of spin foams}

\author[a,c,d]{Alexander F. Jercher,}
\emailAdd{alexander.jercher@uni-jena.de}

\author[a]{Sebastian Steinhaus,}
\emailAdd{sebastian.steinhaus@uni-jena.de}

\author[b]{Johannes Th\"urigen}
\emailAdd{johannes.thuerigen@uni-muenster.de}

\affiliation[a]{Theoretisch-Physikalisches Institut, Friedrich-Schiller-Universit\"{a}t Jena\\ Max-Wien-Platz 1, 07743 Jena, Germany, EU}
\affiliation[b]{Mathematisches Institut der Westf\"alischen Wilhelms-Universit\"at M\"unster\\ Einsteinstr. 62, 48149 M\"unster, Germany, EU}
\affiliation[c]{Arnold Sommerfeld Center for Theoretical Physics, Ludwig-Maximilians-Universit\"{a}t M\"{u}nchen\\ Theresienstrasse 37, 80333 M\"{u}nchen, Germany, EU}
\affiliation[d]{Munich Center for Quantum Science and Technology (MCQST)\\ Schellingstrasse 4, 80799 M\"{u}nchen, Germany, EU}

\begin{abstract}{
It has been shown in~\cite{Steinhaus:2018aav} that a class of restricted spin foam models can feature a reduced spectral dimension of space-time. 
However, it is still an open question how curvature affects the flow of the spectral dimension.
To answer this question, we consider another class of restricted spin foam models, so called spin foam frusta, which naturally exhibit oscillating amplitudes induced by curvature, as well as an extension of the parameter space by a cosmological constant.
Numerically computing the spectral dimension of $1$-periodic frusta geometries using extrapolated quantum amplitudes, 
we find that quantum effects lead to a small change of spectral dimension at small scales and an agreement to semi-classical results at larger scales. 
Adding a cosmological constant~$\Lambda$, we find additive corrections to the non-oscillating result at the diffusion scale $\tau\sim 1/\sqrt{\Lambda}$. 
Extending to $2$-periodic configurations, we observe a reduced effective dimension, the form of which sensitively depends on the values of the gravitational constant $G$ and the cosmological constant $\Lambda$. 
We provide an intuition for our results based on an analytical estimate of the spectral dimension. 
Furthermore, we present a simplified integrable model with oscillating measure that qualitatively explains the features found numerically.
We argue that there exists a phase transition in the thermodynamic limit which crucially depends on the parameters $G$ and $\Lambda$. 
In principle, the dependence on $G$ and $\Lambda$ presents an exciting opportunity to infer phenomenological insights about quantum geometry from measurement of the spectral dimension. 
}
\end{abstract}

\setcounter{tocdepth}{2}

\maketitle

\section{Introduction}

Deriving observable consequences is crucial for understanding the structure of any quantum theory of gravity and of quantum space-time itself, especially when formulated using discrete structures, such as in the approaches of loop quantum gravity~\cite{Rovelli:2011tk,Ashtekar:2004fk}, spin foams~\cite{Perez:2013uz}, tensorial group field theories~\cite{Freidel:2005qe,Oriti:2012wt,Carrozza:2013oiy}, causal dynamical triangulations (CDT)~\cite{Loll:2019rdj,Ambjorn:2012vc} and causal set theory~\cite{Surya:2019ndm,Dowker:aza}. The spectral dimension, derived from the spectrum of the Laplacian on quantum space-time, has proven to be an informative phenomenological quantity.

This observable provides a notion of an effective dimension of emergent space-time on different scales.
On the one hand, it allows for a consistency check that the space-times obtained from quantum gravity exhibit the observed dimension of $\std=4$ at large length scales. 
This is already highly non-trivial as various approaches only yield fractal or two-dimensional geometries \cite{DiFrancesco:1995ih, Gurau:2013th}. On the other hand, new small-scale effects beyond classical continuum gravity, sourced by the quantum nature of space-time, can be examined. 
Most prominently, a dimensional flow to values $0 < D < 4$ at small scales is an effect present in many quantum gravity approaches~\cite{Carlip:2017ik,Crane:1985ba,Crane:1986gf,Ambjorn:2005fh,Coumbe:2015bq,Lauscher:2005kn,Horava:2009ho,Benedetti:2009fo,Alesci:2012jl,Arzano:2014ke,Steinhaus:2018aav,Eckstein:2020gjd,Reitz:2022dbj,Calcagni:2014ep,Calcagni:2015is,Eichhorn:2017djq},
possibly leaving observable traces in gravitational wave astronomy~\cite{Calcagni:2019ngc}. 
In any case, this phenomenon is interesting as it allows to compare conceptually different approaches at small scales. Investigating the existence and structure of such a dimensional flow in spin foam quantum gravity~\cite{Perez:2013uz},
specifically using the restricted setting of spin foam frusta models \cite{Bahr:2017bn,Bahr:2018gwf},
is the main objective of this article.

Spin foam models provide a path integral quantization of discretized geometries where the microscopic gravitational degrees of freedom are encoded in group representation labels and intertwiners. Although model defining amplitudes are defined rigorously and consistently on the quantum level, challenges remain to turn spin foam quantum gravity into a consistent computational formalism. While there is a clear relation to semi-classical discrete gravity on a local level~\cite{Barrett:2009ci,Barrett:2009mw,Dona:2017dvf,Dona:2019dkf}, 
exploring the semi-classical regime of extended discrete structures is a subject of active research~\cite{Asante:2022lnp,Asante:2020qpa,Asante:2020iwm,Han:2020npv,Asante:2021zzh,Han:2021kll,Dona:2019jab}. Moreover, a consistent quantum theory of gravity needs to be independent of any discretization chosen in the quantization procedure. 
Discretization independence of spin foams can be attained either in the group field theory formalism~\cite{Freidel:2005qe,Oriti:2012wt,Carrozza:2013oiy} or via a spin foam renormalization procedure~\cite{Dittrich:2013xwa,Steinhaus:2020lgb,Asante:2022dnj}. Furthermore, in recent years, there has been promising progress in numerical methods~\cite{Dona:2022yyn} that tackle the challenge of performing calculations in spin foams from different angles~\cite{Dona:2019dkf,Dona:2019jab,Gozzini:2021kbt,Dona:2022dxs,Dona:2023myv,Han:2020npv,Han:2021kll,Han:2023cen,Asante:2020qpa,Asante:2020iwm,Allen:2022unb,Asante:2022lnp}. Defining and explicitly computing the spectral dimension in spin foam models makes contact with these open questions and serves as a coarse characterization of the quantum space-times obtained therein.

A first attempt to determine the spectral dimension $\Ds$ of spin foams in~\cite{Steinhaus:2018aav} has been in the setting of spin foam cuboids~\cite{Bahr:2015gxa,Bahr:2016hwc,Allen:2022unb,Ali:2022vhn}, a restricted subclass of the Euclidean Engle-Pereira-Rovelli-Livine/Freidel-Krasnov (EPRL-FK) model~\cite{Engle:2008fj,Freidel:2008fv} defined within the Kaminski-Kisielowski-Lewandowski extension to general 2-complexes~\cite{Kaminski:2010ba}. Therein, a restriction to $\mathcal{N}$-periodic geometric configurations proved crucial for numerically feasible calculations. 
At finite $\mathcal{N}$, the spectral dimension shows an intermediate value which depends on the choice of a face amplitude parameter $\alpha$. 
In the limit $\mathcal{N}\rightarrow\infty$, the analytical results of~\cite{Steinhaus:2018aav} suggest that there is a phase transition from $\Ds = 0$ to $\Ds = 4$ at the point~$\alpha_*$ where the amplitudes become scale-invariant. 

Although an important step towards control of the spectral dimension of spin foam quantum gravity, the results on spin foam cuboids~\cite{Steinhaus:2018aav} come with two  major limitations. 
First, the spectral dimension is computed utilizing semi-classical amplitudes which capture only the large scale behaviour correctly. Second, and most importantly, spin foam cuboids are inherently flat, reflected in the fact that the semi-classical action vanishes on critical points. 
As a consequence, cuboid amplitudes exhibit a simple semi-classical scaling behaviour without curvature-induced oscillations. 
Overcoming these limitations is the main purpose of the present work. 

Since the spectral dimension for the full EPRL-FK model currently exceeds computational capacities, we choose the so-called spin foam frusta model, introduced in~\cite{Bahr:2017bn} and further examined in~\cite{Bahr:2018ewi,Bahr:2018gwf,Allen:2022unb},
to include curvature effects and a non-vanishing cosmological constant. Though the underlying combinatorial structure is a hypercubic lattice, 
the intertwiners of the theory are restricted in such a way that the $4$-dimensional building blocks semi-classically describe $4$-frusta. 
These building blocks can be understood as a $4$-dimensional generalization of regular trapezoids, with $3$-cubes as base and top connected by six boundary $3$-frusta. It has been shown in~\cite{Bahr:2017bn} that semi-classical spin foam frusta exhibit curvature if the $3$-cubes are of different size. Due to their strongly restricted geometry, frusta exhibit a discrete analogue of spatial homogeneity. Thus, frusta are promising candidates for a cosmological subsector of spin foam models~\cite{Bahr:2017bn}. At the same time, this homogeneity leads to simplifications of the Laplace operator, providing a feasible setting for numerical computations of the spectral dimension.

Extrapolating $1$-periodic frusta quantum amplitudes and computing the spectral dimension therewith, we discover a non-trivial flow of the spectral dimension, exhibiting an intermediate value which is controlled by the face amplitude parameter $\alpha$. Compared to a semi-classical analysis, the quantum amplitudes lead to an additive correction to the spectral dimension at small scales. Adding a cosmological constant $\Lambda$ and thus introducing oscillations to the amplitudes, we find additive corrections to the spectral dimension at diffusion scales $\tau\sim1/\sqrt{\Lambda}$. 
Supported by analytical calculations, we can explain these effects from quantum amplitudes and $\Lambda$-oscillations in terms of an effective scaling behaviour of the amplitudes. For $2$-periodic frusta amplitudes we observe an intricate dependence of the dimensional flow on Newton's constant $G$ and $\Lambda$, showing the significant role of curvature. On these grounds, we argue for a phase transition in the large-$\mathcal{N}$ limit that crucially depends on the parameter values of $\alpha$, $G$ and $\Lambda$.

We begin this article by setting up the spin foam frusta model, its semi-classical limit and by defining the Laplace operator on frusta geometries. Thereafter, we present the concept of $\mathcal{N}$-periodicity, which has been introduced already in~\cite{Steinhaus:2018aav} and provide the formulas for computing the quantum spectral dimension. In Sec.~\ref{sec:Quantum amplitudes from extrapolation}, we provide a method to extrapolate $1$-periodic quantum amplitudes. Using the resulting extrapolated amplitudes, we compute the $1$-periodic spectral dimension in Sec.~\ref{subsec:1-periodic spectral dimension}. Proceeding with semi-classical amplitudes, we extend the analysis of the spectral dimension to include a cosmological constant in Sec.~\ref{subsec:Cosmological constant} and finally to $2$-periodic configurations in Sec.~\ref{subsec:2-periodic spectral dimension}. Based on the analytical estimate of the spectral dimension presented in~\ref{subsec:Analytical estimate of the spectral dimension}, we provide a broader discussion of our results in Sec.~\ref{sec:Discussion} and give and conclude our work in Sec.~\ref{Sec:Conclusion} with an outlook on possible future studies.


\section{Spin foam hyperfrusta and the spectral dimension}\label{sec:Spin foam hyperfrusta and the spectral dimension}

In this section we provide a brief introduction to the setting of the present work and, in particular, define the spectral dimension as a spin foam observable.  Giving first the relevant formulas for classical hyperfrusta, we introduce spin foam frusta as restrictions of the Euclidean EPRL-FK model. Thereafter, we set up the classical spectral dimension in a discrete setting and translate these ideas to spin foam hyperfrusta.


\subsection{Spin foam frusta}

In the present work, we employ the Euclidean EPRL-FK model~\cite{Engle:2008fj,Freidel:2008fv} with a Barbero-Immirzi parameter $\bi < 1$. Originally defined on a $2$-complex $\Gamma$ dual to a triangulation, this model was later generalized to arbitrary $2$-complexes~\cite{Kaminski:2010ba,Oriti:2015kv}, which includes in particular the hypercuboidal structure we intend to work with. In the following, we introduce all the defining ingredients of the EPRL-FK model. 

\subsubsection{Definition of spin foam frusta}\label{subsec:Spin foam frusta}

Given a $2$-complex $\Gamma$ with vertices~$v$, edges $e$ and faces $f$, we associate to it the following partition function
\begin{equation}
Z = \sum_{j_f,\iota_e} \prod_f\mathcal{A}_f\prod_e\mathcal{A}_e\prod_v\mathcal{A}_v.
\end{equation}
The variables $j_f$ denote irreducible $\SU$-representations, frequently referred to as spins, and are assigned to the faces $f$ of $\Gamma$. The $\iota_e$ are $\SU$-intertwiners, i.e. elements in the invariant subspace of the tensor product of representations meeting at the edge $e$. $\mathcal{A}_f,\mathcal{A}_e$ and $\mathcal{A}_v$ are the face, edge and vertex amplitudes, respectively, and we define them in the following.

\begin{figure}
    \centering
    \includegraphics[width = 0.6\textwidth]{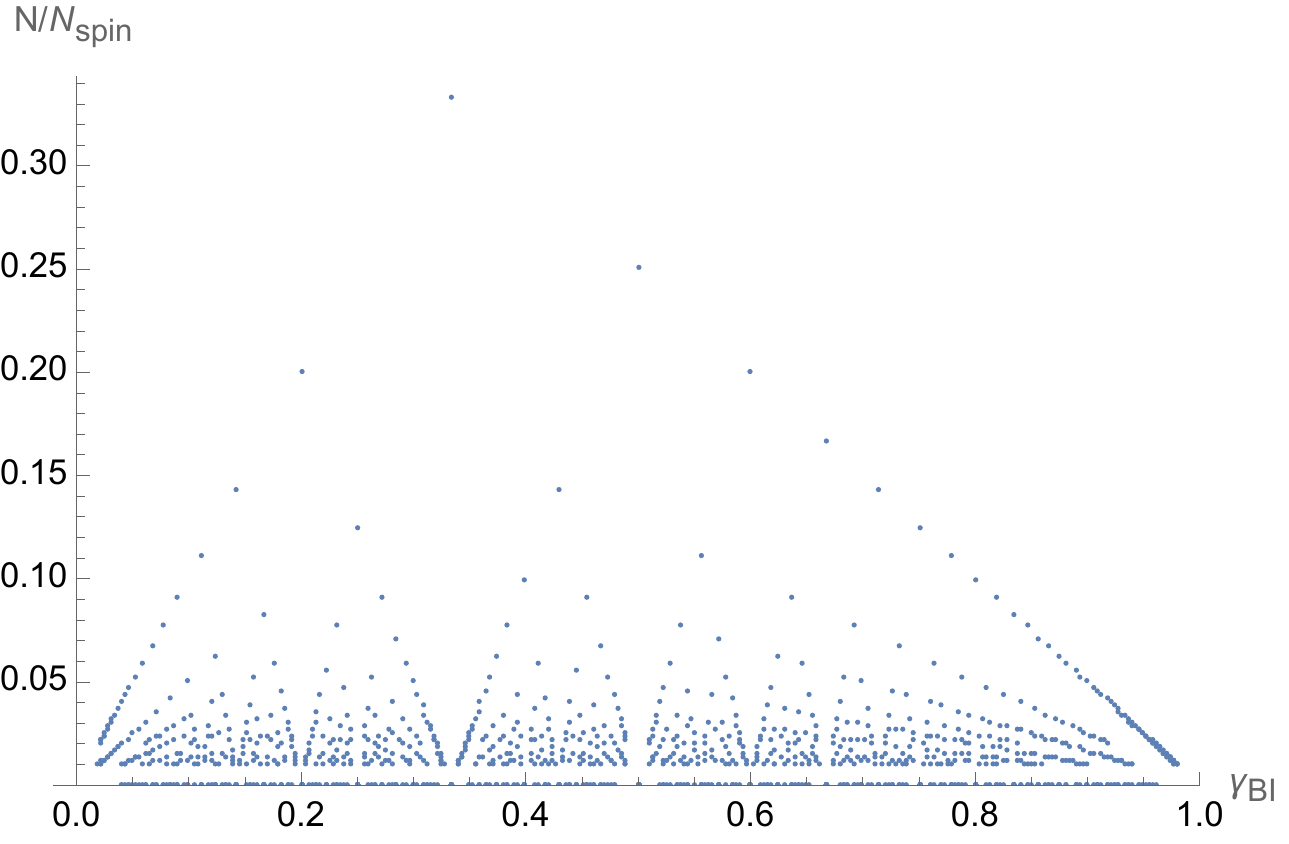}
    \caption{Ratio of the number $N$ of allowed configurations and the total number of spins $N_{\mathrm{spins}}$ for rational Barbero-Immirzi parameters $\bi=\frac{p}{q}$ with $q\in\{1,2,...,50\}$ and $p\in\{1,2,...,q-1\}$.
    }
    \label{fig:EPRL condition}
\end{figure}

Characteristic for Euclidean gravity, the local gauge group is SO$(4)$, or its double cover $\Spin\cong\SU\times\SU$, with $\SU$ being a subgroup. A distinctive property of the Euclidean EPRL-FK model is the simplicity constraint which provides an embedding~$Y_{\bi}$ of $\SU$ into $\Spin$-representations~\cite{Engle:2008fj}. For $\bi < 1$, the explicit relation between $\SU$ spins~$j$ and $\Spin$-labels $(j^+,j^-)$ is given by
\begin{equation}\label{eq:EPRL condition}
j^{\pm} = \frac{\vert1\pm \bi\vert}{2}j\in\frac{\mathbb{N}}{2},
\end{equation}
with the condition that the $j^\pm$ are half-integers. For this map to be non-empty, $\bi$ is required to be rational. Given a certain range of spins with $N_\mathrm{spin}$ configurations in total, only a certain number $N(\bi)$ is allowed.%
\footnote{The ratio $N(\bi)/N_\mathrm{spin}$ for $\bi=\frac{p}{q}\in\mathbb{Q}$ is either $1/q$ or $1/2q$ depending on whether $q\pm p$ is even or odd
    which explains the asymmetry in Fig.~\ref{fig:EPRL condition} (e.g. for $q=5$ we have ratio $1/5$ for $p=1,3$ while $1/10$ for $p=2,4$).   
More precisely, let 
$q\in\N$ and $p\in\{1,2,...,q-1\}$ but $p\perp q$ (coprime, otherwise there are smaller $p',q'$ with $\bi=\frac{p}{q}=\frac{p'}{q'}$ to be considered). 
    Then the EPRL condition \eqref{eq:EPRL condition} yields 
    $\frac{q\pm p}{2q}n\in\N$.
    Write $\frac{q\pm p}{2q}=\frac{r}{s}$ with $r\perp s$.
    Then the condition is fulfilled if $\frac{n}{s}\in \N$, that is for every $s$'s spin. 
    Thus, the ratio is ${N}/{N_\text{spins}}=s$ and there are two cases for given $q$, either $s=q$ or $s=2q$. To see this, note first that $q+p \perp 2q$ iff $2q-(q+p)=q-p \perp 2q$, thus the two cases $\pm\bi$ give the same $s$. 
    Then, $q+p$ and  $2q$ have a common divisor iff $q+p$ is even in which case $s=q$ and this happens for every second $p$ for a given $q$.
    }
A plot for the ratio of $N(\bi)$ and $N_\mathrm{spin}$ is given in Fig.~\ref{fig:EPRL condition}.

Different face amplitudes are proposed in the literature%
\footnote{Following~\cite{Bianchi:2010fj}, the two most common choices for the face amplitude in the Riemannian setting are either the dimension of the $\Spin$-representation or the dimension of the $\SU$-representation. Viewing the Euclidean EPRL-model as a constrained $\Spin$-BF theory, the dimension factor is a result of expanding $\delta$-distributions on $\Spin$ in terms of representations. Choosing the $\SU$-dimension instead can be motivated physically, as argued in~\cite{Bianchi:2010fj}. The two choices correspond to $\alpha = 1$, respectively $\alpha = \frac{1}{2}$. Clearly, the parametrization of~\eqref{eq:face amplitude} constitutes a continuous interpolation between these two choices.},
the choice of which has direct consequences for the critical behaviour of the partition function~\cite{Riello:2013bzw}, as well as the spectral dimension~\cite{Steinhaus:2018aav}. We parametrize this ambiguity by $\alpha\in\R$ as
\begin{equation}\label{eq:face amplitude}
\mathcal{A}_f^{(\alpha)} = \left((2j_f^+ + 1)(2j_f^- + 1)\right)^{\alpha} \, .
\end{equation}
The edge amplitude of the model is introduced as a normalization factor
\begin{equation}\label{eq:general EAmp}
\mathcal{A}_e = \frac{1}{\vert\vert Y_{\bi}\iota_e\vert\vert^2}
\end{equation}
and the vertex amplitude is defined as
\begin{equation}\label{eq:general vampl}
\mathcal{A}_v = \mathrm{Tr}\left(\bigotimes_{e\subset v}Y_{\bi}\iota_e\right),
\end{equation}
where the trace is understood so that either $(Y_{\bi}\iota_e)$ or $(Y_{\bi}\iota_e)^{\dagger}$ are contracted, depending on the edge $e$ being ingoing or outgoing from the vertex $v$, respectively.

\

With the general definition of the model being set up, we now introduce two restrictions on the partition function, being of combinatorial and geometrical kind. First, we assume that the $2$-complex $\Gamma$ and thus also the discretization is hypercubic. As we will see later in Sec.~\ref{subsec:Spectral dimension and semi-classical geometry of hyperfrusta} and Sec.~\ref{subsec:N-periodicity}, respectively, the choice of a hypercubic lattice is convenient for defining a discrete Laplace operator as well as for considering periodic configurations. 
Notice, that this choice of combinatorics does not imply a reduction to hypercubic geometries.

\begin{figure}
    \centering
    \includegraphics[width=0.4\textwidth]{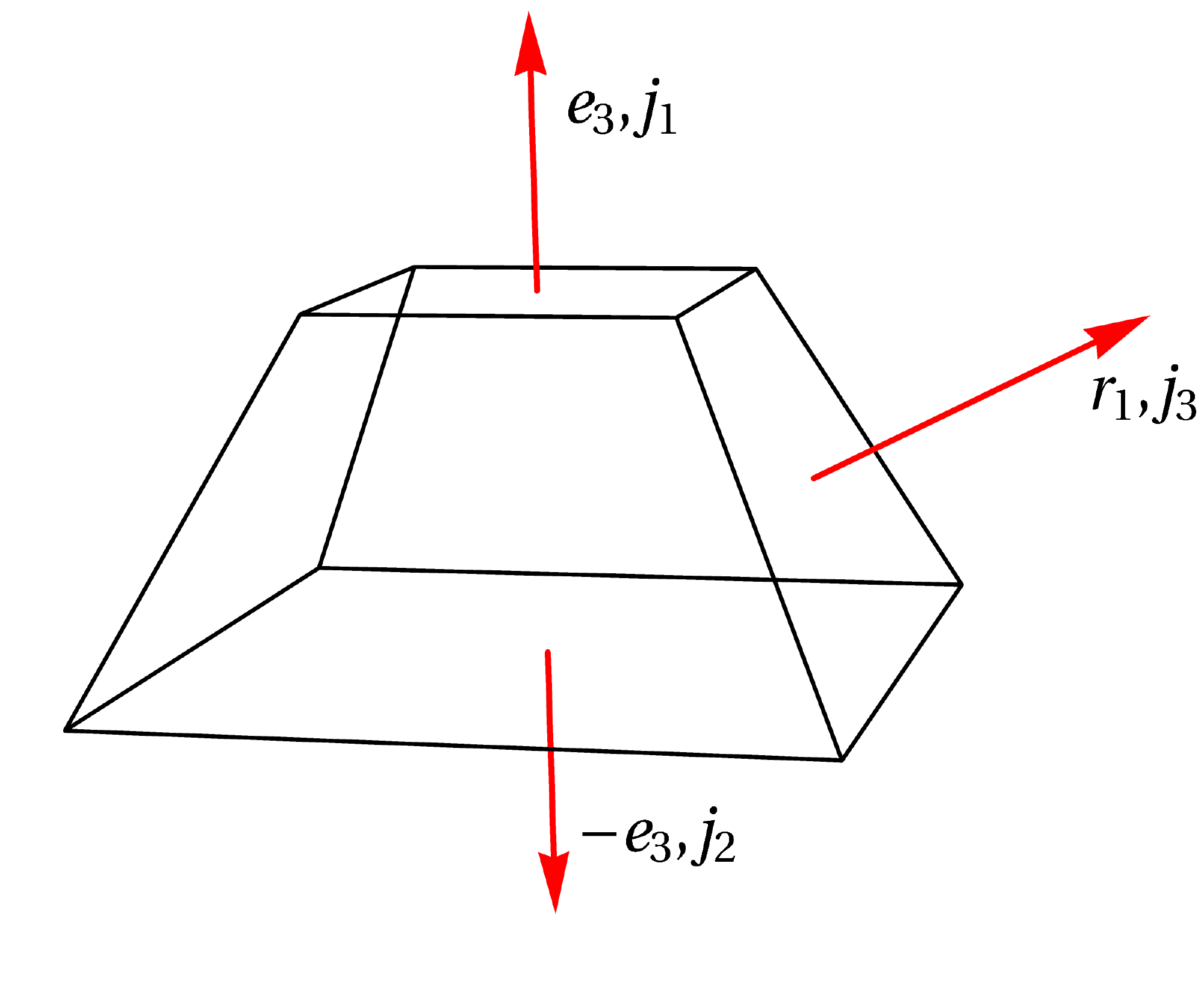}
    \caption{A regular $3$-frustum with squares in the $e_3$-plane. The slope angle $\phi$ is defined between $e_3$ and the $r_l$.}
    \label{fig:3-frustum}
\end{figure}

On the geometrical side, we introduce the restriction to regular $4$-frustum geometries~\cite{Bahr:2017bn}. The boundary of such configurations consist of two cubes of a priori different size and six equal $3$-frusta. It is therefore reasonable to impose a $3$-frustum shape on the Livine-Speziale coherent intertwiners~\cite{Livine:2007bq}, since these objects are naturally peaked on the geometry of $3$-dimensional polyhedra.%
\footnote{Notice that the frustum symmetry reduction is a restriction on the quantum level, as it is imposed on the coherent intertwiners. Such a procedure is distinct from a classical symmetry reduction followed by quantization. We remark further, that lifting this restriction is unfortunately not straightforward. As elaborated in Sec.~\ref{Sec:Conclusion}, a triangulation offers a more manageable setting to study models with less symmetries.
} 
An explicit expression for $3$-frustum intertwiners is given by
\begin{equation}\label{eq:intertwiner}
\iota_{j_1 j_2 j_3} = \int\limits_{\SU}\d{g}\: g\triangleright \left(\Ket{j_1, \hat{e}_3}\otimes\Ket{j_2 -\hat{e}_3}\bigotimes_{l=0}^3 \Ket{j_3, \hat{r}_l}\right),
\end{equation}
where the case $j_1 = j_2 = j_3$ reduces $\iota_{j_1j_2j_3}$ to a cubical intertwiner~\cite{Bahr:2016co}. In~\eqref{eq:intertwiner}, ``$\triangleright$'' denotes the action of an $\SU$-group element on coherent states in representation space, $\hat{e}_3$ is the unit vector in $\R^3$ along the axis $e_3$ and $\hat{r}_l = \e^{-i\frac{\pi}{4}l\sigma_3}\e^{-i\frac{\phi}{2}\sigma_2}\triangleright \hat{e}_3$ for $l\in\{0,1,2,3\}$. A depiction of a regular $3$-frustum is given in Fig.~\ref{fig:3-frustum}. 

A semi-classical $3$-frustum can be understood as the generalization of a regular trapezoid to three dimensions, characterized by three areas $j_1, j_2$ and $j_3$ of the base and top square and of the bounding trapezoids, respectively. Another convenient parametrization is given by $j_1,j_2$ and the slope angle $\phi$. The relation between the spins and $\phi$ is given by
\begin{equation}\label{eq:cosphi}
\cos(\phi) = \frac{j_1-j_2}{4j_3},
\end{equation}
which follows from the closure condition. Clearly, for the slope angle to be well-defined, the condition $-1\leq \frac{j_1-j_2}{4j_3}\leq 1$ is required to hold.

\begin{figure}
    \centering
    \includegraphics[width=0.6\textwidth]{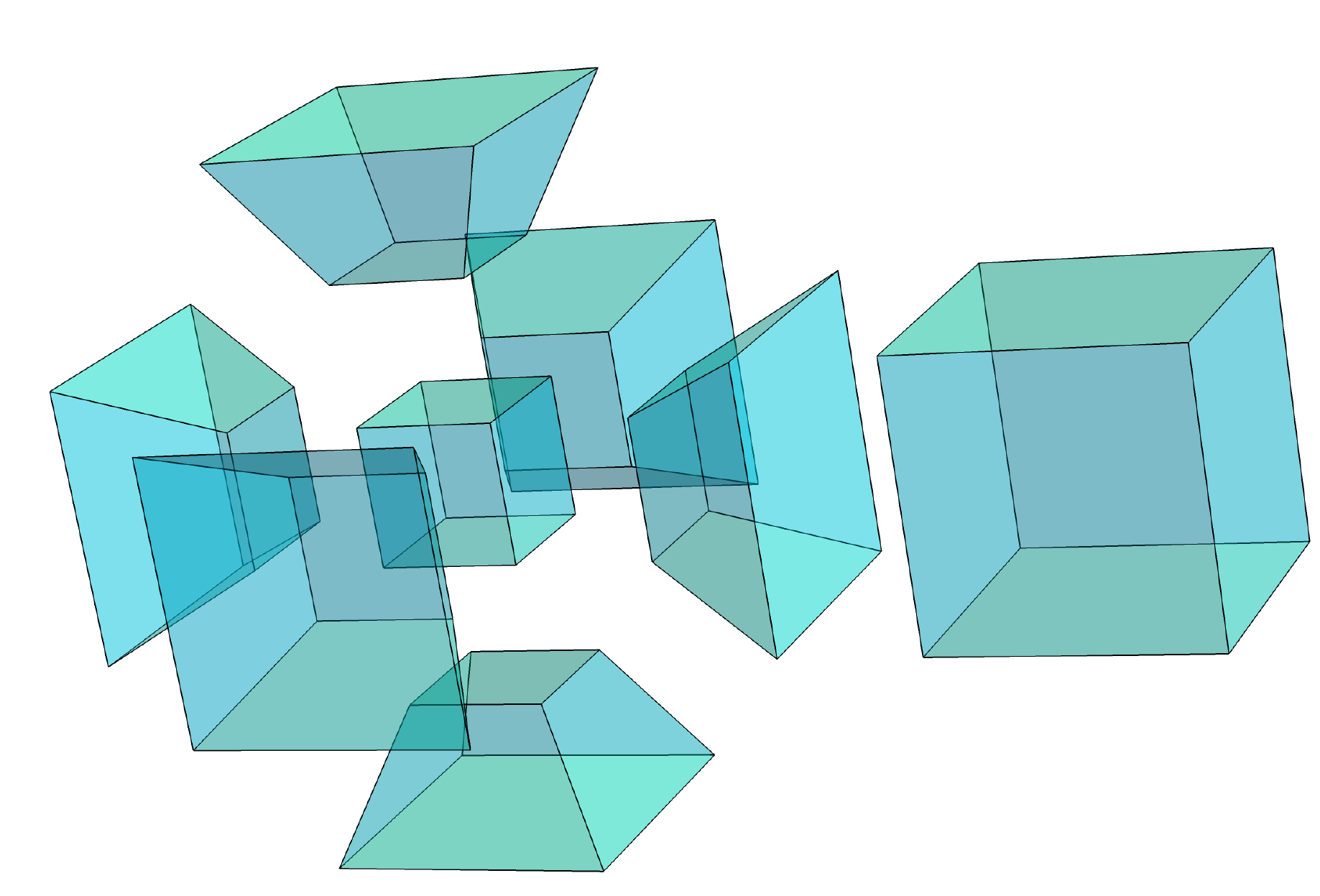}
    \caption{Boundary of a $4$-frustum given by two $3$-cubes of, generically, different size and six equal $3$-frusta.}
    \label{fig:4-frustum}
\end{figure}

Gluing two cubes and six $3$-frusta as indicated above, one obtains a $4$-dimensional hyperfrustum. The two $3$-cubes lie in separated spatial hypersurfaces at the base and top, connected by the six $3$-frusta. A visualization of the unwrapped boundary of a single $4$-frustum is given in Fig.~\ref{fig:4-frustum}.

As a whole complex, the discretization can be understood as a slicing, where the $n$th thick slice is bounded by two spatial hypersurfaces cubulated by $3$-cubes with area $j_n$ and~$j_{n+1}$, respectively, and connected by $3$-frusta with spatio-``temporal'' areas given by $k_n$. Due to the gluing conditions, the spins $j_n, j_{n+1}$ and $k_n$ are constant throughout a whole slice. Therefore, spin foam configurations of hyperfrusta are translation invariant in spatial directions. This implies that the volume of cubes changes only between different spatial hypersurfaces, i.e. in temporal direction. 
This makes hyperfrustum spin foams well-suited for the description of discrete classical cosmology, where we refer to~\cite{Bahr:2017bn} for further details.

\subsubsection{Asymptotics of spin foam frusta}\label{subsec:Asymptotics of spin foam frusta}

In general, the quantum amplitudes defined by~\eqref{eq:general vampl} and~\eqref{eq:general EAmp} are highly involved functions of the representation labels. 
To find a suitable approximation and to obtain an analytical expression for later purposes, we present in the following the results of an asymptotic analysis~\cite{Barrett:2009ci,Barrett:2009mw} of the frustum quantum amplitudes~\cite{Bahr:2017bn}. The regime of large representation labels is interpreted as the semi-classical sector of the theory, thereby connecting it to classical discrete gravity~\cite{Rovelli:2014vg}. 

Introducing a uniform rescaling of spins, $j_i \rightarrow \lambda j_i$, and sending $\lambda\rightarrow\infty$, the quantum face amplitude in~\eqref{eq:face amplitude} is approximated by 
\begin{equation}\label{eq:sc face amplitude}
\mathcal{A}_{f} = \left[\left(1-\bi^2\right)j_n^2\right]^{\alpha}.
\end{equation}
To find the asymptotics of the vertex and the edge amplitude $\mathcal{A}_i$, we notice the factorization property $\mathcal{A}_i = \mathcal{A}^+_i\mathcal{A}^-_i$ with $\mathcal{A}_i^{\pm}$ being the $\SU$-amplitudes evaluated on the spins $j^{\pm}$. This is a result of the $Y_{\bi}$-map entering~\eqref{eq:general EAmp} and~\eqref{eq:general vampl} which holds for $\bi < 1$. 
We rewrite an amplitude $\mathcal{A}_i^{\pm}$ as an integral of a complex exponential of an action which scales linearly in $\lambda$. 
Then one can apply an extended stationary phase approximation~\cite{Barrett:2009ci} in the limit $\lambda\rightarrow\infty$ which consists of evaluating the integrand on the critical points of the action, multiplied by the inverse square root of the Hessian and some factors of $\pi$.  
Following this procedure for the $\SU$-vertex amplitude, one finds~\cite{Bahr:2017bn}
\begin{equation}\label{eq:SU2Vamp}
\mathcal{A}^{\SU}_v(j_n,j_{n+1},k_n) = \frac{1}{2^{21/2}\pi^{7/2}}\left(\frac{\e^{\frac{i}{G}S_R}}{\sqrt{-D}} + \frac{\e^{-\frac{i}{G}S_R}}{\sqrt{-D^*}}\right),
\end{equation}
where $D$ is the determinant of the Hessian, defined as
\begin{equation}\label{eq:determH}
\begin{aligned}
D &\defeq 16 j_n^3 j_{n+1}^3 k_n^{15} K\left(K-i K^2+iQ\right)^3 \times \\[7pt]
& \times \left(1+K^2-2 Q\right)^3(K+i)^6 (K-3 i)^2 \left(1+3 K^2-2Q-2 i K (Q-1)\right)^3,
\end{aligned}
\end{equation}
with
\begin{equation}\label{eq:Q theta K}
Q \defeq 2+\frac{j_n+j_{n+1}}{2k_n},\qquad \theta \defeq \arccos \frac{1}{\tan\phi},\qquad K \defeq \sqrt{-\cos 2\phi}.
\end{equation}
Furthermore, $S_R$ is the Regge action~\cite{Regge:1961ct} of a single hyperfrustum and given by
\begin{equation}\label{eq:Regge action}
 S_R = 6(j_n - j_{n+1})\left(\frac{\pi}{2}-\theta\right) +12k_n\left(\frac{\pi}{2}-\arccos(\cos^2(\theta))\right)
\end{equation}
and $G$ is the gravitational constant, which has been added by hand as in~\cite{Bahr:2018gwf}.\footnote{In units where $\hbar = c =1$, the gravitational constant has the dimension of area. Considering the $j_n$ as mere representation labels, only $G j_n$ has the interpretation of an area. For the majority of the spin foam literature, the spins are implicitly understood to have the dimension of area and therefore implicitly depend on $G$~\cite{Bahr:2018gwf}.}
Given~\eqref{eq:SU2Vamp}, the semi-classical $\Spin$-vertex amplitude is 
\begin{equation}\label{eq:Spin4Vamp}
\mathcal{A}_v 
= 
\frac{1}{\pi^7(1-\bi^2)^{21/2}}\left(\frac{\e^{ \frac{i}{G}S_{R} }}{-D}+\frac{\e^{-\frac{i}{G}S_{R} }}{-D^{*}}+2\frac{\cos(\frac{ \bi} {G} S_{R}-\frac{\Lambda}{G}V^{(4)})}{\sqrt{D D^{*}}}\right),
\end{equation}
where we have introduced a cosmological constant $\Lambda$, following the work of~\cite{Han:2011aa} and~\cite{Bahr:2018vv}. Therein, the quantum amplitudes are deformed in a heuristic fashion by a real parameter which, in a semi-classical limit, can be related to a cosmological constant of either sign. It enters with the $4$-volume $V^{(4)}$ of a hyperfrustum, defined in~\eqref{eq:4Volume} below, which is strictly positive and thus does not correspond to a signed volume.\footnote{Notice, that this positive quantity is the same quantity that enters the definition of the Laplace operator in \eqref{eq:defLaplace}. The cosmological constant term enters with a cosine in the quantum amplitudes since the critical points correspond to the two orientations.}

Repeating the analysis, the semi-classical $\SU$-edge amplitude is given by
\begin{equation}\label{eq:SU2EAmp}
\mathcal{A}_e^{\SU}(j_n,j_{n+1},k_n) = 2\sqrt{2\pi}\sqrt{\frac{k_n\sin^2(\phi)}{2}\Big{(}j_n+j_{n+1}+2k_n\big{(}1-\cos^2(\phi)\big{)}\Big{)}^2}.
\end{equation}
Consequently, the semi-classical approximation to the $\Spin$-edge amplitude, \eqref{eq:general EAmp}, is
\begin{equation}\label{eq:Spin4Eamp}
\mathcal{A}_e = \pi(1-\bi^2)^{3/2}\frac{k_n\sin^2(\phi)}{2}\Big{(}j_n+j_{n+1}+2k_n\big{(}1-\cos^2(\phi)\big{)}\Big{)}^2.
\end{equation}
For general discretizations an edge is shared by two vertices. Special to a hypercubic lattice, every face is shared by four vertices. These two facts allow to define a dressed vertex amplitude
\begin{equation}\label{eq:dVAmp}
\hat{\mathcal{A}}_v = \prod_{f\supset v}\mathcal{A}_f^{1/4}\prod_{e\supset v}\mathcal{A}_e^{1/2}\mathcal{A}_v,
\end{equation}
such that the amplitude of the whole complex can be written as the product of dressed vertex amplitudes.

Even in the presence of oscillations, originating from a non-vanishing cosmological constant or a non-vanishing Regge curvature, the dressed vertex amplitudes behave under uniform rescaling as
\begin{equation}
\hat{\mathcal{A}}_v(\lambda j_n,\lambda j_{n+1},\lambda k_n) \sim \lambda^{12\alpha -9}.
\end{equation}
Comparing $\hat{\mathcal{A}}_v$ to the dressed vertex amplitude of spin foam cuboids~\cite{Bahr:2016co}, we see that the this scaling is the same. However, $\hat{\mathcal{A}}_v$ is \emph{not} a homogeneous function in the spins, due to the two types of oscillations in~\eqref{eq:dVAmp}, which are not present in the case of cuboids. As a consequence, we cannot deduce that $\alpha = \frac{3}{4}$ is a point of scale-invariance of the frusta model.

\subsection{Spectral dimension of spin foam frusta}

In this section, we introduce the notion of spectral dimension, first classically on semi-classical frusta geometries. After restricting to so-called $\mathcal{N}$-periodic configurations, we set up the spectral dimension of spin foam frusta as a quantum expectation value. 

\subsubsection{Spectral dimension and semi-classical geometry of hyperfrusta}\label{subsec:Spectral dimension and semi-classical geometry of hyperfrusta}

The spectral dimension serves as an effective measure of the dimension of a space. In the continuum, consider a Riemannian manifold $(\mathcal{M},g)$ together with the heat kernel $K(x,x_0;\tau)$ which is a solution of the heat equation~\cite{Carlip:2017ik}
\begin{equation}
\partial_{\tau}K(x,x_0;\tau) = \Delta K(x,x_0;\tau).
\end{equation}
Here, $x,x_0\in\mathcal{M}$ and $\tau$ provides a measure of the size of the probed region, often referred to as diffusion time. The classical spectral dimension $\Ds^{\mathrm{cl}}(\tau)$ is then extracted from the scaling of the return probability
\begin{equation}\label{eq:classical retprob}
P(\tau) \defeq  \int_{\mathcal{M}}\d{x}\sqrt{g}~  K(x,x;\tau),
\end{equation}
by the relation
\begin{equation}
\Ds^{\mathrm{cl}}(\tau) \defeq -2\frac{\d \log P(\tau)}{\d \log\tau}.
\end{equation}
Clearly, the return probability implicitly depends on the metric $g$ and  therefore is a functional of the geometry. Thus, in  a quantum gravity path integral picture with partition function
\begin{equation}\label{eq:continuum partition function}
Z = \int\d{g}\;\e^{iS_{\mathrm{EH}}[g]},
\end{equation}
where $S_{\mathrm{EH}}[g]$ is the Einstein-Hilbert action, the expectation value of the return probability is
\begin{equation}
\braket{P(\tau)} = \frac{1}{Z}\int\d{g}\; P(\tau)e^{iS_{\mathrm{EH}}[g]}.
\end{equation}
Consequently, one defines the spectral dimension as
\begin{equation}
\Ds(\tau) \defeq -2\frac{\d \log \braket{P(\tau)}}{\d \log\tau}.
\end{equation}
Notice, that we do not compute $\langle \Ds^\mathrm{cl}(\tau)\rangle$ but define the quantum spectral dimension as the scaling of $\langle P(\tau)\rangle$. 

To translate these continuum notions to the context of spin foams, we introduce now a discrete formulation of the Laplace operator, the return probability and ultimately the spectral dimension. 
Assuming that the space $\mathcal{M}$ is discretized on a hypercubic lattice, we denote vertices in the dual graph $\Gamma$ by $\vec{n}\in\mathbb{Z}^4$. Interpreting the return probability in~\eqref{eq:classical retprob} as the trace over the heat kernel, the discrete return probability is simply given by~\cite{\COTa,Thurigen:2015uc}
\begin{equation}
P(\tau) = \sum_{\vec{n}\in\Gamma}K(\vec{n},\vec{n},\tau).
\end{equation}
Then, one can rewrite the return probability as \cite{Thurigen:2015uc}
\begin{equation}\label{eq:discrete retprob}
P(\tau)
=
\sum_{\lambda\in\mathrm{spec}(\Delta)}e^{-\tau\lambda},
\end{equation}
where $\mathrm{spec}(\Delta)$ is the spectrum of the discrete Laplacian. Consequently, to obtain the spectral dimension, knowledge of the full spectrum on the whole lattice is required.

Following~\cite{Calcagni:2013ku}, the notion of an exterior derivative and a Laplace operator can be defined on general cellular complexes, using methods of discrete exterior calculus~\cite{Desbrun:2005ug}. To that end, one introduces a scalar test field%
\footnote{Importantly, test fields should not be confused with physical fields, minimally coupled to spin foams~\cite{Ali:2022vhn}. In particular, a test field does not have to satisfy spatial homogeneity even if the spin foam is fixed to spatially homogeneous configurations. Also, $\mathcal{N}$-periodicity, which we are going to introduce down below does not need to be fulfilled by $\phi_{\vn}$.} 
discretized on the complex, which can be either placed on the vertices of the discretization or on its dual vertices. The strategy we follow here is to consider the scalar field $\phi_{\vn}$ on dual%
\footnote{Other choices work equally well. E.g. in.~\cite{Ali:2022vhn}, a scalar field is placed on the vertices of the discretization.
Beyond scalar fields, other tensor or $p$-form fields might yield different results for ``generalized'' spectral dimensions, though \cite{Reitz:2022dbj}.}
vertices of the complex $\Gamma$. 
From the continuum perspective, this case can be obtained by integrating a smooth scalar field over the region of a $4$-cell. Finally the discrete Laplacian is defined by its action on the scalar test field $\phi_{\vn}$~\cite{Thurigen:2015uc}
\begin{equation}\label{eq:defLaplace}
-(\Delta \phi)_{\vn} = -\sum_{\vm\sim\vn}\Delta_{\vn\vm}\left(\phi_{\vn} - \phi_{\vm}\right) = \frac{1}{V^{(4)}_{\vn}}\sum_{\vm\sim\vn}\frac{V^{(3)}_{\vn\vm}}{l^*_{\vn\vm}}\left(\phi_{\vn}-\phi_{\vm}\right),
\end{equation}
where the sum runs over all adjacent vertices $\vm$, indicated by ``$\sim$''. $\Delta_{\vn\vm}$ are the coefficients of the discrete Laplacian, which can be split into a diagonal part and a part which is proportional to the adjacency matrix of the complex $\Gamma$. $V^{(3)}_{\vn\vm}$ and $l^*_{\vn\vm}$ indicate the $3$-volume, respectively the dual edge length, of $(\vn\vm)$. The positive $4$-volume of the frustum dual to $\vn\in\Gamma$ is denoted by $V^{(4)}_{\vn}$ and is the same $4$-volume that accompanies the cosmological constant in the vertex amplitude of Eq.~\eqref{eq:Spin4Vamp}. Going beyond the scope of this work, we do not consider the possibility of working with oriented spin foam amplitudes and with signed $4$-volumes in the Laplace operator.

As discussed in detail in~\cite{Thurigen:2015uc}, the definition of volume and dual edge length and thus that of a discrete Laplacian is not unique in general. For constructing the dual $2$-complex, we pursue the following strategy. In $4$-dimensional Euclidean space, we orient a $4$-frustum along the $t$-axis such that the two cubes lie in a constant-$t$ hypersurface. Vertices of the dual lattice are then obtained by forming the average of the corner points of the $4$-frustum. Consequently, these points lie on the $t$-axis on half of the $4$-frustum height. Dual edges, ``spacelike'' and ``timelike'', are chosen to be orthogonal on the faces of the $4$-frustum such that their lengths are minimal.%
\footnote{Defining the dual edges in this way is convenient because of spatial homogeneity. In a more general setting, defining half dual edges by connecting midpoints of the $4$-frusta with midpoints of boundary hexahedra appears more natural.} 
Notice, that these definitions do not correspond to a barycentric construction. The distance between barycentric vertices would yield the dual edge lengths we obtain below, rescaled by a constant factor. 
For the return probability, such a constant pre-factor can be absorbed into the scale $\tau$ and thus has no physical effect.

In the following, we define the geometric quantities that enter~\eqref{eq:defLaplace}, based on the construction of the dual lattice discussed above. Thereby, we adopt the notation for which, in the $n$th ``slice'', the spins $j_n, j_{n+1}$  refer to the area of the lower and upper cube, respectively, and where $k_n$ labels the spatio-temporal area of boundary $3$-frusta. 

First, we notice that the four-dimensional dihedral angles imply the following restriction on the possible spin values for each slice~\cite{Allen:2022unb},
\begin{equation}\label{eq:spin ineq}
-\frac{1}{\sqrt{2}} \leq \frac{j_n - j_{n+1}}{4 k_n} \leq \frac{1}{\sqrt{2}}.
\end{equation}
posing a stronger condition than closure in~\eqref{eq:cosphi}. Volume and height of a $4$-frustum in the $n$th slice are respectively given by
\begin{equation}\label{eq:4Volume}
V^{(4)}_{n} = \frac{1}{2} k_n (j_n + j_{n+1}) \sqrt{1 - \frac{(j_n - j_{n+1})^2}{8 k_n^2}},
\end{equation}
and
\begin{equation}\label{eq:4Height}
H_{n} =\frac{2 k_n}{\sqrt{j_n} + \sqrt{j_{n+1}}} \sqrt{1 - \frac{(j_n - j_{n+1})^2}{8 k_n^2}}.
\end{equation}
Since the boundary of a $4$-frustum consists of two types of building blocks, cubes and $3$-frusta, we make a distinction in the following. This can be interpreted as separating the cases where $\vn$ and $\vm$ having ``spacelike'' or ``timelike'' separation.\footnote{This designation needs to be used cautiously, since we work in a Riemannian context where there are no notions of causality.}

\paragraph{``Timelike'' dual edges}

\begin{figure}
    \centering
    \includegraphics[width=0.35\textwidth]{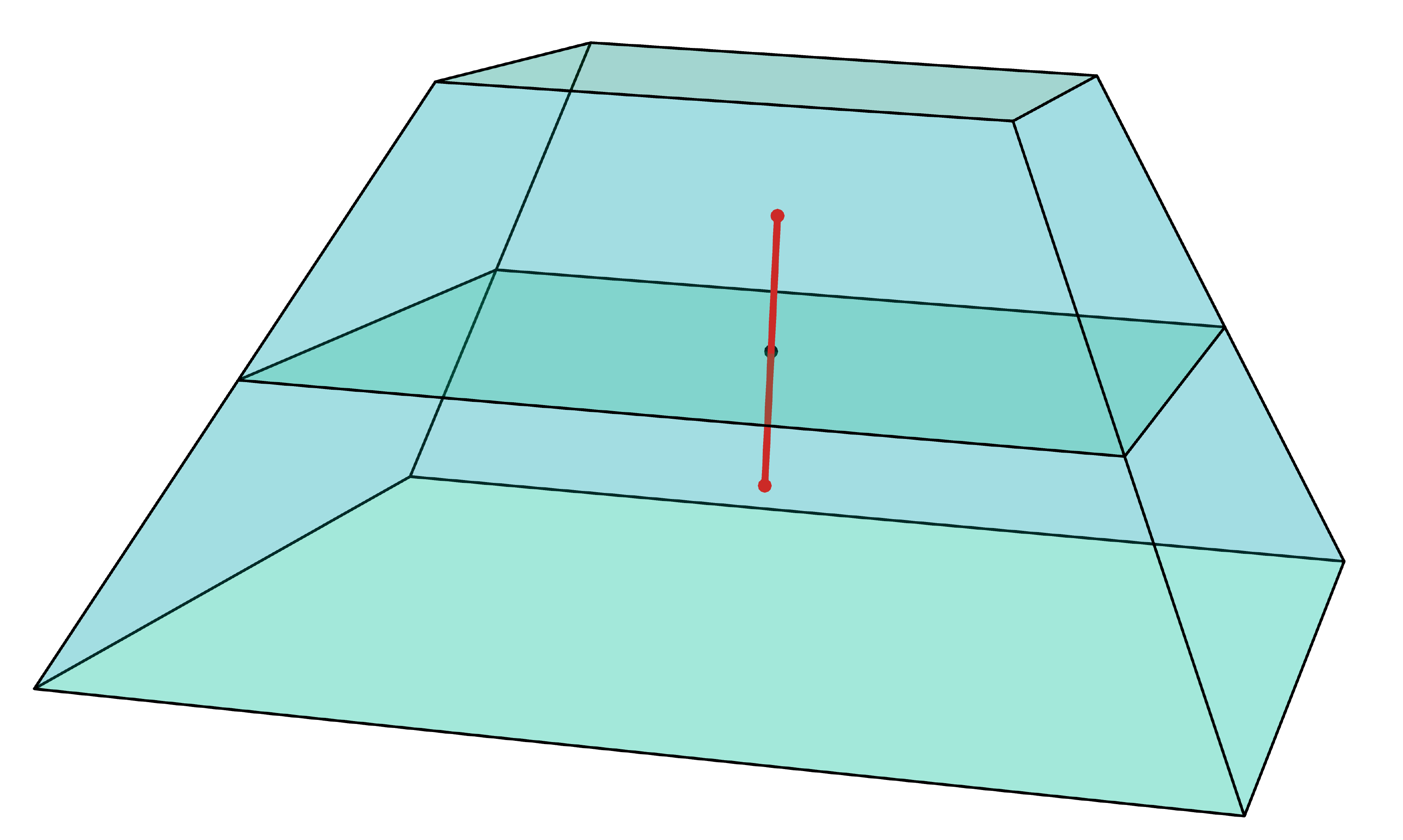}
    \caption{Three-dimensional representation of a ``timelike'' dual edge, drawn in red, connecting the midpoints of two timelike separated frusta. From this representation, the duality of timelike edges and cubes (here squares) is immediate.}
    \label{fig:dualedge_timelike}
\end{figure}

$3$-dimensional volumes between the vertices $\vm$ and $\vn$, which connect the $(n-1)$th and $n$th slice, are simply given by the volume of cubes with area $j_n$
\begin{equation}\label{eq:timelike 3-volume}
V_{\vm \vn}^{(3)} = j_n^{\frac{3}{2}}.
\end{equation}
Following the construction of a dual lattice outlined above, the length of dual edges is given as the half of the sum of heights of ``past'' and ``future'' hyperfrusta
\begin{equation}\label{eq:timelike dual edge length}
\begin{aligned}
l_\star^{\vm \vn} &= \frac{1}{2} (H^{n-1} + H^{n})\\[7pt] 
&=
\frac{ k_{n-1}}{\sqrt{j_{n-1}} + \sqrt{j_{n}}} \sqrt{1 - \frac{(j_{n-1} - j_{n})^2}{8 k_{n-1}^2}} + \frac{ k_n}{\sqrt{j_n} + \sqrt{j_{n+1}}} \sqrt{1 - \frac{(j_n - j_{n+1})^2}{8 k_n^2}}.
\end{aligned}
\end{equation}
A visualization of the dual edge and its length in a $3$-dimensional analogue is given in Fig.~\ref{fig:dualedge_timelike}. 

This defines all the ingredients of the Laplace operator $\Delta_{\vm\vn}$ for neighbouring vertices $\vm,\vn$ which have a ``timelike'' separation, and we use the notation $\Delta_{n-1 n}$  in the following.

\paragraph{``Spacelike'' dual edges}

\begin{figure}
    \centering
    \includegraphics[width=0.4\textwidth]{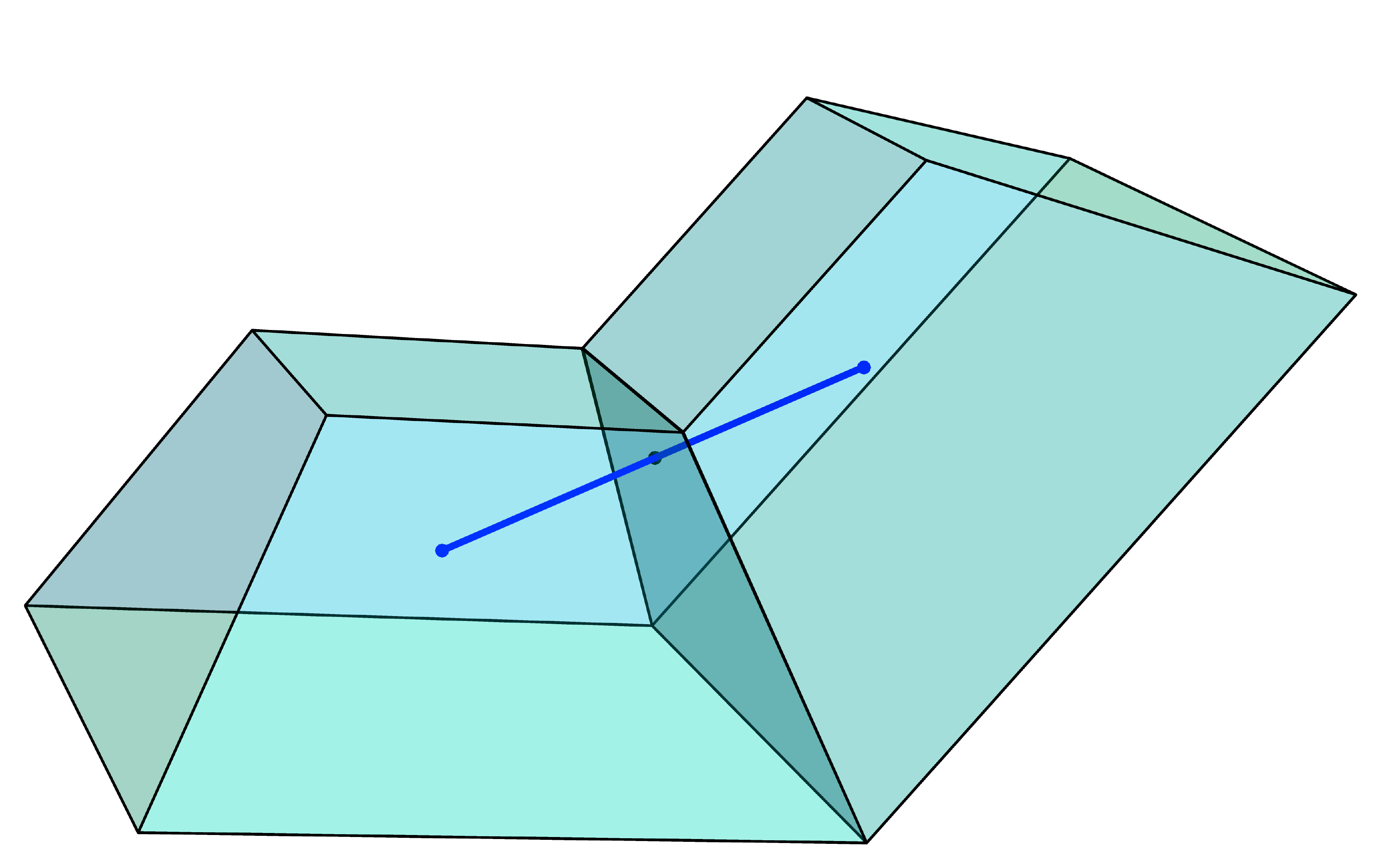}
    \caption{Three-dimensional representation of a ``spacelike'' dual edge, drawn in blue, connecting the midpoints of two spacelike separated frusta. From this representation, the duality of spacelike edges and $3$-frusta (here trapezoids) is immediate.}
    \label{fig:dualedge_spacelike}
\end{figure}

In ``spacelike'' direction, so within a given slice $n$, $4$-frusta are connected with each other via boundary $3$-frusta. The corresponding $3$-volumes are given by
\begin{equation}\label{eq:spacelike 3-volume}
V^{(3)}_{\vm\vn}  = \frac{2k_n(j_n+\sqrt{j_nj_{n+1}}+j_{n+1})}{3(\sqrt{j_{n+1}}+\sqrt{j_n})}\sqrt{1-\frac{(j_{n+1}-j_n)^2}{16k_n^2}}.
\end{equation}
To obtain the length of ``spacelike'' dual edges, for which a visualization is given in~Fig.~\ref{fig:dualedge_spacelike}, it suffices to project the geometry onto the plane spanned by the $t$-axis and the dual edge. 
In this picture, the dual edge connects two glued trapezoids and is orthogonal with respect to the connecting face. For $\Theta$ the dihedral angle between the $(n+1)$-cube and the boundary $3$-frustum, defined as~\cite{Bahr:2017bn}
\begin{equation}
\Theta_n = \arccos\left(\frac{1}{\tan(\phi_n)}\right),
\end{equation}
the dual edge length is then given by
\begin{equation}\label{eq:spacelike dual edge length}
l_\star^{\vm \vn} = \frac{1}{2} \left(\sqrt{j_n} + \sqrt{j_{n+1}}\right) \cos\left(\frac{\pi}{2} - \Theta_n\right).
\end{equation}
We denote the components of the Laplacian $\Delta_{\vn\vm}$ with vertices $\vn$ and $\vm$ having a ``spacelike'' separation by $\Delta_{nn}$.

Note that the definition of the discrete Laplacian requires a semi-classical interpretation of the geometry, as interpreting the spins $j_n,k_n$ as areas is only valid in a semi-classical regime. However, in the computation of the spectral dimension, we assume that the definition $\Delta(j_n,k_n)$ holds for arbitrarily small spins, which can be seen as a continuation of the semi-classical Laplacian and provides \emph{one} possible definition of $\Delta$ in the quantum regime. While there may be many valid definitions of microscopic Laplacians, they must converge to the semi-classical definition provided above.

\subsubsection{$\mathcal{N}$-periodicity}\label{subsec:N-periodicity}

Following the discussion in~\cite{Steinhaus:2018aav}, evaluating the spectral dimension even in the setting of restricted spin foams is challenging because of two main reasons. 
First, the computation of the Laplacian's spectrum for many classical configurations is very costly. 
Second, evaluating the spin foam partition function scales exponentially with the number of lattice sites. 
We elaborate on these numerical challenges in the following.

Let us consider the spectrum of the Laplacian for a given spin configuration first. 
As~\eqref{eq:discrete retprob} shows, the return probability, and therefore the spectral dimension, requires full knowledge of the  spectrum of the discrete Laplacian.\footnote{Although simplifications for certain values of $\tau$ are conceivable, the viability of these would depend on the geometric configurations $(j_n,k_n)$, obscuring a straightforward implementation.} 
Given that the lattice contains $L$ sites in each direction, this amounts to setting up and diagonalizing an $L^4\times L^4$ matrix, the complexity of which grows exponentially in $L$. At the same time, $L$ determines the scale $\tau_{\mathrm{comp}}$ at which boundary effects become dominant. 
To avoid fixed data on the boundary, we henceforth assume periodic boundary conditions, equipping the lattice with a toroidal topology. The resulting compactness leads to a fall-off in spectral dimension for $\tau>\tau_{\mathrm{comp}}$~\cite{Thurigen:2015uc}. Consequently, a large lattice size is required to observe a non-trivial spectral dimension between the smallest lattice scale $\tau\sim j_{\mathrm{min}}$ and $\tau_{\mathrm{comp}}$, which is in clear conflict with computational feasibility.  

Apart from the computational effort to diagonalize the discrete Laplacian for a given spin configuration, the evaluation of the partition function becomes increasingly difficult with larger lattice size. That is because the number of spin configurations scales exponentially with the lattice sites. Due to spatial homogeneity there are in total $2L$ different spins, each of which takes $N = 2(j_\mathrm{max}-j_{\mathrm{min}})+1$ values. Here, $j_{\mathrm{min}}$ and $j_{\mathrm{max}}$ are the lower, respectively upper cut-offs of the spins, the meaning of which we discuss in detail in Sec.~\ref{subsec:1-periodic spectral dimension}. 
Imposing the Riemannian EPRL condition of~\eqref{eq:EPRL condition} and the inequality of~\eqref{eq:spin ineq}, $N$~allowed values for the spins $(j_n,k_n)$ remain. The total number of configurations on the whole lattice is then $N^{2L}$. 
Consequently, the Laplacian needs to be diagonalized for each configuration to compute the return probability. This part is most demanding in numerical resources. Once the return probability for every spin configuration is obtained, the expectation value can be computed rather efficiently. To do so, one writes the amplitudes as a vector and the return probabilities as a matrix, with one index representing the configurations and the other one the diffusion time $\tau$. The expectation value of the return probability is then obtained by a simple matrix-vector multiplication which is a highly optimized numerical operation.

To solve both of the above problems simultaneously, we adopt the assumption of $\mathcal{N}$-periodicity proposed in~\cite{Steinhaus:2018aav}, based on results of~\cite{Sahlmann:2010bb}. The key idea is to assume that the geometry of the spin foam is $\mathcal{N}$-periodic in every direction, meaning that the geometric labels repeat after $\mathcal{N}$ steps in every direction. As a consequence, one can perform a Fourier transform of the discrete Laplacian in the Brillouin zone which amounts to diagonalizing an $\mathcal{N}^4\times\mathcal{N}^4$-matrix. The spectrum of the Laplacian is then given in terms of four momenta~$p_{\mu}$, which are either discrete or continuous, depending on whether the lattice is respectively finite or infinite. The return probability in~\eqref{eq:discrete retprob} is then obtained as a sum, respectively an integral, over the momenta. Also at the level of the amplitudes, periodicity reduces the computational complexity, as the number of independent spin variables reduces to $2\mathcal{N}$. 

\subsubsection{Spectrum of the discrete Laplacian}

The first step in deriving the spectrum of the discrete Laplacian is to introduce the Fourier transform of the test field and to notice that the homogeneity of the geometry effectively reduces the Laplace coefficients $\Delta_{\vn\vm}$ to an $L\times L$ matrix. To make this explicit, let us first introduce some notation: We indicate the ``spatial'' component of $\vn\in\mathbb{Z}_L^4$ as $\mathbf{n}\in\mathbb{Z}_L^3$. 
Taking $\mathcal{N}$-periodicity into account, we indicate a slice 
as $n_0 + z\mathcal{N}$ where $n_0\in\mathbb{Z}_{\mathcal{N}}$ and $z\in\mathbb{Z}_{L/\mathcal{N}}$. Thus, the variable $z$ labels the $\mathcal{N}$-cell in which the slice is located and $n_0$ denotes the $n_0$-th slice within a given $\mathcal{N}$-cell. Using this notation, the scalar test field is written as $\phi_{n_0,\mathbf{n}}^{(z)}$ with $\phi_{n_0+\mathcal{N},\mathbf{n}}^{(z)}\equiv \phi^{(z+1)}_{n_0,\mathbf{n}}$.

To write~\eqref{eq:defLaplace} in Fourier space, we consider a similar ansatz to the proposal of~\cite{Sahlmann:2010bb}, given by
\begin{equation}\label{eq:test field ansatz}
\phi_{n_0,\mathbf{n}}^{(z)} = c_{n_0}\e^{ip_0z}\e^{i\mathbf{n}\cdot\mathbf{p}},
\end{equation}
where $c_{n_0}$ is an $\mathcal{N}$-dimensional vector. A phase with spatial momentum $p_i$ is picked whenever changing  one lattice site in spatial direction $i$. In contrast, a phase with temporal momentum $p_0$ is only picked up when changing to another $\mathcal{N}$-cell. This pattern of picking up phases will be reflected in Eqs.~(\ref{eq:Deltac1}),~(\ref{eq:Deltac2}) and~(\ref{eq:Deltac3}). For a finite lattice of size $L^4$ with periodic boundary conditions, the momenta $p_{\mu}$ take values
\begin{equation}\label{eq:discrete momenta}
p_{\mu} = \frac{2\pi}{L} k_{\mu},\quad k_{\mu}\in\mathbb{Z}_L.
\end{equation}
In the limit $L\rightarrow\infty$, the momenta lie in the Brillouin zone $p_\mu\in [-\pi,\pi]$. Inserting the ansatz of~\eqref{eq:test field ansatz} into~\eqref{eq:defLaplace}, we obtain for $n \notin \{0,\mathcal{N}-1\}$
\begin{equation}\label{eq:Deltac1}
-(\Delta c)_n = -\left[\Delta_{nn+1}(c_n-c_{n+1}) + \Delta_{nn-1}(c_n - c_{n-1}) + 2\Delta_{nn} c_n\sum_{i = 1}^3(1-\cos(p_i)) \right].
\end{equation}
$\Delta_{nn+1}$ are the components of the Laplace operator on dual edges connecting slices $n$ and $n+1$, defined by Eqs.~(\ref{eq:4Volume}),~(\ref{eq:timelike 3-volume}) and~(\ref{eq:timelike dual edge length}). $\Delta_{nn}$ are the components of the Laplace operator within a slice, defined by Eqs.~(\ref{eq:4Volume}),~(\ref{eq:spacelike 3-volume}) and ~(\ref{eq:spacelike dual edge length}), associated to ``spacelike'' separated vertices. 
Due to spatial homogeneity, $\Delta_{nn}$ is independent of the spatial direction and therefore factorizes from the spatial momenta. 
For the slices $n = 0, \mathcal{N}-1$ connecting neighbouring $\mathcal{N}$-cells, exponential factors of $\e^{\pm i p_0}$ are picked up
\begin{align}
-(\Delta c)_0               &= -\left[w_{0}(c_0-c_{1})+w_{\mathcal{N}-1}(c_0-c_{\mathcal{N}-1}e^{-ip_0})+W_0c_0\right],\label{eq:Deltac2}\\
-(\Delta c)_{\mathcal{N}-1} &= -\left[w_{\mathcal{N}-1}(c_{\mathcal{N}-1}-c_{0}e^{ip_0})+w_{\mathcal{N}-2}(c_{\mathcal{N}-1}-c_{\mathcal{N}-2})+W_{\mathcal{N}-1}c_{\mathcal{N}-1}\right],\label{eq:Deltac3}
\end{align}
where for brevity, we introduced the notation 
\begin{equation}\label{eq:w and W}
w_n \defeq \Delta_{nn+1}
\quad , \quad 
W_n \defeq 2\Delta_{nn}\sum_i (1-\cos(p_i)) \, .
\end{equation}
Exploiting $\mathcal{N}$-periodicity and spatial homogeneity, we observe that the action of the Laplace operator reduces to a vector equation in $\mathcal{N}$ dimensions. Altogether, Eqs.~(\ref{eq:Deltac1}), (\ref{eq:Deltac2}) and~(\ref{eq:Deltac3}) are captured by
\begin{equation}
-(\Delta c)_m = -\sum_{n = 0}^{\mathcal{N}-1} M_{mn}c_n,
\end{equation}
with the matrix $M$ being defined as
\begin{equation}\label{eq:Laplace_momentum}
M \defeq 
\begin{pmatrix}w_{\mathcal{N}-1}+w_0 +W_0 & -w_0 & &\dots &  -w_{\mathcal{N}-1}e^{-ip_0} \\
-w_0 & w_0 + w_1 + W_1  &  &&\\
\vdots & & &  \ddots && \\
 -w_{\mathcal{N}-1}e^{ip_0}  & \dots & &-w_{\mathcal{N}-2} & w_{\mathcal{N}-2} + w_{\mathcal{N}-1} + W_{\mathcal{N}-1}
\end{pmatrix}.
\end{equation}
For a given spin configuration, the spectrum in momentum space is then given by the eigenvalues of the matrix $M(p_0,p_1,p_2,p_3)$, which must be computed for every combination of momenta. 

\subsubsection{Expectation value of the return probability}\label{subsec:Expectation value of the return probability}

Following~\eqref{eq:discrete retprob}, the return probability of a given spin configuration is obtained by either summing or integrating over the spectrum of the Laplacian. Let $\omega_i(\{p_{\mu}\})$ denote the $\mathcal{N}$ momentum-dependent eigenvalues of the Laplace matrix in~\eqref{eq:Laplace_momentum}. Then, the $\mathcal{N}$-periodic return probability $P_{\mathcal{N}}(\tau)$ on a lattice of length $L$ is given by
\begin{equation}\label{eq:return_prob_N L finite}
P_{\mathcal{N}}(\tau) = \sum_{i=1}^{\mathcal{N}}\prod_{\mu=0}^3\sum_{k_{\mu}\in\mathbb{Z}_L}\e^{-\tau\omega_i(\{p_{\mu}\})},
\end{equation}
where the integers $k_{\mu}$ and the lattice momenta $p_{\mu}$ are related by~\eqref{eq:discrete momenta}. In the limit $L\rightarrow\infty$, the summation is replaced by an integration over the Brillouin zone, yielding
\begin{equation} \label{eq:return_prob_N L infinite}
    P_\mathcal{N}(\tau) = \sum_{i = 1}^{\mathcal{N}} \prod_{\mu=0}^3\int_{-\pi}^{\pi}\mathrm{d}{p_\mu} \; e^{-\tau \omega_i(\{p_\mu\})} \quad .
\end{equation}
Notice, that the eigenvalues $\omega_i(\{p_{\mu}\})$ depend on the geometry of the entire lattice, turning the return probability into a highly non-local quantity. 
These different eigenvalues are usually called branches, e.g. the ``acustic'' branch in which $\omega_i \rightarrow{0}$ for all $p_\mu \rightarrow{0}$. 

Already for the $2$-periodic case, the summation, respectively integration in~\eqref{eq:return_prob_N L finite} and~\eqref{eq:return_prob_N L infinite} cannot be performed individually, since the eigenvalues $\omega_i(\{p_{\mu}\})$ do not split into a sum of terms for each momentum component $p_{\mu}$. This severely affects the computational effort to compute the return probability at a given configuration. 
Summation is implemented by $L^4$ nested \texttt{for}-loops. For the numerical integration on the other hand, one cannot perform a product of one-dimensional integrals but needs to consider instead a single $4$-dimensional integration. Using the \texttt{Cuba}-package\footnote{See \url{http://www.feynarts.de/cuba/} for the original version and \url{https://github.com/giordano/Cuba.jl} for the \href{https://arxiv.org/pdf/Julia}{\texttt{Julia}} package.}, such higher-dimensional integrations are possible but more costly and issues of convergence due to an exponential decay are more likely to arise.

The expectation value of the return probability for a finite $\mathcal{N}$-periodic lattice in this setting is
\begin{equation}\label{eq:return_prob_exp pre}
 \langle P_{\mathcal{N}}(\tau) \rangle = \frac{1}{Z} \sum_{\{j_i,k_i\}}\left(\prod_{n=1}^{\mathcal{N}}\hat{\mathcal{A}}(j_n,j_{n+1},k_n)^{L^3}\right)^{L/\mathcal{N}}P_{\mathcal{N}}(\tau,\{j_i,k_i\}),
\end{equation}
where the spin labels are in the range $j_{\min} \leq j_i,k_i \leq j_{\max}$.\footnote{We discuss the cut-off dependence of our results in Sec.~\ref{sec:results}.} Note that a priori, $\mathcal{A}$ denotes the full quantum amplitude of $4$-frusta. However, as we argue in Sec.~\ref{subsec:1-periodic spectral dimension}, the asymptotics of the dressed vertex amplitude already captures the behaviour of the spectral dimension sufficiently for $\tau\gtrsim 10^2$. 

As argued in Sec.~\ref{subsec:N-periodicity}, the lattice size $L$ is required to be large in order to resolve a non-trivial spectral dimension between the minimal scale $\tau\sim j_{\mathrm{min}}$ and the compactness scale $\tau_{\mathrm{comp}}\sim L$. Consequently, the amplitudes enter~\eqref{eq:return_prob_exp pre} with large powers, requiring the utilization of arbitrary precision floating point numbers. 
Arithmetics with this format is costly in memory and computation time. To circumvent this issue, we truncate the total number of amplitudes by assuming that the amplitudes of a single $\mathcal{N}$-cell sufficiently capture the relevant information of the whole spin foam. Within this approximation, the expectation value of the return probability is finally given by
\begin{equation} \label{eq:return_prob_exp}
    \langle P_{\mathcal{N}}(\tau) \rangle =  \frac{1}{Z} \sum_{\{j_i,k_i\}}\prod_{n=1}^{\mathcal{N}} \hat{\mathcal{A}}(j_n,j_{n+1},k_n)^{\mathcal{N}^3} \; P_{\mathcal{N}}(\tau,\{j_i,k_i\}).
\end{equation}
In Sec.~\ref{subsec:The thermodynamic limit} we discuss the limit of $\mathcal{N}\rightarrow\infty$, corresponding to an infinite lattice with infinitely many degrees of freedom.

To summarize, we compute the return probability for the spin foam quantum space-time by summing up the return probability for all possible spin foam frusta geometries, weighted by the quantum or semi-classical spin foam amplitudes $\hat{\mathcal{A}}$ for various diffusion times $\tau$. From this expectation value, we derive the spectral dimension $\Ds$ as
\begin{equation}\label{eq:expval specdim}
    D^{\mathcal{N}}_S := - 2 \frac{\partial \ln \langle P_{\mathcal{N}}(\tau) \rangle}{\partial \ln \tau}.
\end{equation}
Notice, that we do not consider $\Ds$ as an observable itself. Rather, the ``quantum spectral dimension'' characterizes the \emph{scaling} of a quantum expectation value, that is here the heat trace.

\section{1-periodic quantum amplitudes from extrapolation}\label{sec:Quantum amplitudes from extrapolation}

The quantum amplitudes of frusta spin foams, introduced in Sec.~\ref{subsec:Spin foam frusta}, are the necessary ingredients to determine expectation values. 
In essence, there are two ways to compute these. 
First, the intertwiners in~\eqref{eq:intertwiner} can be explicitly computed via $\SU$-integrations. Then, as the formulas in~\eqref{eq:general EAmp} and~\eqref{eq:general vampl} suggest, quantum amplitudes are straightforwardly computed by contracting $3$-frustum and $3$-cube intertwiners accordingly.  Notice, that due to the higher valence of the intertwiners, the frustum computation is more costly than an analogous computation on a triangulation. 
Another conceivable strategy is to derive a spin network expression of the vertex amplitude via $\SU$-recoupling theory, in the spirit of \cite{Dona:2017dvf}. 
The result is then to be contracted with the overlap of intertwiners in the spin network and coherent state bases. Still in this case, the computation is more costly than in a triangulated setting. 

Choosing to follow the first strategy, we face the following numerical challenge. Since the range of magnetic indices grows with the spin size, numerical contractions of intertwiners either demand increasing memory when using the \texttt{TensorOperations}\footnote{See \url{https://jutho.github.io/TensorOperations.jl/stable/} package for a documentation.} package, 
or increasing time  using nested \texttt{for}-loops. 
This sets numerical limits to the computation of the vertex amplitude already at low spins $j\sim 4$~\cite{Allen:2022unb}, in conflict with resolving a non-trivial flow of the spectral dimension that requires $\frac{j_{\mathrm{max}}}{j_{\mathrm{min}}}\gg 1$, as argued below in Sec.~\ref{subsec:1-periodic spectral dimension}. Consequently, the number of configurations for which the exact quantum amplitudes are available, is insufficient to compute $\langle P(\tau)\rangle$. 

While resorting to semi-classical amplitudes as done in~\cite{Steinhaus:2018aav} is a sensible choice, the approximation deviates significantly for small spins $j\lesssim 10$~\cite{Allen:2022unb}. 
In order to find amplitudes which are closer to the actual quantum amplitudes for small spins, while still showing a convergence to the semi-classical amplitudes in the limit of large spins, we present in this section a method to extrapolate quantum amplitudes in the simplest case of periodicity $\mathcal{N} = 1$ using the \texttt{FindFit}-function in \texttt{Mathematica}~\cite{Mathematica}. The restriction to $1$-periodic spin foam frusta reduces the model to a specific subclass of spin foam cuboids which satisfy geometricity~\cite{Bahr:2016co,Steinhaus:2018aav,Allen:2022unb}. 
A particular feature of cuboidal amplitudes is that they exhibit a pure scaling behaviour without any oscillations, therefore simplifying the analysis drastically.

As mentioned above, the amplitudes of the Riemannian EPRL model factorize into $\SU$-amplitudes for $\bi < 1$. 
Thus, we first present an extrapolation method for the $\SU$-vertex and edge amplitude in Sec.~\ref{subsec:SU2-quantum amplitudes}. 
Thereafter, we combine these results of that to set up an extrapolation of the dressed quantum vertex amplitude in Sec.~\ref{subsec:Dressed quantum vertex amplitude}, which will be used later on to compute the spectral dimension.

\subsection{SU(2)-quantum amplitudes}\label{subsec:SU2-quantum amplitudes}

One-periodic spin foam frusta are characterized by the amplitudes of a single $4$-frustum, the initial and final $3$-cubes of which carry the same spin $j$. The connecting $3$-frusta are uniquely characterized by the area of the trapezoidal faces, labelled by $k$. 
Hence, the pair $(j,k)$ fully determines the geometry. 
In this case, the numerical limitations allow a computation of the exact quantum vertex and edge amplitudes for spins in the range $\frac{1}{2}\leq j,k\leq 4$~\cite{Allen:2022unb}, thus yielding $64$ data points.

Best suited for fitting, we compute the relative error
\begin{equation}\label{eq:relerr}
\varepsilon_i(j,k)\defeq \frac{\vert\mathcal{A}_i^{\mathrm{sc}} - \mathcal{A}_i^{\mathrm{qu}}\vert}{\mathcal{A}_i^{\mathrm{sc}}}
\end{equation}
with respect to the semi-classical amplitudes, which are given in~\eqref{eq:SU2Vamp} and~\eqref{eq:SU2EAmp}. Here, $i$ indicates the relative error of vertex $(i=v)$ and edge $(i=e)$, respectively. Within a \textit{homogeneous limit} $(\lambda j,\lambda k)$ with $\lambda\rightarrow\infty$, the relative error $\varepsilon_i$ converges to zero. Importantly, when keeping one variable fixed and only scaling the other one, we expect $\varepsilon$ to converge to a small but non-zero value.\footnote{Due to the restrictive nature of the frusta model in the $1$-periodic case, coupling rules are always satisfied. Furthermore, the boundary states are a function of the spins and at a critical point by construction.} Still, this convergence will ensure that we obtain the convergence to the semi-classical amplitude in a homogeneous limit, as we demonstrate down below.

Now, for every $j$ fixed, we find a fit of the relative error $\varepsilon_i$ in $k$-direction. The resulting fitting function is then used to extrapolate $\varepsilon_i$ up to a chosen value of $k = 5000$ for every $j$. With the results, we proceed similarly to extrapolate in $j$-direction for fixed $k$. Ultimately, we obtain a $10000\times 10000$ matrix for the relative error $\varepsilon_i(j,k)$, from which the amplitudes can be derived via~\eqref{eq:relerr}. Since the fitting procedure is only performed with one-dimensional functions, where either $k$ or $j$ is assumed to be fixed, it is a non-trivial consistency check to show the convergence of the quantum to the semi-classical amplitude in a homogeneous limit $(\lambda j, \lambda k)$ with $\lambda\rightarrow\infty$. In the following, we show that this is indeed the case.

\subsubsection{Extrapolated vertex amplitude}\label{subsubsec:Extrapolated vertex amplitude}

\begin{figure}
    \centering
    \begin{subfigure}{0.5\textwidth}
    \centering
    \includegraphics[width = \linewidth]{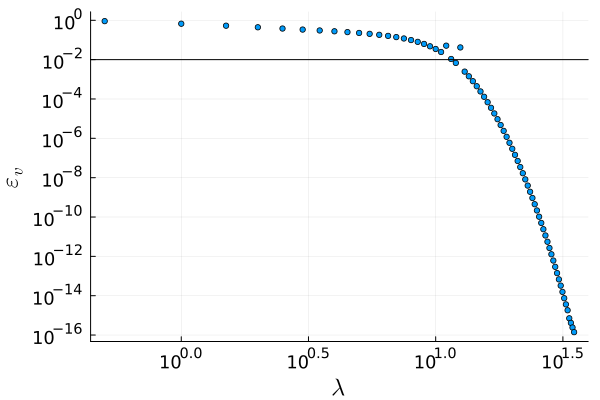}
    \end{subfigure}%
    \begin{subfigure}{0.5\textwidth}
    \centering
    \includegraphics[width = \linewidth]{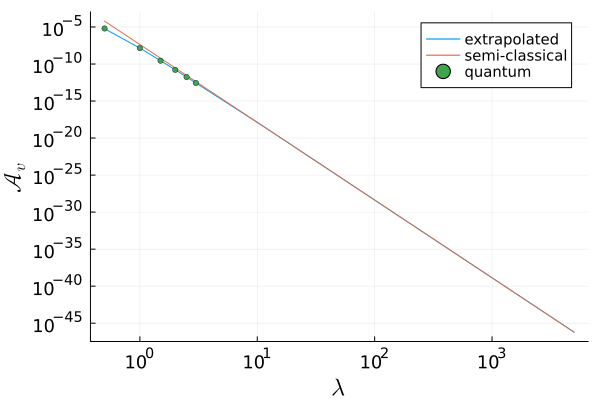}
    \end{subfigure}
    \caption{Left: Relative error $\varepsilon_v$ between extrapolated and semi-classical $\SU$-vertex amplitude.  Right: A direct comparison of the diagonal entries of the extrapolated, semi-classical and quantum $\SU$-vertex amplitude. We remind the reader that there are only few quantum data points depicted since we are considering the diagonal case with $j = k$. For the fitting procedure, $64$ data points have been utilized.}
    \label{fig:SU2_Vamp_relerr}
\end{figure}

On the left panel of Fig.~\ref{fig:SU2_Vamp_relerr}, we show that the relative error $\varepsilon_v(j,k)$ converges to zero in the limit where $(j,k) = (\lambda,\lambda)$ are homogeneously scaled up. 
For values of $\lambda \geq 13$, the relative error is smaller than $1\%$, indicated by the black line in Fig.~\ref{fig:SU2_Vamp_relerr}. 
Notice, that we have plotted $\varepsilon_v$ only up to $\lambda = 35$ since, from this value on, the difference between both amplitudes is already so small that it cannot be resolved with usual \texttt{double precision}-numbers. Notice that the two outlying points are a consequence of the fitting procedure and occur, when there is a sign change in the fitting parameters. A side-by-side comparison of the extrapolated and semi-classical amplitude is presented on the right panel of Fig.~\ref{fig:SU2_Vamp_relerr}. 
Clearly, the semi-classical amplitude is an over-estimate of the extrapolated quantum amplitude for small spins 
which is alignment with the results of~\cite{Steinhaus:2018aav}.

\begin{figure}
    \centering
    \includegraphics[width = 0.5\textwidth]{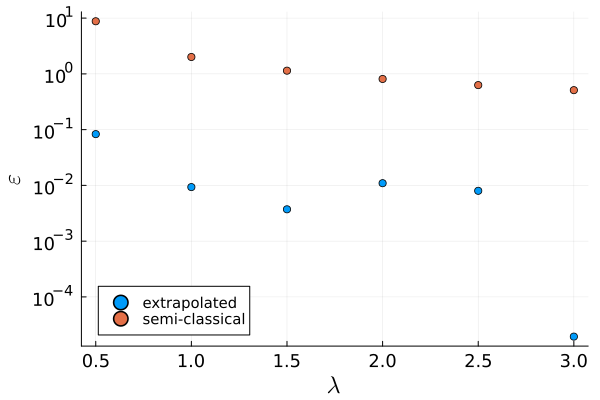}
    \caption{Relative error of the extrapolated and semi-classical vertex amplitude with respect to the exact $\SU$-quantum vertex amplitude.}
    \label{fig:SU2_Vamp_relerr_exact}
\end{figure}

These results indicate that the extrapolated amplitude shows the correct behaviour for large spins. Also for small spins, the extrapolated amplitude provides a better approximation of the actual quantum amplitude than the semi-classical. To show this explicitly, we present in Fig.~\ref{fig:SU2_Vamp_relerr_exact} a plot of the relative error between the quantum amplitude and the extrapolated, respectively the semi-classical amplitude. 
As visible, the extrapolated amplitude is closer to the quantum amplitude by several orders of magnitude, measured by the relative error.

\subsubsection{Extrapolation edge amplitudes}\label{subsubsec:Extrapolation edge amplitudes}

\begin{figure}
    \centering
    \begin{subfigure}{0.5\textwidth}
        \includegraphics[width = \linewidth]{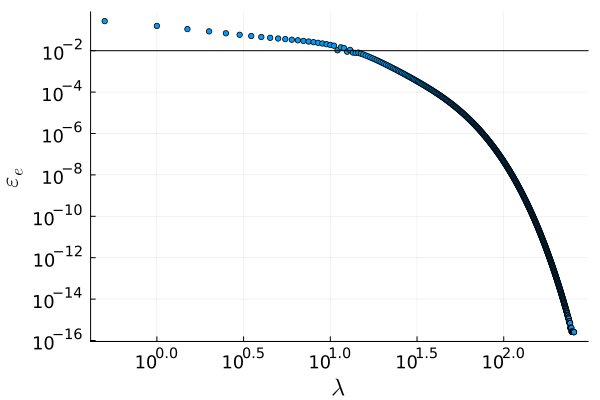}
    \end{subfigure}%
    \begin{subfigure}{0.5\textwidth}
        \includegraphics[width = \linewidth]{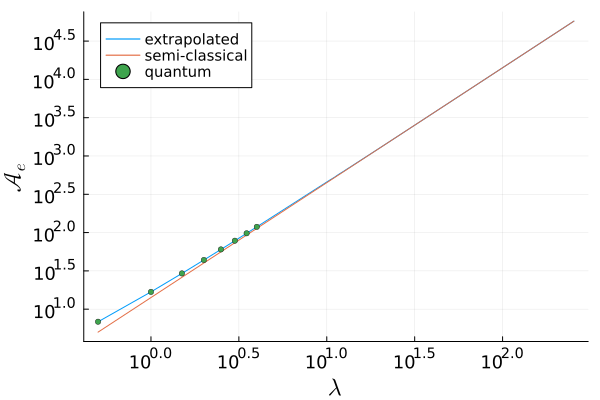}
    \end{subfigure}
    \caption{Left: Relative error $\varepsilon_e$ between extrapolated and semi-classical $\SU$-edge amplitude. Right: A direct comparison of the diagonal entries of the extrapolated, semi-classical and quantum $\SU$-edge amplitude.}
    \label{fig:SU2 Eamp relerr}
\end{figure}

The relative error $\varepsilon_e$ between extrapolated quantum and semi-classical $\SU$-edge amplitudes is depicted on the left panel of Fig.~\ref{fig:SU2 Eamp relerr}. Similar to the vertex amplitude, we observe a rather fast convergence with $\varepsilon_e <1\%$ at spins $\lambda > 13$. The plot is only drawn for $\lambda < 254$, since for larger spins, the difference between extrapolated and semi-classical amplitude cannot be resolved.  On the right panel of Fig.~\ref{fig:SU2 Eamp relerr}, we see that the semi-classical edge amplitude under-estimates the extrapolated amplitude for small spins. That is, because the edge amplitudes in~\eqref{eq:general EAmp} are defined as the inverse of the intertwiner norm.

\begin{figure}
    \centering
    \includegraphics[width=0.5\textwidth]{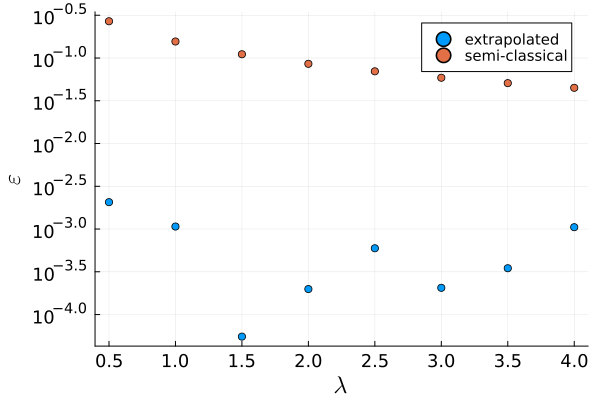}
    \caption{Relative relative error of extrapolated and semi-classical edge amplitude with respect to the exact $\SU$-quantum edge amplitude.}
    \label{fig:SU2_Eamp_relerr_exact}
\end{figure}

Following this consistency check, we show in Fig.~\ref{fig:SU2_Eamp_relerr_exact}, that the extrapolated edge amplitude serves as a better estimate for the exact quantum amplitude at low spins. Indeed, the relative error between extrapolated and quantum amplitudes is below $1\%$, while it is of the order $1$ for the semi-classical amplitude.

To summarize, the extrapolation of quantum amplitudes in the simplified case of $1$-periodic frusta spin foams provides a good approximation to the exact quantum amplitudes. For small spins in particular, the extrapolated amplitudes do not deviate as much from the quantum amplitudes in comparison to the semi-classical approximation. We emphasize that the whole procedure hinges on the assumption of $1$-periodicity. For higher periodicities $\mathcal{N} > 1$, the vertex amplitudes oscillate. Since only few data points of the quantum amplitudes are available, it is to be expected that a simple fitting procedure cannot capture the behaviour of the amplitudes sufficiently. To do so, one would probably have to assume the semi-classical oscillation behaviour which, for small spins, yields the wrong phase~\cite{Allen:2022unb}. Furthermore, this assumption might not be valid for a non-homogeneous scaling, where all but one spin are kept fixed. 

\subsection{Dressed quantum vertex amplitude}\label{subsec:Dressed quantum vertex amplitude}

In this section, we utilize the extrapolated $\SU$-amplitudes of the previous section to compute an approximation of the quantum dressed vertex amplitude $\hat{\mathcal{A}}$ according to~\eqref{eq:dVAmp}. 
Depending on the Barbero-Immirzi parameter $\bi$, the $\SU$-spins $j$ are mapped to different $\Spin$-representations $(j^+,j^-)$. Consequently, the components $\hat{\mathcal{A}}(j,j,k)$ are composed out of different components of $\SU$-amplitudes $\mathcal{A}_i(j^\pm,j^\pm,k^\pm)$, where $i$ indicates either face, edge or vertex. Here and in the following section, we choose the least excluding value of $\bi = \frac{1}{3}$ (see Fig.~\ref{fig:EPRL condition} for details) unless indicated otherwise. In this case $j\in\frac{3}{2}\mathbb{N}$ is mapped to $(\frac{2}{3}j,\frac{1}{3}j)\in\mathbb{N}\times\frac{1}{2}\mathbb{N}$, satisfying the EPRL-condition in~\eqref{eq:EPRL condition}. 
We first analyse the convergence of the resulting amplitude to the semi-classical approximation in the limit $\lambda\rightarrow\infty$. Thereafter, we show that the extrapolated dressed vertex amplitude provides a better approximation than the semi-classical amplitude, which is to be expected from the results of Sec.~\ref{subsec:SU2-quantum amplitudes}. 
We conclude by presenting the effective scaling of the extrapolated dressed vertex amplitude, which is an important factor for the spectral dimension~\cite{Steinhaus:2018aav}.

\begin{figure}
    \centering
    \includegraphics[width = 0.5\textwidth]{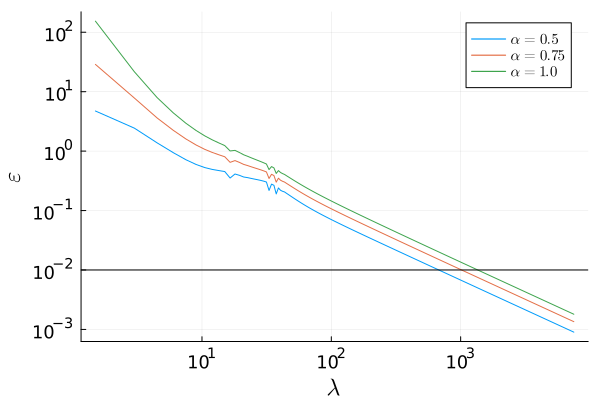}
    \caption{Relative error between extrapolated and semi-classical dressed amplitude, depending on $\alpha \in\{0.5,\; 0.75,\; 1.0\}$.}
    \label{fig:dVamp_relerr}
\end{figure}

While the semi-classical $\SU$-amplitudes provide a good approximation already at small spins of $\lambda\gtrsim 15$, the semi-classical $\Spin$-amplitude is expected to have a slower convergence because of two reasons. First, since various powers of edge and vertex $\SU$-amplitudes enter the dressed amplitude, defined in~\eqref{eq:dVAmp}, the relative errors $\varepsilon_e$ and $\varepsilon_v$ add up. Second, the face amplitude, given in~\eqref{eq:face amplitude}, introduces a third deviation $\varepsilon_f(\alpha)$, which depends on the $\alpha$-parameter. 
By definition of the semi-classical and quantum face amplitude, a larger $\alpha$ leads to a stronger deviation. 
The plots of Fig.~\ref{fig:dVamp_relerr} support these arguments, where we consider the relative error between extrapolated and semi-classical amplitude for $\alpha\in\{0.5,\; 0.75,\; 1.0\}$. 
We find a good agreement of the two amplitudes for $\varepsilon<1\%$, which is the case for spins $\lambda\sim 1000$. 

\begin{figure}
    \begin{subfigure}{0.5\textwidth}
        \includegraphics[width=\linewidth]{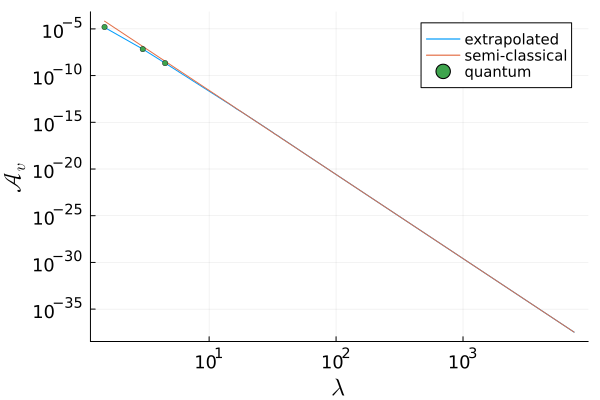}
        \caption{$\alpha = 0$}
        \label{subfig:dVamp a}
    \end{subfigure}%
    \begin{subfigure}{0.5\textwidth}
        \includegraphics[width=\linewidth]{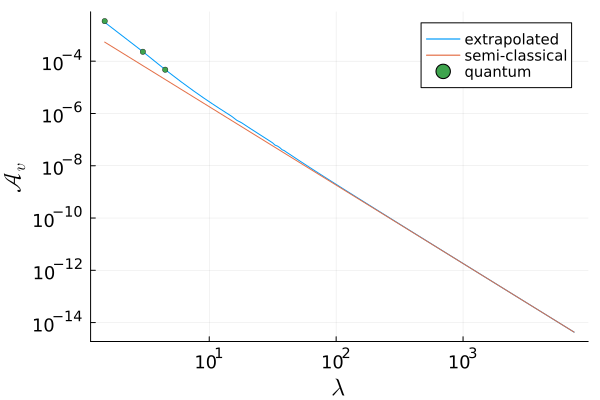}  
        \caption{$\alpha = 0.5$}
        \label{subfig:dVamp b}
    \end{subfigure}\\
    \begin{subfigure}{0.5\textwidth}
        \includegraphics[width=\linewidth]{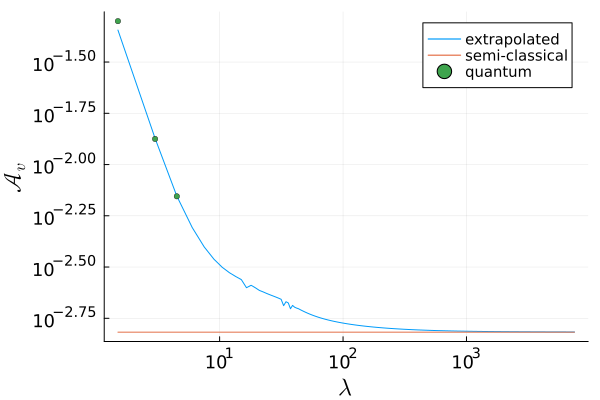}
        \caption{$\alpha = 0.75$}
        \label{subfig:dVamp c}
    \end{subfigure}%
    \begin{subfigure}{0.5\textwidth}
        \includegraphics[width=\linewidth]{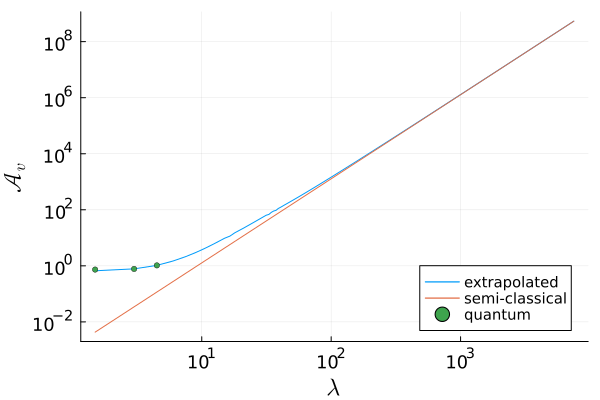}
        \caption{$\alpha = 1$}
        \label{subfig:dVamp d}
    \end{subfigure}
    \caption{Comparison of extrapolated and semi-classical dressed vertex amplitudes for various values of $\alpha$.}
    \label{fig:dVamp}
\end{figure}

To get a more detailed picture of its behaviour and its dependence on $\alpha$,  Fig.~\ref{fig:dVamp} shows the extrapolated dressed amplitude in comparison to the semi-classical approximation for $\alpha\in\{0,\; 0.5,\; 0.75,\; 1.0\}$. Fig.~\ref{subfig:dVamp a} depicts the dressed amplitudes for a trivial face amplitude. Here, the semi-classical amplitude is an over-estimate for small spins. Increasing $\alpha$ to $0.5$, which is a value of interest in Sec.~\ref{sec:results}, the extrapolated amplitude is in fact larger than the semi-classical one, as Fig.~\ref{subfig:dVamp b} shows. A short numerical check reveals that the transition from the amplitude being smaller to being larger than the semi-classical approximation takes place for $\alpha\lesssim 0.24$. At the value $\alpha = 0.75$, the semi-classical $1$-periodic amplitude becomes scale invariant. Following Fig.~\ref{subfig:dVamp c}, the extrapolated dressed amplitude reaches the scale-invariant behaviour asymptotically from above. Above scale-invariance the dressed amplitude diverges in the limit $\lambda\rightarrow\infty$. Visualized in Fig.~\ref{subfig:dVamp d}, the extrapolated amplitude is larger than the semi-classical amplitude with a large relative error at small spins.   

\begin{figure}
    \centering
    \includegraphics[width=0.5\textwidth]{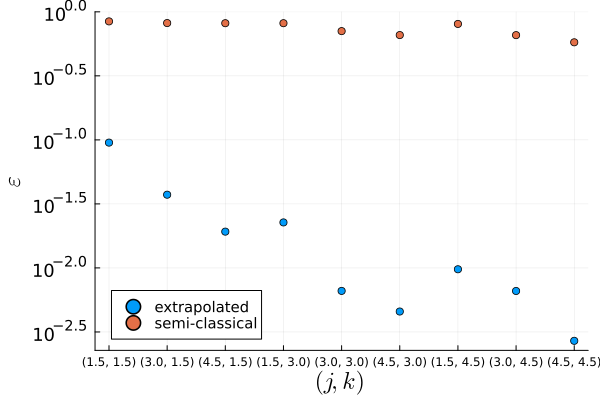}
    \caption{Relative error of extrapolated (blue) and semi-classical dressed vertex amplitudes with respect to the exact quantum dressed vertex amplitude. The values of $\varepsilon$ are plotted against the tuple of spins $(j,k)$ for which the exact quantum amplitude can be computed.}
    \label{fig:dVamp_relerr_exact}
\end{figure}

Only few components of the dressed $\Spin$-amplitude can be computed, since the exact $\SU$-quantum vertex amplitude is only available for spins $j,k\leq 3$. Consequently, we can compute $3\times 3$ entries of the exact dressed amplitude, listed in Fig.~\ref{fig:dVamp_relerr_exact}. There, the relative errors of extrapolated and semi-classical dressed amplitudes with respect to the exact quantum dressed amplitude are depicted for $\alpha = 0.5$. As the plots indicate, the extrapolated dressed amplitude provides a better approximation to the quantum amplitudes for all tuples $(j,k)$.

\begin{figure}
    \centering
    \includegraphics[width=0.5\textwidth]{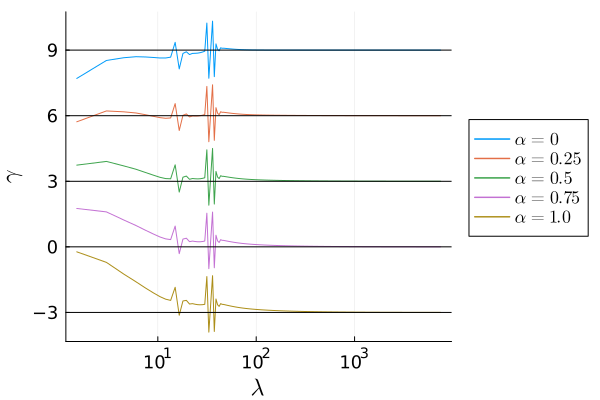}
    \caption{Effective scaling behaviour $\sfs$ of the extrapolated dressed amplitude for $\alpha\in\{0,\; 0.25,\; 0.5,\; 0.75,\; 1\}$. Semi-classical constant values are indicated by the black lines. Fluctuations in the range $10^1 < \lambda < 10^2$ are due to numerical imprecisions in the fitting procedure, where the fitting parameters change sign.}
    \label{fig:dVAmp_scaling}
\end{figure}

As a last point, we compute the effective scaling $\sfs$ of the extrapolated dressed amplitude, defined as
\begin{equation}
\sfs = -\frac{\lambda}{\hat{\mathcal{A}}}\frac{\partial\hat{\mathcal{A}}}{\partial\lambda}.
\end{equation}
Demonstrated for cuboids in~\cite{Steinhaus:2018aav}, the scaling of the amplitudes is a quintessential factor for the behaviour of the spectral dimension. As we show later on in Sec.~\ref{subsec:1-periodic spectral dimension} and Sec.~\ref{subsec:Analytical estimate of the spectral dimension}, this holds also true for semi-classical as well as for extrapolated frusta amplitudes. A change in scaling directly translates to a change of the spectral dimension. Our results for the effective scalings with $\alpha\in\{0,\; 0.25,\; 0.5,\; 0.75,\; 1\}$ are presented in Fig.~\ref{fig:dVAmp_scaling}. For comparison the semi-classical scaling exponents are depicted as black horizontal lines. These are constant because semi-classical amplitudes exhibit a simple polynomial decay. Since the extrapolated amplitudes approach the semi-classical limit from above for $\alpha > 0.24$, the corresponding effective scaling is below the constant semi-classical value. As expected from the previous arguments given above, exactly the opposite behaviour can be observed for $\alpha = 0$. 

\section{Spectral dimension from spin foam frusta}\label{sec:results}


With the ingredients introduced in Sec.~\ref{sec:Spin foam hyperfrusta and the spectral dimension} and Sec.~\ref{sec:Quantum amplitudes from extrapolation}, we present in this section our numerical and analytical studies of the quantum spectral dimension for spin foam frusta. In Sec.~\ref{subsec:1-periodic spectral dimension}, we present numerical results using quantum and semi-classical $1$-periodic amplitudes. Proceeding with semi-classical amplitudes, we introduce a cosmological constant in Sec.~\ref{subsec:Cosmological constant} and thereafter generalize to $2$-periodic configurations in Sec.~\ref{subsec:2-periodic spectral dimension}. We close this section with an analytical estimate of the spectral dimension in Sec.~\ref{subsec:Analytical estimate of the spectral dimension}.

\subsection{1-periodic spectral dimension}\label{subsec:1-periodic spectral dimension}

Assuming that the geometry of frusta is $1$-periodic, the momentum space Laplace operator, defined in~\eqref{eq:Laplace_momentum}, reduces to a single component matrix
\begin{equation}
M = 2w_0(1-\cos(p_0)) + W_0(p_1,p_2,p_3) = \sum_{\mu=0}^3\omega^{(\mu)}(p_{\mu}),
\end{equation}
which decomposes into components $\omega^{(\mu)}(p_\mu)$. Consequently, the classical return probability on an infinite lattice, given in~\eqref{eq:return_prob_N L infinite}, can be written as a product of integrals
\begin{equation}\label{eq:P1}
P_1(\tau) = \prod_{\mu = 0}^3\int_{[-\pi,\pi]}\d{p_{\mu}}\e^{-\tau\omega^{(\mu)}(p_{\mu})} = \left(\int\d{p_0}\e^{-\tau\omega^{(0)}(p_0)}\right)\left(\int\d{p_3}\e^{-\tau\omega^{(3)}(p_3)}\right)^3,
\end{equation}
where in the last step, we exploited spatial homogeneity. Factorization into one-dimensional integrals is advantageous, as the convergence of numerical integration impairs for increasing dimensionality. 

Employing the {extrapolated amplitudes} of the previous section, we compute the expectation value of the return probability via~\eqref{eq:return_prob_exp}. 
In the case of $\mathcal{N} = 1$, the formula reduces to
\begin{equation}
\langle P_1(\tau)\rangle = \frac{1}{Z}\sum_{j,k = j_{\mathrm{min}}}^{j_{\mathrm{max}}}\hat{\mathcal{A}}(j,j,k)P_1(\tau; j,k).
\end{equation}

\begin{figure}
    \centering
    \begin{subfigure}{0.5\textwidth}
    \includegraphics[width=\linewidth]{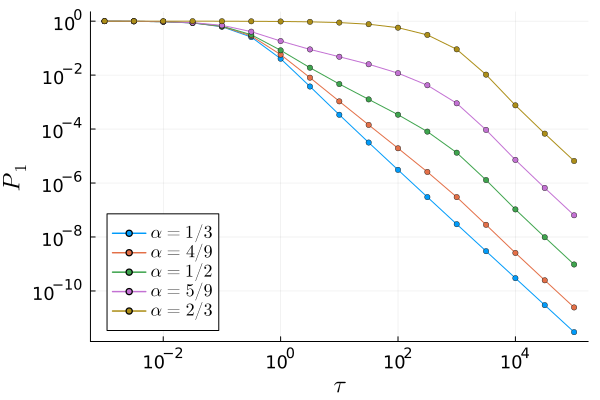}
    \end{subfigure}%
    \begin{subfigure}{0.5\textwidth}
    \includegraphics[width=\linewidth]{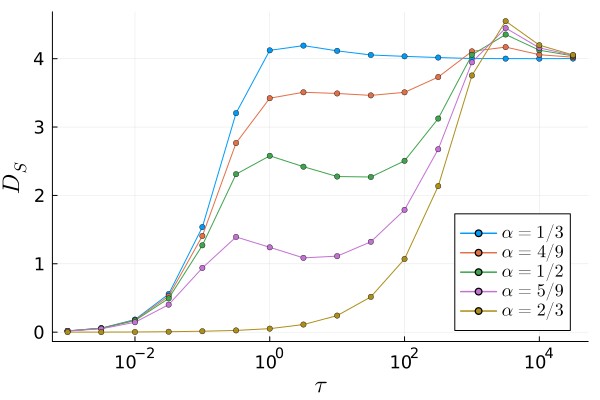}
    \end{subfigure}
    \caption{Left: Expectation value of the $1$-periodic return probability. Right: Spectral dimension computed from the expectation value of the return probability. Both sets of data are computed with cut-offs $j_{\mathrm{min}} = \frac{1}{2}$ and $j_{\mathrm{max}} = 7500$.}
    \label{fig:specdim_quantum}
\end{figure}

The numerical results for the expectation value of the return probability and the spectral dimension are presented in Fig.~\ref{fig:specdim_quantum} for different values of $\alpha$. 
Probing space-time at scales below the lowest lattice scale, $\tau\ll j_{\mathrm{min}}$, $\Ds$ is zero. Above the largest scale, i.e. $\tau\gg j_{\mathrm{max}}$, every classical configuration exhibits a spectral dimension of four and hence the quantum spectral dimension is four as well. Similar to the findings of~\cite{Steinhaus:2018aav}, we observe a non-trivial dimensional flow between $0$ and $4$ for $\alpha$ in a certain interval $[\alpha_{\mathrm{min}},\alpha_{\mathrm{max}}]$. 
We discuss in detail the various influence factors of the spectral dimension in the following paragraphs.

\paragraph{$\pmb{\alpha}$-parameter} The most salient factor driving the spectral dimension is the value of $\alpha$. As Fig.~\ref{fig:specdim_quantum} shows, only a small range of $\alpha\in[\alpha_{\mathrm{min}},\alpha_{\mathrm{max}}]$ leads to an intermediate spectral dimension $\Ds<\std$. For any $\alpha < \alpha_{\mathrm{min}}$, the spectral dimension attains the value $\std=4$ without any non-trivial behaviour before. 
Similarly, $\alpha >\alpha_{\mathrm{max}}$ suppresses an intermediate dimension $\Ds\ne 0$ and $\Ds$ takes a finite non-zero value only at scales $\tau\gtrsim j_{\mathrm{max}}$. 
In Sec.~\ref{subsec:Analytical estimate of the spectral dimension}, we support the statements on the role of $\alpha$ with analytical considerations and present an estimate of the interval $[\alpha_{\mathrm{min}},\alpha_{\mathrm{max}}]$.

\paragraph{Cut-offs} An intermediate spectral dimension between $0$ and $4$ is only resolved if the range of spins between $j_{\mathrm{min}}$ and $j_{\mathrm{max}}$ is sufficiently large. 
To be more precise, it is the ratio $\frac{j_{\mathrm{max}}}{j_\mathrm{min}}$ that is required to be sufficiently large. As numerical tests have shown, the ratio of cut-offs is required to be at least of order $\sim 10^2$ in order to resolve the first sign of a plateau. This signature in $\Ds$ presents itself as two points of inflection which are absent if less configurations are taken into account.

Due to the inherent minimal length scale in the theory, geometry cannot be probed below $\tau\sim j_{\mathrm{min}}$ such that this regime, where the spectral dimension flows to zero for $\tau\rightarrow 0$, is not of physical interest. Similarly, for scales $\tau \gg j_{\mathrm{max}}$, a dimension of four is inevitably reached, since all superimposed spin geometries possess a spectral dimension of four. While the existence of a minimal spin is a quintessential feature of most spin foam models, the upper cut-off is introduced for numerical purposes.\footnote{Following~\cite{Bahr:2018vv,Han:2010pz,Han:2011aa}, apart from what is considered in this work, a cosmological constant can be added to spin foams by replacing the group $\SU$ by its quantum deformation $\SU_q$. Consequently, an upper cut-off $j_{\mathrm{max}}$ is introduced, related to the cosmological constant via $j_{\mathrm{max}}\sim\frac{\pi}{\Lambda l_{\mathrm{P}}^2}$~\cite{Rovelli:2014vg}. However, this value is expected to be much larger than what can could numerically implemented. For instance, a cosmological constant of order $\sim 10^{-122}$ would imply a maximal cut-off of order $j_{\mathrm{max}}\sim 10^{122}$.} 
In order to recover a physical interpretation of our results, we therefore need to consider the limit $j_{\mathrm{max}}\rightarrow\infty$. 
As numerical tests with different $j_{\mathrm{max}}$ as well as the results of~\cite{Steinhaus:2018aav} show, the intermediate regime extends to infinity in the limit of infinitely large upper cut-off. 

\paragraph{Barbero-Immirzi parameter} In the case of $1$-periodic spin foam frusta, the Barbero-Immirzi parameter $\bi$ controls the spacing of allowed $\SU$-spins according to the $Y_\bi$-map defined in~\eqref{eq:EPRL condition}. Thus, changing the value of $\bi$ results in a rescaling of the allowed spins, which in turn can be absorbed into the diffusion scale $\tau$. This holds for quantum as well as for semi-classical amplitudes.\footnote{For $j\in\mathbb{N}/2$ that do not satisfy the EPRL-condition, the amplitudes are zero. We exploit this exclusion for the computation of the return probability in that we only compute it for the allowed configurations, given a value of $\bi$. All other components would be multiplied with zero in the expectation value.} In contrast, the value of $\bi$ has a non-trivial effect on amplitudes of $\mathcal{N}>2$. As the semi-classical vertex amplitude in~\eqref{eq:Spin4Vamp} suggests, $\bi$ controls the relative phase of the oscillations. Therefore, it is in general to be expected that the spectral dimension is non-trivially affected by the value of $\bi$. 

\paragraph{Compactness effects} Considering a finite lattice of length $L$ with periodic boundary conditions rather than an infinite lattice,  compactness effects are introduced to the spectral dimension which set in at scales $\tau>\tau_{\mathrm{comp}}(L)$.\footnote{For a comparison to the continuum spectral dimension of a torus, we refer to~\cite{Calcagni:2014ep}, where the compactness effects are clearly visible.} Following~\cite{Thurigen:2015uc}, the return probability of a classical configuration with spins $j$ is constant for $\tau > j$ and thus, the spectral dimension reaches zero at $\tau\sim j$. For the quantum spectral dimension, this implies that for $L$ sufficiently large, i.e. such that $\tau_{\mathrm{comp}}(L)\gg j_{\mathrm{max}}$, a dimensional flow between zero, a possible intermediate value and four will be resolved. At $\tau$ close to compactness, $\Ds$ will then flow to zero and remain zero for all $\tau>\tau_{\mathrm{comp}}$. For $j_{\mathrm{min}}\lesssim\tau_{\mathrm{comp}}\lesssim j_{\mathrm{max}}$ and an intermediate regime $\Ds^\alpha$ existent, the spectral dimension will flow to this intermediate value and then back to zero after the compactness scale is reached. For $\tau_{\mathrm{comp}}\ll j_{\mathrm{min}}$, the spectral dimension is zero everywhere.  

\begin{figure}
    \centering
    \includegraphics[width = 0.5\textwidth]{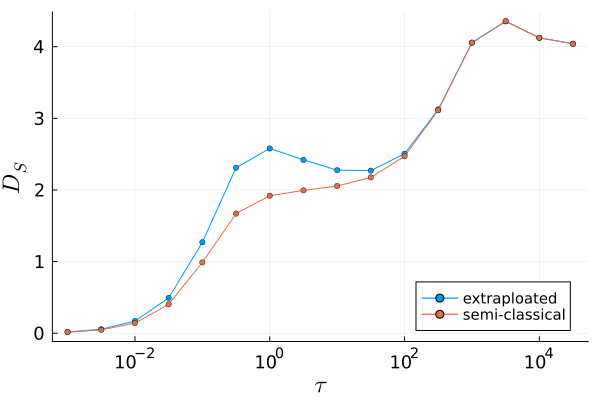}
    \caption{Spectral dimension computed with extrapolated (blue) and semi-classical (red) amplitudes at $\alpha = 0.5$.}
    \label{fig:specdim1_sc_vs_qu}
\end{figure}

\paragraph{Semi-classical amplitudes} Employing the semi-classical amplitude for computing the expectation value of the return probability and the spectral dimension, we observe a behaviour similar to that obtained with extrapolated amplitudes. A direct comparison is presented in Fig.~\ref{fig:specdim1_sc_vs_qu} for $\alpha = 0.5$. With semi-classical amplitudes, the spectral dimension is constant in the intermediate regime. In particular, in the limit $j_{\mathrm{max}}\rightarrow\infty$, where the upper cut-off is removed, this plateau extends to $\tau\rightarrow\infty$.   

Considering the results of Sec.~\ref{subsec:Dressed quantum vertex amplitude} and the analytical explanations of~\cite{Steinhaus:2018aav}, the deviation between the two curves is a consequence of the different scaling behaviours of the amplitudes for small spins. 
From Fig.~\ref{fig:dVAmp_scaling}, it follows that larger effective inverse scaling $\sfs$ of the amplitudes implies a larger spectral dimension. 
Also, since the effective scaling of the extrapolated amplitudes is non-constant, there is a non-constant flow of the spectral dimension to the semi-classical constant value at larger scales.\footnote{This effect is only partially resolved in Fig.~\ref{fig:specdim1_sc_vs_qu}, since effects of the upper cut-off appear for large $\tau$.} As this flow is visible at scales $\tau > 10$, it is not a mere discreteness artifact but a physical effect. 

The different behaviour of the spectral dimension due to the different amplitudes appears in the regime $10^{-2} < \tau < 10^2$ and is of quantitative nature. Although providing an increasingly bad approximation at low spins, this suggests that the semi-classical amplitude is sufficient for extracting the spectral dimension on large scales. In particular, there is agreement with the quantum amplitude results for scales $\tau > 10^2$, even in the limit of infinite upper cut-off. Therefore, we are going to employ semi-classical amplitudes for the rest of this work. 

Using the semi-classical amplitudes comes with the following three advantages. 
First, having an analytical expression for the amplitudes available allows for numerical integration, which is based on assuming the spins to be continuous variables. In that way, the results we have obtained so far can be compared to the findings of~\cite{Steinhaus:2018aav}, which are based on continuous integration. 
Second, the semi-classical setting allows for a straightforward inclusion of a cosmological constant via an ad hoc deformation of the amplitudes~\cite{Han:2011aa,Bahr:2018vv}. 
Third, the extrapolation method for quantum amplitudes, discussed in Sec.~\ref{sec:Quantum amplitudes from extrapolation}, strongly relies on the assumption of $1$-periodicity. Since we do not expect this method to be straightforwardly applicable for $\mathcal{N} > 1$, quantum amplitudes beyond small spins are not in reach for $\mathcal{N} > 1$. Due to these technical limitations, resorting to semi-classical amplitudes allows the study of the spectral dimension at higher periodicities. In the following, we take advantage of the possibilities that the semi-classical amplitudes offer, and discuss these cases in greater detail.

\paragraph{Discrete summation vs. numerical integration} Up to this point, we computed the expectation value of the return probability via a discrete sum over all spin configurations in the range $j_{\mathrm{min}}\leq j,k \leq j_{\mathrm{max}}$. In the case where $\frac{j_{\mathrm{max}}}{j_{\mathrm{min}}}\gg 1$, the sum can be replaced by an integral, which is the strategy employed in~\cite{Steinhaus:2018aav}. As a consequence, the expectation value of the return probability is obtained as an integral over the configurations with the amplitudes and the return probability being continuous functions of the spins. Note that, for this strategy, an analytical expression of the amplitudes is required. Determining the corresponding functions for quantum amplitudes is currently out of reach. Therefore, one needs to resort to the semi-classical approximation, where the spins are simply understood as continuous variables.\footnote{Notice, that the EPRL-condition of~\eqref{eq:EPRL condition} cannot be implemented, since it would yield a function of measure zero.} We have numerically checked that both methods yield very similar results.
For small and large $\tau$, we observe a convergence, while the spectral dimension differs quantitatively in the regime where the intermediate value $D_S^\alpha$ is reached. This is, because according to~\cite{Thurigen:2015uc}, continuous configuration variables lead to a smoothening of discreteness peaks. 

\subsection{Cosmological constant}\label{subsec:Cosmological constant} 

A way to add a cosmological constant to the simplicial EPRL-FK model was introduced in~\cite{Han:2011aa} and generalized to arbitrary $4$-dimensional polyhedra in~\cite{Bahr:2018vv}. 
In essence, the vertex amplitudes of the model are deformed while keeping the boundary Hilbert space fixed. 
Relating the deformation parameter with the cosmological constant, the asymptotic vertex amplitude yields the Regge action with a cosmological constant.\footnote{Another, mathematically more rigorous way of introducing a cosmological constant to spin foams is to replace the group $\SU$ by its quantum deformation $\SU_q$~\cite{Turaev:1992hq,Major:1995yz,Han:2010pz,Fairbairn:2010cp,Dittrich:2018dvs,Dupuis:2020ndx,Dupuis:2013lka}. The deformation parameter $q$ is related to $\Lambda$ via $q = \exp(2\pi i/(k+2))$ with $k =1/(\hbar G \sqrt{\Lambda})$, see e.g.~\cite{Dittrich:2018dvs}. For more methods on implementing a cosmological constant in 4d spin foams see e.g.~\cite{Haggard:2014xoa,Han:2021tzw}.} Given that the Regge curvature, defined in~\eqref{eq:Regge action}, vanishes for $1$-periodic configurations, the introduction of a cosmological constant allows to consider oscillations even when $\mathcal{N} = 1$. In this setting, the sign of the cosmological constant is irrelevant for the amplitudes, as~\eqref{eq:cc oscillations} below shows. Following from the definition of the $\Spin$-vertex amplitude defined in~\eqref{eq:Spin4Vamp}, these oscillations are of a particular type in comparison to cases of vanishing Regge curvature. 
First, $\Lambda$ oscillations allow only for simple cosine type shape. In contrast, the Regge term describes a superposition of cosines with different phases controlled by Barbero-Immirzi parameter. Notice that this superposition is a peculiar feature of the Riemannian EPRL-model, absent in the Lorentzian setting \cite{Barrett:2009mw}. 
Second, since $\Lambda$ enters the action via a $4$-volume, these oscillations scale quadratically in spins, whereas curvature terms scale linearly. 
Despite its simple form, we consider the cosmological constant in order to get a first glimpse of the effects of oscillations on the spectral dimension.

\begin{figure}
    \centering
    \begin{subfigure}{0.5\textwidth}
    \centering
        \includegraphics[width=\linewidth]{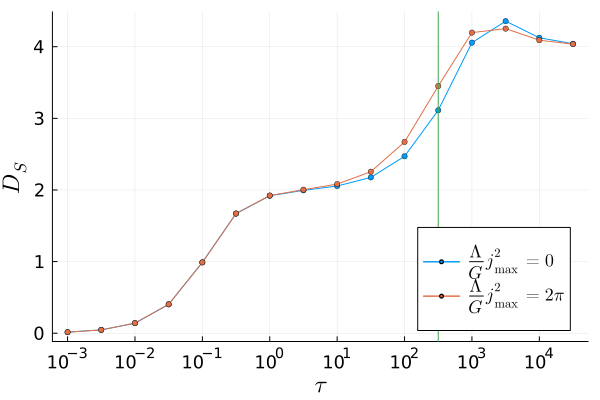}
        \caption{$\frac{\Lambda}{G}j_{\mathrm{max}}^2 = 2\pi$}
    \end{subfigure}%
    \begin{subfigure}{0.5\textwidth}
    \centering
        \includegraphics[width=\linewidth]{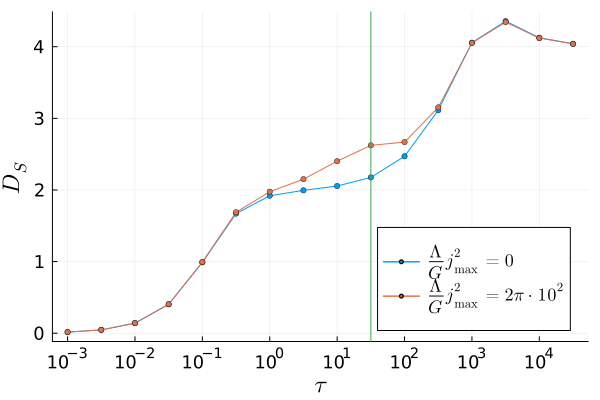}
        \caption{$\frac{\Lambda}{G}j_{\mathrm{max}}^2 = 2\pi 10^2$}
    \end{subfigure}\\
    \begin{subfigure}{0.5\textwidth}
        \includegraphics[width=\textwidth]{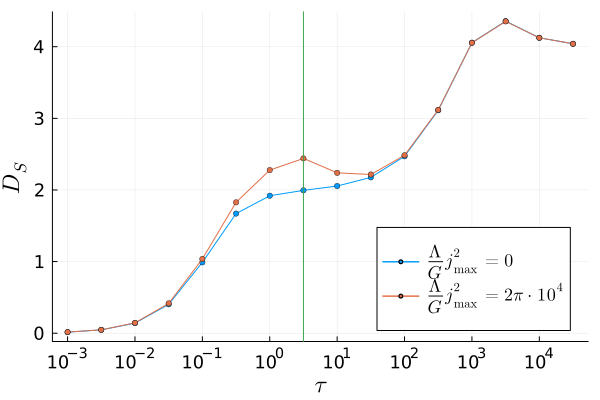}
        \caption{$\frac{\Lambda}{G}j_{\mathrm{max}}^2 = 2\pi 10^4$}
    \end{subfigure}%
    \begin{subfigure}{0.5\textwidth}
        \includegraphics[width=\textwidth]{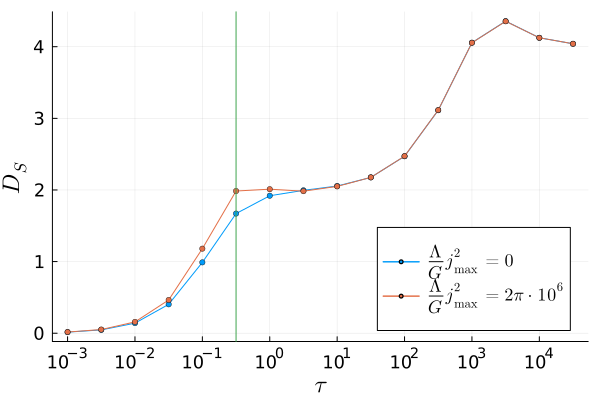}
        \caption{$\frac{\Lambda}{G}j_{\mathrm{max}}^2 = 2\pi 10^6$}
    \end{subfigure}
    \caption{Spectral dimension at $\alpha = 0.5$ with different values of the cosmological constant $\Lambda-G$. Notice that in general, the values of $\Lambda$ are independent of $j_{\mathrm{max}}$.}
    \label{fig:specdim1_Lambda}
\end{figure}

Explicitly, the oscillating term is given by
\begin{equation}\label{eq:cc oscillations}
\cos\left(\frac{\Lambda}{G} jk\right)-\frac{\mathfrak{Re}\{D\}}{\vert D\vert},
\end{equation}
where $D$ is the determinant of the Hessian, defined in~\eqref{eq:determH}. Consequently, for a given upper cut-off $j_{\mathrm{max}}$, the amplitudes and hence the return probability are not altered for $\frac{\Lambda}{G}\ll \frac{1}{j_{\mathrm{max}}^2}$ since the argument of the cosine is small for all possible values of spins. However, for $\frac{\Lambda}{G}\gtrsim \frac{1}{j_{\mathrm{max}}^2}$ the spectral dimension is affected by the oscillations of the amplitudes, as Fig.~\ref{fig:specdim1_Lambda} shows. 

\begin{figure}
    \centering
    \begin{subfigure}{0.5\textwidth}
    \includegraphics[width=\linewidth]{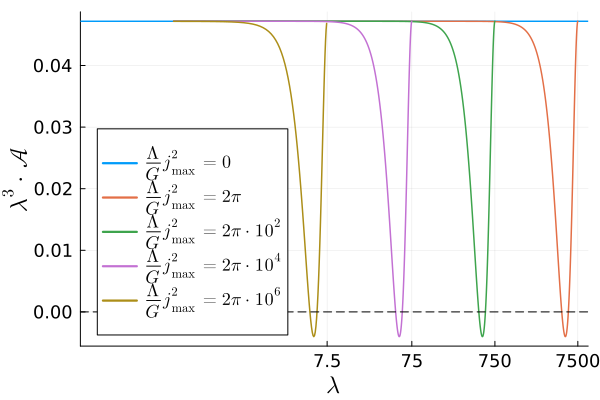}
    \end{subfigure}%
    \begin{subfigure}{0.5\textwidth}
    \includegraphics[width=\linewidth]{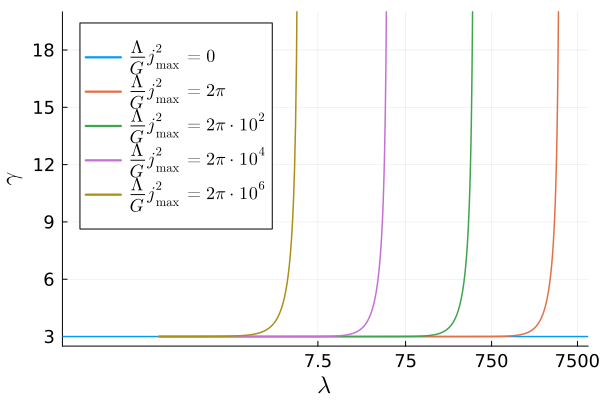}
    \end{subfigure}
    \caption{Left: The first complete oscillation of the re-scaled amplitudes for different values of $\Lambda/G$, given $\alpha = 0.5$. The dashed line indicates zero, which is crossed for the first time at $\lambda = \lambda_0^{(\Lambda)}$. Right: Scaling $\sfs$ of the amplitudes for $\lambda < \lambda_0^{(\Lambda)}$ with the same parameters as on the left. Notice that both of the plots assume the spin variable $\lambda$ to be continuous, ignoring the EPRL-condition.}
    \label{fig:dVamp_Lambda}
\end{figure}

The strongest deviation from the $\Lambda = 0$ case is localized at the scale at which the first oscillation 
takes place, marked by a green vertical line.
If this scale is in the intermediate regime, the spectral dimension is larger than in the case of vanishing $\Lambda$. 
To be more precise, it is the regime below the first zero of the amplitudes, given by
\begin{equation}
\lambda_0^{(\Lambda)} = \sqrt{\frac{G}{\Lambda}\arccos\left(\frac{\mathfrak{Re}\{D\}}{\vert D\vert}\right)}
\end{equation}
which we expect to be most influential. To visualize this, Fig.~\ref{fig:dVamp_Lambda} shows the first oscillation of the (rescaled) amplitudes as well as the effective scaling $\sfs$ of the amplitude for all $\lambda < \lambda_0^{(\Lambda)}$. Since the shift in~\eqref{eq:cc oscillations} is constant, the oscillations are not symmetric with respect to the $\lambda$-axis. Still, since $\Big{\vert}\frac{\mathfrak{Re}\{D\}}{\vert D\vert}\Big{\vert} < 1$, the amplitudes attain negative values when crossing zero, indicated by the dashed line. The zeros as well as the extrema of the amplitudes lead to an abrupt change in sign of the scaling $\sfs$. Consequently, for $\lambda > \lambda_0^{(\Lambda)}$, the scaling oscillates rapidly, which however does not affect the spectral dimension. 
Similar to the comparison of quantum and semi-classical amplitudes, a value $\sfs>9-12\alpha$ larger than the semi-classical scaling implies a larger spectral dimension. 
This effect is in particular resolved in the regime $\lambda < \lambda_0^{(\Lambda)}$. Notice that an increase of $\Ds$ is a feature present for all values of $\Lambda\neq 0$, since the amplitudes exhibit a scaling $\sfs > 9-12\alpha$ for $\lambda < \lambda_0^{(\Lambda)}$.

\subsection{2-periodic spectral dimension}\label{subsec:2-periodic spectral dimension}

The results of the previous section have shown, that the semi-classical amplitude serves as a sufficient approximation to extract the qualitative behaviour of the spectral dimension in the $1$-periodic case. 
It is therefore reasonable to employ the semi-classical amplitudes to compute the spectral dimension with higher periodicity, being $\mathcal{N} = 2$ here. We discuss possible quantum corrections to our results at the end of this section. 

Following~\eqref{eq:Laplace_momentum}, the Laplace operator at $\mathcal{N} = 2$  becomes a $2\times 2$ matrix in momentum space
\begin{equation}
M = 
\begin{pmatrix}
w_0+w_1+W_0 & -w_0-w_1 e^{-ip_0} \\
-w_0-w_1 e^{ip_0} & w_0+w_1+W_1
\end{pmatrix},
\end{equation}
where $W_0$ and $W_1$, defined in~\eqref{eq:w and W}, depend on the spatial momenta $p_i$. The corresponding eigenvalues are given by
\begin{equation}
\omega_{\pm}(p_\mu) = w_0+w_1+\frac{W_0+W_1}{2}\pm\sqrt{w_0^2+w_1^2+2w_0w_1\cos(p_0)+\left(\frac{W_0-W_1}{2}\right)^2}.
\end{equation}
Compared to the $1$-periodic case in~\eqref{eq:P1}, this expression is more involved due to the intermingling of the $p_0$ and $p_i$ terms. As a consequence, the return probability from momentum integration,
\begin{equation}
P_2(\tau) = \sum_{\epsilon = \pm}\int\prod_{\mu = 0}^3\d{p_\mu}\e^{-\tau\omega_{\epsilon}(\{p_\nu\})},
\end{equation}
cannot be written as the product of $1$-dimensional integrals. Instead, full $4$-dimensional integration is required to compute $P_2$, leading to larger numerical computation times. 

Under the assumption that the amplitudes of a single $\mathcal{N}$-cell, here a $2$-cell consisting of 16 hyperfrusta, already capture the relevant information, the full expression for the expectation value of the return probability is given by
\begin{equation}\label{eq:ev2}
\langle P_2(\tau)\rangle = \frac{1}{Z}\sum_{j_1,j_2,k_1,k_2}\left(\hat{\mathcal{A}}(j_1,j_2,k_1)\right)^8\left(\hat{\mathcal{A}}(j_2,j_1,k_2)\right)^8 P_2(\tau;j_1,j_2,k_1,k_2).
\end{equation}
Again, the range of all of the spins is given by $j_{\mathrm{min}}\leq j_i,k_i \leq j_{\mathrm{max}}$. For the numerical results presented below, we have chosen $j_{\mathrm{min}}=\frac{1}{2}$ and $j_{\mathrm{max}}=201$. Since in the $2$-periodic case, the $3$-cubes are not restricted to equal size, frustum geometries arise which lead to non-vanishing Regge curvature. Consequently, the vertex amplitudes $\hat{\mathcal{A}}$ exhibit oscillations even in the case of vanishing cosmological constant, $\Lambda = 0$, which we assume from now on if not stated otherwise. Given three fixed spins $(j,j',k)$, a single dressed vertex amplitude scales as
\begin{equation}\label{eq:2per oscillations}
\hat{\mathcal{A}}(\lambda j,\lambda j',\lambda k) \sim \lambda^{12\alpha - 9}\left[\cos\left(\frac{\bi}{G}\lambda S_{\mathrm{R}}(j,j',k)\right)-\cos\left(\frac{1}{G}\lambda S_{\mathrm{R}}(j, j', k)-\varphi(j,j',k)\right)\right],
\end{equation}
where $S_R$ is the Regge action and $\varphi$ is the phase of the determinant $D$ of the Hessian, both being evaluated on the spins $(j,j',k)$. Notice that $\varphi$ is invariant under a homogeneous scaling of all spins. Since the expectation value of the $2$-periodic return probability, \eqref{eq:ev2}, contains high powers of cosine-functions, the amplitudes are narrowly peaked on the maxima of oscillation. Furthermore, since the powers in~\eqref{eq:ev2} are even, no negative values occur. 

\begin{figure}
    \centering
    \begin{subfigure}{0.5\textwidth}
        \centering
        \includegraphics[width=\linewidth]{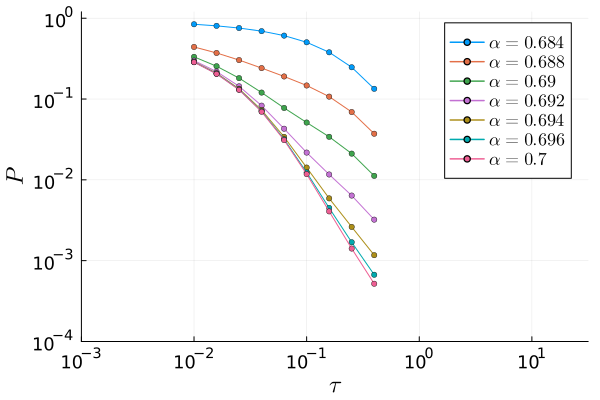}
    \end{subfigure}%
    \begin{subfigure}{0.5\textwidth}
        \centering
        \includegraphics[width=\linewidth]{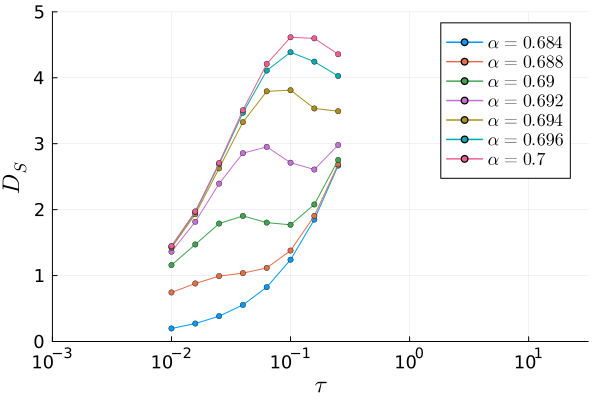}
    \end{subfigure}
    \caption{Left: Expectation value of the $2$-periodic return probability. Right: $2$-periodic spectral dimension. Parameters $G,\bi$ and $\Lambda$ are set to $G = 10^{10}$, $\bi=1/3$ and $\Lambda = 0$ with cut-offs $j_{\mathrm{min}}=1/2$ and $j_{\mathrm{max}}=201$.}
    \label{fig:specdim2_alphas}
\end{figure}

\paragraph{Large Newton's constant} As a first computation, we consider the limiting case of $G\rightarrow\infty$, where curvature oscillations become negligible. In Fig.~\ref{fig:specdim2_alphas} we present the results for $G =10^{10}$. This already captures the large $G$ behaviour for the chosen $j_\text{max}$ as we have checked that for larger $G$ the results do not change anymore.

Just as for the $1$-periodic case, we observe that there exists an intermediate flow of the spectral dimension for $\alpha$-values in a certain interval $\alpha\in[\alpha_{\mathrm{min}},\alpha_{\mathrm{max}}]$, where $\alpha_{\mathrm{min}}\approx 0.68$ and $\alpha_{\mathrm{max}}\approx 0.7$. From the results Sec.~\ref{subsec:1-periodic spectral dimension}, we conclude that the intermediate spectral dimension is again a result of the amplitudes exhibiting a scaling behaviour. Compared to the $1$-periodic case, the size of the interval of $\alpha$ is smaller which is in accordance with the findings of~\cite{Steinhaus:2018aav}. As we are going to discuss in more detail in Sec.~\ref{subsec:Analytical estimate of the spectral dimension}, this is expected. In the following we discuss the effects of a finite value of $G$.

\begin{figure}
    \centering
    \includegraphics[width=0.5\textwidth]{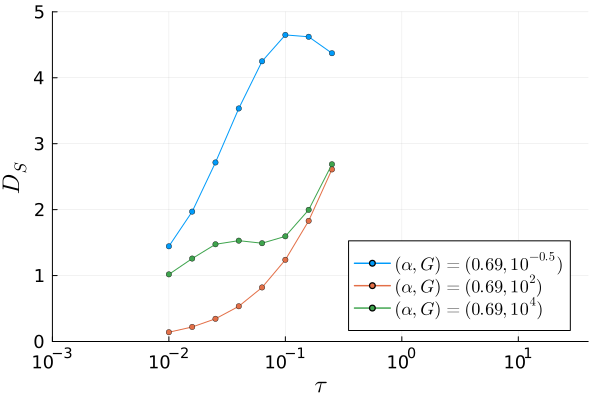}
    \caption{$2$-periodic spectral dimension at $\alpha = 0.69$ for three exemplary values of $G$.}
    \label{fig:specdim2_Gs}
\end{figure}

\paragraph{Regge curvature oscillations} Given that in principle an intermediate spectral dimension $\Ds^\alpha$ can be observed in the case of $\mathcal{N}=2$, we study next the influence of oscillating amplitudes by varying $G$. Following from the form of the amplitudes in~\eqref{eq:2per oscillations}, Regge curvature oscillations are expected to become relevant for $G$ being comparable to $S_R$. In the light of the results of Sec.~\ref{subsec:Cosmological constant}, we expect that the oscillations, now induced by Regge curvature, have an effect on the spectral dimension. Our numerical results show that the spectral dimension is indeed perceptive to the value of $G$. Since small changes in $G$ lead to very different flows of $\Ds$, we conclude that the spectral dimension is in fact highly sensitive to Newton's constant. In Fig.~\ref{fig:specdim2_Gs}, we present the spectral dimension at fixed $\alpha$ for three exemplary values of $G$, showing that one can have either a positive or a negative correction to the case $G\rightarrow\infty$, or no correction at all. 

In contrast to $1$-periodic amplitudes oscillating with a cosmological constant, the flow of the $2$-periodic spectral dimension is not straightforwardly understood by considering the scaling behaviour of the amplitudes. Main reason for that is the strong dependence of the amplitudes on the spins $(j,j',k)$, given explicitly in~\eqref{eq:2per oscillations}.

\begin{figure}
    \centering
    \begin{subfigure}{0.5\textwidth}
        \includegraphics[width=\linewidth]{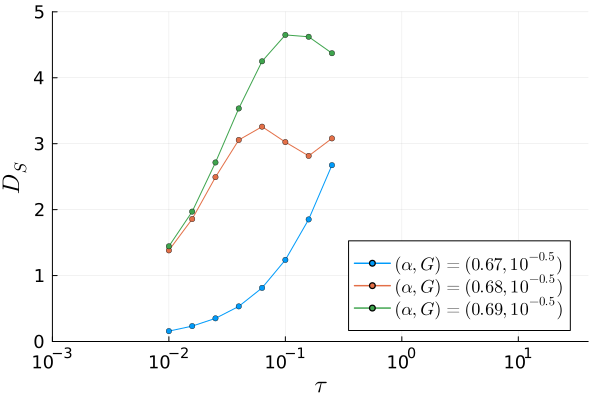}
    \end{subfigure}%
    \begin{subfigure}{0.5\textwidth}
        \includegraphics[width=\textwidth]{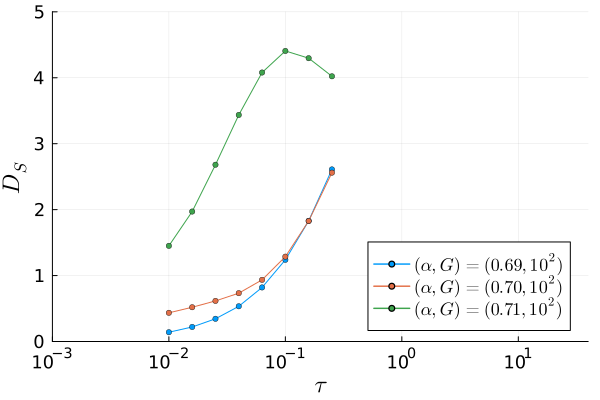}
    \end{subfigure}
    \caption{$2$-periodic spectral dimension for $G = 10^{-0.5}$ (left) and $G = 10^2$ (right) showing that the window of an intermediate regime is $G$-dependent.}
    \label{fig:specdim2_alphavary}
\end{figure}

\paragraph{$\pmb{G}$-dependence of intermediate regime} Given the plots of Fig.~\ref{fig:specdim2_Gs}, the blue and red curves might indicate that the interval $[\alpha_{\mathrm{min}},\alpha_{\mathrm{max}}]$ in which an intermediate spectral dimension occurs depends on the value of $G$. 
We test this possibility by computing the spectral dimension for $G = 10^{-0.5}$ and $G = 10^2$ and a wide range of $\alpha$. 
Our results are depicted in Fig.~\ref{fig:specdim2_alphavary}. 
Indeed, we observe that for different values of $G$ the boundaries $\alpha_{\mathrm{min}}$ and $\alpha_{\mathrm{max}}$ change. 
While at $G = 10^{-0.5}$, these 
are $\alpha_{\mathrm{min}} \approx 0.67$ and $\alpha_{\mathrm{max}}\approx 0.69$, their values at $G = 10^2$ are $\alpha_{\mathrm{min}}\approx 0.69$ and $\alpha_{\mathrm{max}} \approx 0.71$. 

An additional feature we observe is the following. Within the interval $[\alpha_{\mathrm{min}},\alpha_{\mathrm{max}}]$, the spectral dimension of purely scaling amplitudes is approximately a decreasing linear function of $\alpha$. In contrast, Fig.~\ref{fig:specdim2_alphavary} suggests that the slope of $\Ds$ as a function of $\alpha$ can be positive or negative, depending on $G$. 
We will discuss interpretations of these phenomena and their entailing consequences for the limit of large periodicities $\mathcal{N}$ in Sec.~\ref{subsec:The thermodynamic limit}.

\begin{figure}
    \centering
    \includegraphics[width=0.5\linewidth]{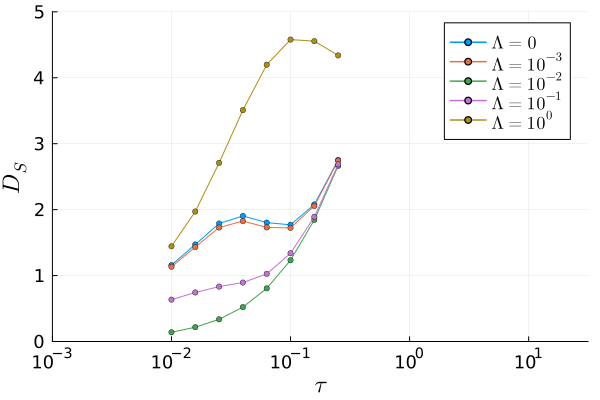}
    \caption{$2$-periodic spectral dimension for varying $\Lambda$, with $G = 1$ and $\alpha = 0.69$.}
    \label{fig:specdim2_Lambdas}
\end{figure}

\paragraph{Cosmological constant} Since the Regge curvature does not vanish for $\mathcal{N}>1$, the effects of a non-vanishing cosmological constant, $\Lambda \neq 0$, on the amplitudes is not as apparent as in the $1$-periodic case. Following~\eqref{eq:Spin4Vamp},  $\Lambda$ introduces a phase shift to the cosine term containing the parameter $\bi$. Entering with the $4$-volume of the frusta, such a term scales quadratically in the spins, leading to an intricate behaviour in combination with the Regge curvature contributions. Fixing $G = 1$ and $\alpha = 0.69$, we present the $2$-periodic spectral dimension for a selection of $\Lambda$-values in Fig.~\ref{fig:specdim2_Lambdas}. 

Effects of $\Lambda$ become important only for $\Lambda$ not much smaller than $G$ as expected from the form of the amplitudes. 
The resulting phase shift in the oscillating part of the amplitudes has significant impact on the flow of the spectral dimension, as Fig.~\ref{fig:specdim2_Lambdas} shows. 
In comparison to the $1$-periodic case discussed in Sec.~\ref{subsec:Cosmological constant}, we observe the following additional features. 
First, high frequency oscillations due to large values of $\Lambda$ do not appear negligible, at least in the small range of $\tau$ that we can observe. 
Second, the region of $\tau$'s, where $\Lambda$ leads to a deviation in the flow of $\Ds$, is not as clearly localized as in Fig.~\ref{fig:specdim1_Lambda}. This is apparent by observing that in the region $\tau\in[10^{-2},10^{-0.6}]$, the spectral dimension is affected for various orders of magnitude of $\Lambda$. Third, the direction in which the spectral dimension is corrected by the presence of $\Lambda$, so to lower or larger values than for $\Lambda = 0$, is obscured compared to the $1$-periodic case. That is, because $\Lambda$ does not solely control the position of the first root of the amplitudes when $\mathcal{N}=2$ but leads to a phase shift, the consequences of which are not as straightforward to analyse.

In the presence of a non-vanishing Regge curvature, the amplitudes are not invariant under sign change of $\Lambda$. However, numerical tests have shown that, at least for the small region of $\tau$'s depicted in Fig.~\ref{fig:specdim2_Lambdas}, different signs of $\Lambda$ have a negligible effect in $\Ds^\alpha$. Hence, the magnitude of the cosmological constant appears to be the significant quantity.

Similar to the above, we find that the range of $\alpha$, for which an intermediate dimension exists, depends on the value of $\Lambda$. In summary, $\alpha_{\mathrm{min}}$ and $\alpha_{\mathrm{max}}$ are functions of both, Newton's constant $G$ as well as the cosmological constant $\Lambda$. 
We will pursue this discussion in Sec.~\ref{subsec:The thermodynamic limit}.


\subsection{Analytical estimate of the spectral dimension}\label{subsec:Analytical estimate of the spectral dimension}

Building up on the ideas of~\cite{Steinhaus:2018aav}, we present in this section a strategy to extract information from the spectral dimension of oscillating amplitudes using analytical methods. Tackling first the spectral dimension of general $\mathcal{N}$-periodic spin foam frusta, we obtain a qualitative expression for the spectral dimension which is, however, still too intricate to compute explicitly. Nevertheless it serves as a support for the numerical results as well as a guidance for the limit $\mathcal{N}\rightarrow\infty$, discussed in Sec.~\ref{subsec:The thermodynamic limit}. In the second part of this section, we consider an explicitly integrable model which qualitatively explains the cosmological constant results of Sec.~\ref{subsec:Cosmological constant}.

\subsubsection{An analytical argument}

For the analysis of the spectral dimension, it is advantageous to introduce the average spin variable $r^2\defeq\frac{1}{n}\sum j_f^2$, where $n$ is the total number of degrees of freedom, being $n = 2\mathcal{N}$ in the case of $\mathcal{N}$-periodic spin foam frusta. Technically, the variable $r$ is a radial coordinate in the space of configurations $j_f$. Likewise, the remaining variables can be seen as an angular part, and we therefore denote them by $\Omega$ in the following. As the definition of the Laplace operator in Sec.~\ref{subsec:Spectral dimension and semi-classical geometry of hyperfrusta} and the arguments of~\cite{Sahlmann:2010bb} and~\cite{Calcagni:2015is,Steinhaus:2018aav} show, it is reasonable to assume under the spin foam measure
\begin{equation}\label{eq:Laplace scaling}
\Delta(j_f)\sim\frac{1}{r}\Delta,
\end{equation}
where $\Delta$ is the Laplace matrix on the equilateral hypercubic lattice. 

Within this assumption, let us have a closer look on the expectation value of the return probability with respect to semi-classical amplitudes. As we have argued in Sec.~\ref{subsec:1-periodic spectral dimension}, for $\frac{j_{\mathrm{max}}}{j_{\mathrm{min}}}\gg 1$, the summation over configurations can be approximated by an integral. Following that and performing a change to spherical coordinates as described above, we obtain
\begin{equation}
\langle P(\tau)\rangle = \frac{1}{Z}\int\d{\Omega}\int\limits_{j_{\mathrm{min}}}^{j_{\mathrm{max}}}\d{r}\;r^{n-1}\prod_v\mathcal{A}_v(r,\Omega)\Tr\left(\e^{\frac{\tau\Delta}{r}}\right).
\end{equation}
Forming the logarithmic derivative of this expression yields
\begin{equation}
\frac{\tau}{\langle P\rangle}\frac{\partial \langle P\rangle}{\partial \tau} = \frac{1}{\langle P\rangle Z}\int\d{r}\d{\Omega}\;r^{n-1}\prod_v\mathcal{A}_v\Tr\left(\frac{\tau\Delta}{r}\e^{\frac{\tau\Delta}{r}}\right).
\end{equation}
Since $\Tr\left(\e^{\frac{\tau\Delta}{r}}\right)$ is in fact a function of the ratio $\tau/r$, we can trade the derivative with respect to $\tau$ with an $r$ derivative,
\begin{equation}\label{eq:tdt = -rdr}
\Tr\left(\frac{\tau\Delta}{r}\e^{\frac{\tau\Delta}{r}}\right) = -r\frac{\partial}{\partial r}\Tr\left(\e^{\frac{\tau\Delta}{r}}\right).
\end{equation}
Using the $r$ derivative, we can integrate by parts,
\begin{equation}
\frac{\tau}{\langle P\rangle}\frac{\partial \langle P\rangle}{\partial \tau} = \frac{1}{\langle P\rangle Z}\left[\partial I(\tau) + \int\d{r}\d{\Omega}r^{n-1}\prod_v\mathcal{A}_v\left(n + \sum_v\frac{r}{\mathcal{A}_v}\frac{\partial\mathcal{A}_v}{\partial r}\right)\Tr\left(\e^{\frac{\tau\Delta}{r}}\right)\right].
\end{equation}
Here, $\partial I$ denotes the boundary term in the partial integration, explicitly given by
\begin{equation}
\partial I(\tau) = -\int\d{\Omega} \;r^n \prod_v\mathcal{A}_v\Tr\left(\e^{\frac{\tau\Delta}{r}}\right)\bigg|_{r = j_{\mathrm{min}}}^{j_{\mathrm{max}}}.
\end{equation}
The other terms inside the brackets stem from the $r$-derivative acting first on $r^n$ and then on the product of amplitudes. Notice that $-\frac{r}{\mathcal{A}_v}\frac{\partial\mathcal{A}_v}{\partial r}$ is exactly the effective scaling $\gamma$ of $\mathcal{A}_v$ which we have discussed previously in Sec.~\ref{subsec:Dressed quantum vertex amplitude}.

Before tackling more general cases, we consider the simplified scenario in which the amplitudes satisfy a uniform scaling behaviour, i.e. $\mathcal{A}_v=h_v(\Omega)r^{-\sfsc}$. 
For a scaling $\sfsc = 9-12\alpha$, this describes $\mathcal{N}$-periodic cuboids (and thus $1$-periodic frusta) with a vanishing cosmo\-logical constant, $\Lambda = 0$. 
Importantly, the radial and angular parts factorize as a consequence of the scaling assumption of $\Delta$, simplifying the following equations significantly. As the effective scaling of $\mathcal{A}_v$ with respect to the radial coordinate is the constant $\sfsc$, the spectral dimension is
\begin{equation}
\Ds(\tau) = -2\frac{\tau}{\langle P\rangle}\frac{\partial \langle P\rangle}{\partial \tau} = 2(\sfsc V - n)-2\frac{\partial I(\tau)}{\langle P\rangle Z}.
\end{equation}
Our results from $1$-periodic semi-classical frusta, together with the findings of~\cite{Steinhaus:2018aav} suggest that, if $\alpha$ allows for an intermediate spectral dimension $0 < \Ds^\alpha < 4$, the boundary term vanishes there. Consequently, $\Ds^\alpha$ is given by
\begin{equation}\label{eq:intermediate D}
\Ds^{\alpha} = 2\left((9-12\alpha)V-n\right).
\end{equation}
If the value of $\Ds^\alpha$ lies outside the interval $[0,4]$, the boundary term counteracts to yield either zero or four. Due to the spatial homogeneity of frusta as well as the assumptions we introduced, the number of vertices $V$ and the number configurations $n$ are respectively given by $V = \mathcal{N}^4$ and $n = 2\mathcal{N}$. Plugging these expressions into~\eqref{eq:intermediate D} for $\mathcal{N} = 1$, the analytical estimate is compatible with the numerical findings of Sec.~\ref{subsec:1-periodic spectral dimension} 
as well as the previous results in \cite{Steinhaus:2018aav}.

Quantum amplitudes as well as semi-classical amplitudes for $\mathcal{N}>1$ do not show a simple scaling behaviour. However, all these more general cases have in common that they factorize into a scaling part and a non-trivial residual part. To capture this, we write
\begin{equation}
\mathcal{A}_v 
= r^{-\sfsc}\mathcal{C}_v(r,\Omega),
\end{equation}
where $\mathcal{C}_v(r,\Omega)$ can be understood as a \emph{correction} term to the scaling part with constant~$\sfsc$. Splitting the amplitudes into this form allows to re-express the spectral dimension as
\begin{equation}\label{eq:approximation D}
\Ds = 2(\sfsc V - n)-2\sum_v\frac{\int\d{\Omega}\d{r}\;r^{n-1}\frac{r}{\mathcal{C}_v}\frac{\partial\mathcal{C}_v}{\partial r}\prod_{v'}\mathcal{A}_{v'}\Tr\left(\e^{\frac{\tau\Delta}{r}}\right)}{\int\d{\Omega}\d{r}\;r^{n-1}\prod_{v'}\mathcal{A}_{v'}\Tr\left(\e^{\frac{\tau\Delta}{r}}\right)}-2\frac{\partial I}{\langle P\rangle Z}.
\end{equation}
For oscillating correction terms that attain many zeros, and therefore lead to divergences of the effective scaling, the integration domain of the above needs to be restricted accordingly. Aspects of well-definedness and convergence need to be addressed for each given $\mathcal{C}_v$ individually. Put into this form,~\eqref{eq:approximation D} suggests that the pure scaling value of $\Ds^\alpha = 2((9-12\alpha)V-n)$ is corrected by a term arising from the effective scaling $-\frac{r}{\mathcal{C}_v}\frac{\partial \mathcal{C}_v}{\partial r}$ of the correction term $\mathcal{C}_v$ as well as the boundary term.
Assuming that the boundary term is negligible in the regime of an intermediate spectral dimension, we compare the analytical estimate in~\eqref{eq:approximation D} with the results of $1$-periodic quantum amplitudes, a cosmological constant for $\mathcal{N}=1$ and finally the $2$-periodic case.

For $\alpha\gtrsim 0.24$, extrapolated quantum amplitudes exhibit an effective scaling larger than the semi-classical value, as Fig.~\ref{fig:dVAmp_scaling} shows. This implies that 
\begin{equation}
-\frac{r}{\mathcal{C}_v}\frac{\partial\mathcal{C}_v}{\partial r} > 0,
\end{equation}
and hence, that the spectral dimension is corrected to a larger value. This is exactly what we observe in Fig.~\ref{fig:specdim1_sc_vs_qu}. 

In the presence of a cosmological constant $\Lambda$, the correction term in~\eqref{eq:cc oscillations} is a shifted cosine with a quadratically scaling phase. Since the constant shift is smaller than one, the correction terms hits many zeros, such that the effective scaling diverges at these points. Consequently,~\eqref{eq:approximation D} is only valid on a restricted domain. Still, let $r_0^{(\Lambda)}$ denote the first zero of $\mathcal{C}_v$ depending on~$\Lambda$. Then the effective scaling of $\mathcal{C}_v$ is larger than zero in the interval $[j_{\mathrm{min}},r_0^{(\Lambda)}]$. This suggests that for growing $\tau$, the first correction to the spectral dimension is to a larger value, which can be observed numerically in Sec.~\ref{subsec:Cosmological constant}. For the remaining integration range $r\in[r_0^{(\Lambda)},j_{\mathrm{max}}]$, we conjecture that the rapidly changing scaling behaviour is averaged out within the scales that are emphasized by $\Tr\left(\e^{\frac{\tau\Delta}{r}}\right)$. However, it is currently not in reach to substantiate this statement further.

Frusta amplitudes of periodicity $\mathcal{N}>1$ exhibit a highly non-trivial correction term $\mathcal{C}_v(r,\Omega)$, as~\eqref{eq:2per oscillations} indicates. Given that $\mathcal{C}_v$ hits many zeros, leading to divergences of its effective scaling, the integral in~\eqref{eq:approximation D} has a highly restricted domain of validity which depends sensitively on the values of $(G,\bi,\Lambda)$. Due to these obstacles for understanding the correction term, let us first consider a regime where $\mathcal{C}_v$ is approximately constant, obtained for large values of $G$, and comment on the more general case afterwards. In the absence of the correction and the boundary term, the intermediate spectral dimension is given by~\eqref{eq:intermediate D} for $V = 16$ and $n = 4$. The admissible values of $\alpha$ to observe an intermediate spectral dimension approximately lie in the interval $[0.72,0.73]$, as predicted by the analytical considerations. 

Our numerical results of Sec.~\ref{subsec:2-periodic spectral dimension} are in partial accordance with this prediction. Qualitatively, we find an intermediate spectral dimension which is controlled solely by $\alpha$, as Fig.~\ref{fig:specdim2_alphas} shows. Also, we find that the window of admissible $\alpha$ for such a regime is smaller compared to the $1$-periodic case. However, the qualitative predictions of the analytical arguments, notably~\eqref{eq:intermediate D}, do not agree with the numerical results. In particular, the values of $\alpha_{\mathrm{min}}$ and $\alpha_{\mathrm{max}}$ as well as the values of $\Ds^\alpha$ of the analytical derivation appear to be shifted with respect to the numerical ones. Recall that the whole analytical argument hinges on the assumption that the Laplacian exhibits a scaling behaviour with a negligible angular dependence, given in~\eqref{eq:Laplace scaling}. While this might be justified for $\mathcal{N}=1$, the angular dependence of $\Delta$ for $\mathcal{N}>2$ is more intricate. Therefore, we suspect this assumption to be the main source of the discrepancy between the analytical predictions and the numerical results. As a result, changes of the intermediate dimension which are not captured by~\eqref{eq:intermediate D} are conceivable. Still, we recall that the window of intermediate scales $j_{\mathrm{min}} < \tau < j_{\mathrm{max}}$ for the results of Sec.~\ref{subsec:2-periodic spectral dimension} is small, necessitating further analysis of the qualitative inconsistency between analytical and numerical results.

When correction terms cannot be neglected, i.e. when the value of the action $S_R$ is of order $G$, the spectral dimension is sensitive to the values of $G$ and $\Lambda$, as the numerical results of Sec.~\ref{subsec:2-periodic spectral dimension} show. In the intermediate regime $j_{\mathrm{min}}
<\tau<j_{\mathrm{max}}$ 
one observes corrections to $\Ds^\alpha$ for some $(G,\Lambda)$ to smaller as well as to larger values. Moreover, the window $[\alpha_{\mathrm{min}},\alpha_{\mathrm{max}}]$ is shifted depending on $G$ and $\Lambda$. This suggests that for higher periodicities, the correction term is predominant, introducing an intricate dependence of the intermediate dimension on the parameter values $(\alpha,G,\Lambda)$.

\subsubsection{An integrable model with oscillations}\label{sec:integrable model}

Faced with the obstacle of computing the corrections of oscillating terms explicitly, we consider in the following a simplified integrable system. The results we derive support the intuition we have attained in the preceding part of this section. 

\begin{figure}
    \centering
    \includegraphics[width=0.45\linewidth]{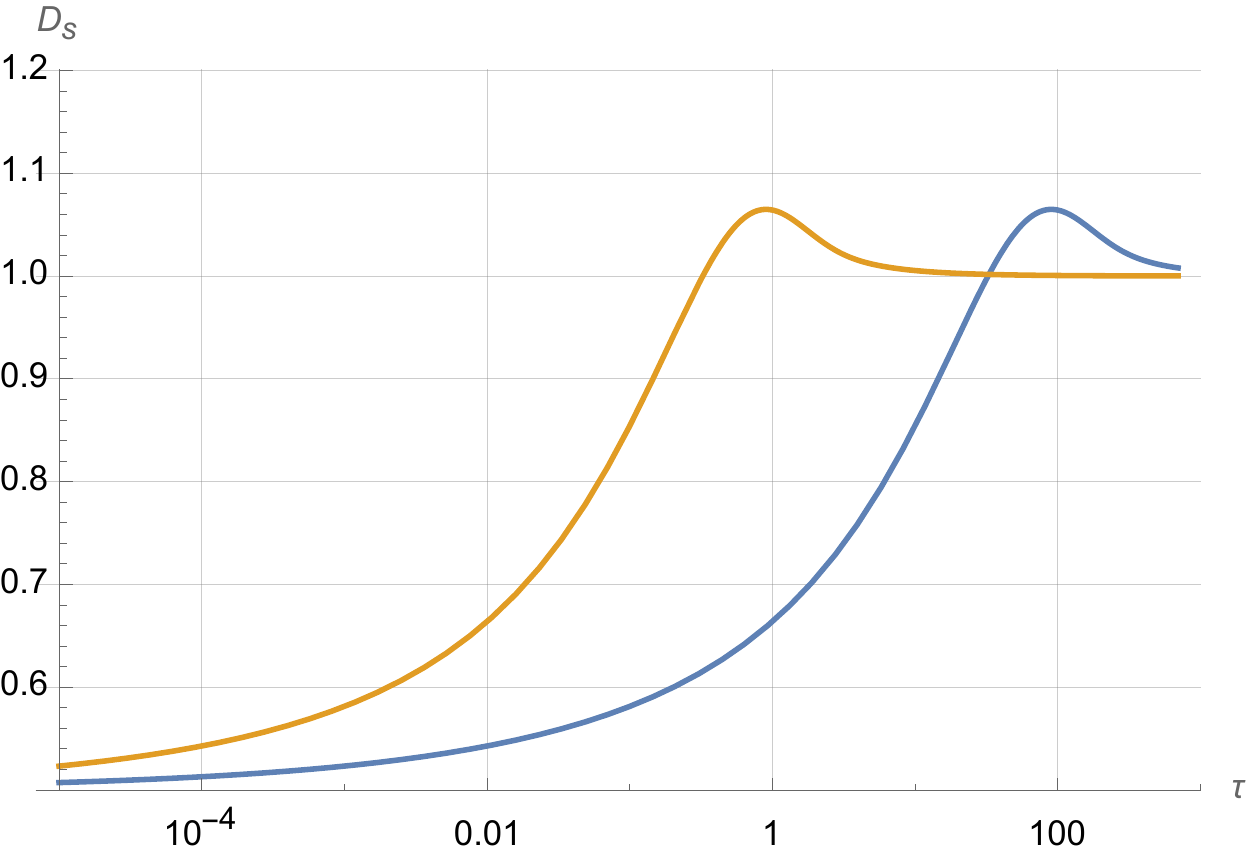}
    \includegraphics[width=0.45\linewidth]{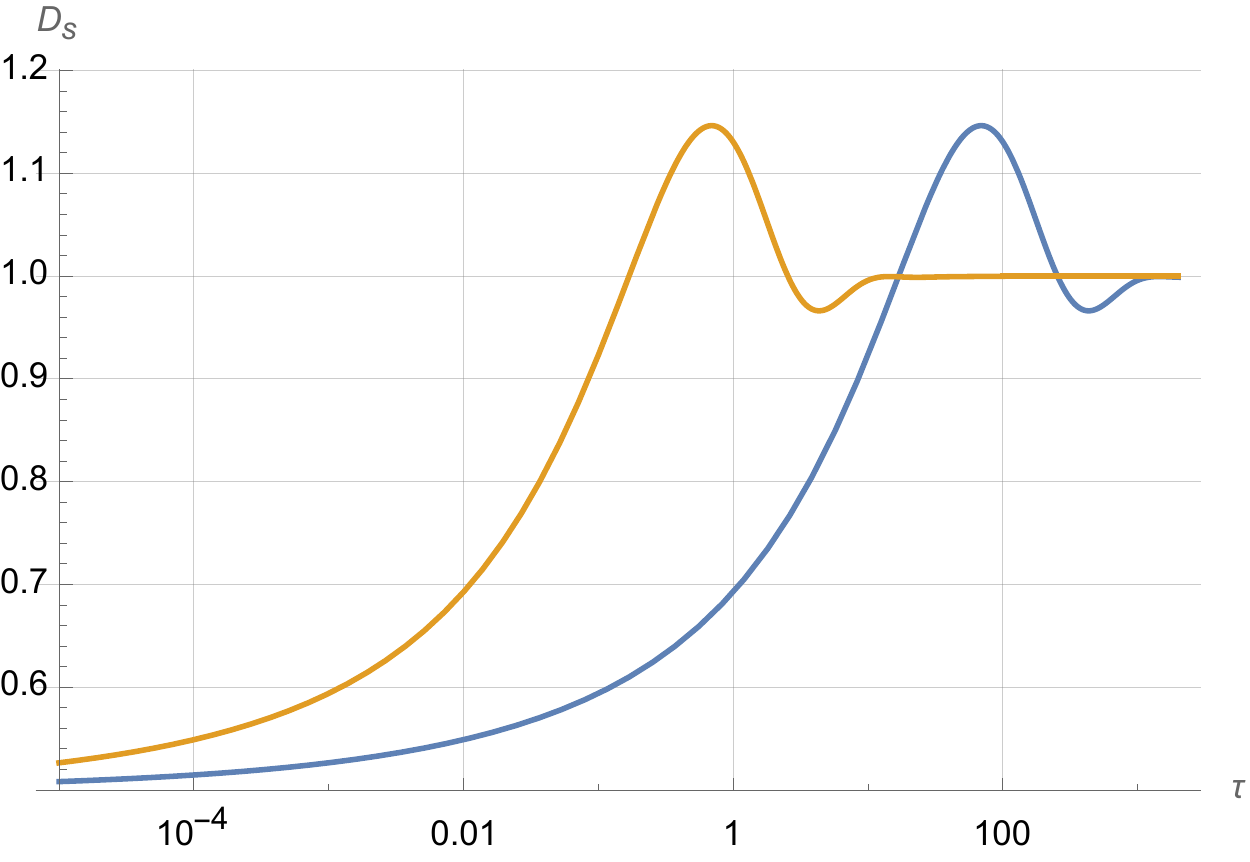}
    \caption{
    Left: Spectral dimension of superposed one-dimensional lattice geometries with purely scaling measure, $\sfs = \frac{5}{4}$, integrated from $\jmin=0$ to $\jmax=1$ (yellow) and $\jmax=100$ (blue). 
    Right: The same 
    for an oscillating measure integrated from $\jmin=0$ to $\jmax=\infty$, with $\sfs = \frac{5}{4}$ and $\omega=1$ (yellow) and $\omega=1/100$ (blue).}
    \label{fig:ds1cos}
\end{figure}

The simplified model is an equilateral lattice.
We find closed expressions for the heat trace in one dimension, though the results should extend to hypercubic lattices of any dimension since the lattice heat trace factorizes \cite{Thurigen:2015uc}.
On the one-dimensional lattice, it is possible to integrate the heat trace 
\[
P_{1\textsc{d}}(x) := \int\limits_0^\pi \d p \,\e^{- x (1-\cos p)} = \pi \e^{-x} I_0(x)
\]
where $I_0$ is the modified Bessel function.
For this, integration with a purely scaling measure (constant $\sfs=\sfsc$) gives
\[\label{eq:1d heat trace scaling}
    \expec{P(\tau)}_\sfs :=
    \int\limits_\jmin^\jmax \d r \,r^{-\sfs}  P_{1\textsc{d}}(\tau/r)
    = \frac{\pi}{1-\sfs}r^{1-\sfs} 
    \, {}_2F_2\bigg(\begin{matrix}
        \frac{1}{2},& \sfs-1\\
        1,& \sfs
    \end{matrix}; -{2\tau}/{r} \bigg) \bigg|_\jmin^\jmax \, ,
\]
where 
$
{}_pF_q\bigg(\begin{matrix}
        a_1,& ..., & a_p\\
        b_1,& ..., & b_q
    \end{matrix}; z \bigg)  
$
is the generalized hypergeometric function.\footnote{For an explicit definition of the generalized hypergeometric function ${}_p F_q$, see for instance~\cite{Slater}.} 
In the regime $\jmin\ll \tau \ll \jmax$, this example gives the expected spectral dimension $\Ds = 2(\sfs-1)$, presented in Fig.~\ref{fig:ds1cos}. 
In particular, one finds this value as the exact constant result for the integral with $\jmin=0$ and $\jmax=\infty$.

This example can still be integrated with an oscillating measure over the positive reals,
\[
\expec{P(\tau)}_{\sfs,\omega} 
    := \int_0^\infty r^{-\sfs} \cos(\omega r) \d r  \, P_{1\textsc{d}}(\tau/r)
\]
Interestingly, the result turns out to be a function solely in the combined variable $\omega\tau$ up to a factor $\omega^{\sfs-1}$,
\begin{align}\label{eq:1d heat trace oscillations}
    \expec{P(\tau)}_{\sfs,\omega} 
    =& \sqrt{\pi} \frac{\Gamma \left(\frac{3}{2}-\sfs \right) \Gamma (\sfs -1)}{ \Gamma (2-\sfs )} (2\tau)^{1-\sfs }
    \, _2F_5\left( \begin{matrix}
        \frac{3-2\sfs}{4},\frac{5-2\sfs}{4} \\
        \frac{1}{2},\frac{2-\sfs }{2},\frac{2-\sfs}{2},\frac{3-\sfs}{2},\frac{3-\sfs}{2}
    \end{matrix}
    ;-{(\omega\tau)^2}/{4}\right) \nonumber\\
    &-\omega^{\sfs-1} \pi \Gamma (-\sfs ) \Bigg[
    \cos(\pi\sfs/2) \, \omega\tau \, _2F_5\left(\begin{matrix} 
        \frac{3}{4},\frac{5}{4}\\
        1,\frac{3}{2},\frac{3}{2},\frac{1+\sfs}{2},\frac{2+\sfs}{2}
    \end{matrix};-{(\omega\tau)^2}/{4}\right)     \nonumber\\
    &\quad\quad\quad\quad\quad\quad\quad +\sfs \sin(\pi\sfs/2)\, _2F_5\left(\begin{matrix}
        \frac{1}{4},\frac{3}{4}\\
        \frac{1}{2},\frac{1}{2},1,\frac{1+\sfs}{2},\frac{\sfs}{2}
    \end{matrix};-{(\omega\tau)^2}/{4}\right)
    \Bigg] \nonumber \, .
\end{align}
As a consequence, the spectral dimension is also a function in $\omega\tau$.
Surprisingly, even though there are no boundary terms, this spectral dimension shows a flow from $2(\sfs-1)$ to $\Ds=1$ with an intermediate maximum at $\tau\approx1/\omega$ directly followed by a local minimum already close to $\Ds=1$, see Fig \ref{fig:ds1cos}.
For a finite upper integration boundary $\jmax>\omega\tau$ we would therefore expect no significant difference.

\begin{figure}
    \centering
    \includegraphics[width=0.6\linewidth]{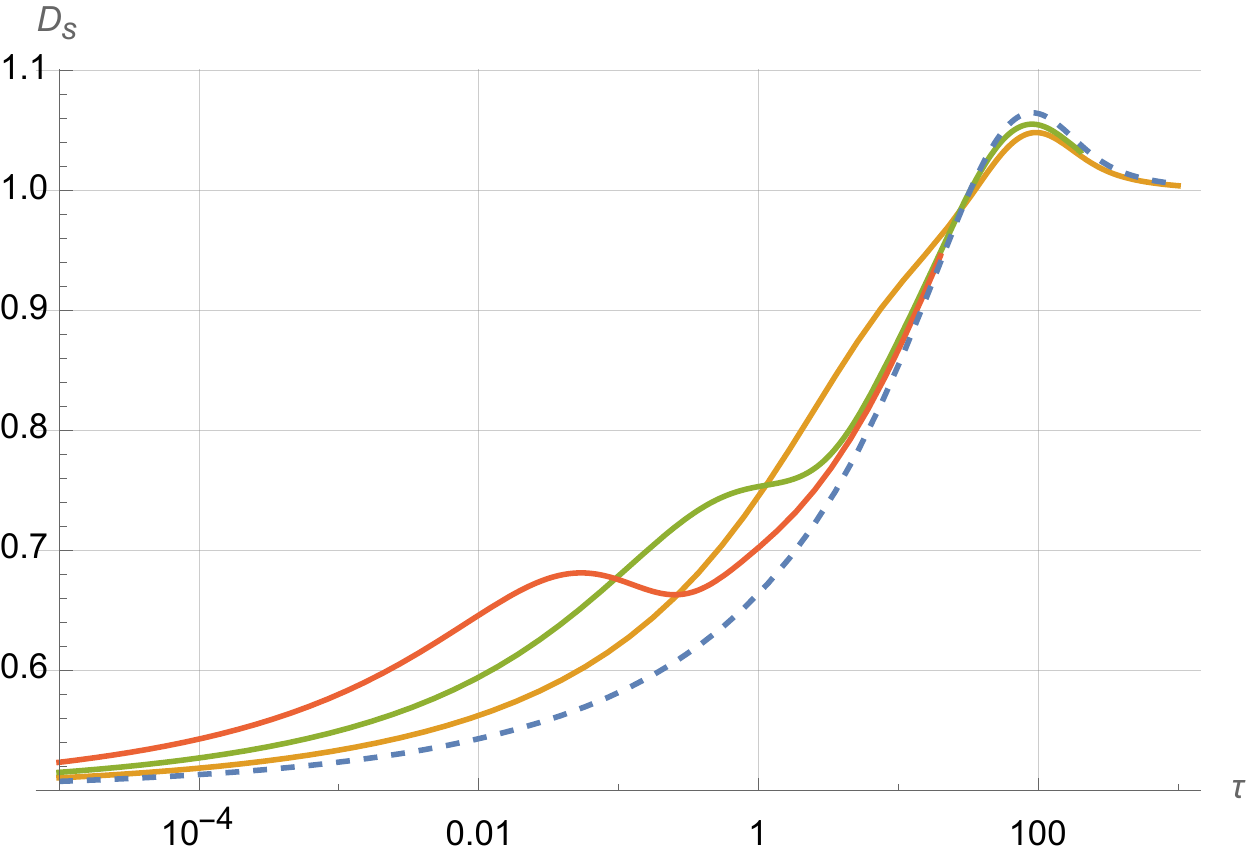}
    \caption{Spectral dimension of superposed one-dimensional lattice geometries with a combined measure, \eqref{eq:1d heat trace shift oscillations}, with $\jmin=0$, $\jmax=100$ (in the scaling part), a relative factor $a=1.5$ and oscillations $\omega = 10, 1, 1/10$ (with resulting peaks at $\tau\sim1/\omega$ from left to right); the purely scaling case is given for comparison as dashed curve.}
    \label{fig:ds1scaling+cos}
\end{figure}

The spin foam measure in~\eqref{eq:cc oscillations} that we want to model has a correction which is a linear combination of a constant term and an oscillating one.
Thus, we have to consider the spectral dimension of the combination $a \expec{P(\tau)}_{\sfs} + \expec{P(\tau)}_{\sfs,\omega} $,
that is, the quantity
\[\label{eq:1d heat trace shift oscillations}
\Ds(\tau) = -2\tau\frac{a \partial_\tau \expec{P(\tau)}_{\sfs}+\partial_\tau \expec{P(\tau)}_{\sfs,\omega}}{a \expec{P(\tau)}_{\sfs} + \expec{P(\tau)}_{\sfs,\omega}} \, .
\]
From earlier work \cite{Thurigen:2015uc} we know that a linear combination of two heat trace expectation values $\expec{P(\tau)}_{\sfs}$ and $\expec{P(\tau)}_{\sfs'}$ with different scalings $\sfs>\sfs'$ leads to a spectral dimension of value $2(\sfs-1)$ followed by a value $2(\sfs'-1)$ at larger scale $\tau$.
Something similar happens in the linear combination here 
when $\jmax\gg1/\omega$: 
As Fig.~\ref{fig:ds1scaling+cos} shows, we see the oscillation peak of the $\expec{P(\tau)}_{\sfs,\omega}$ part at $\omega\tau\approx 1$ but after that the $\expec{P(\tau)}_{\sfs}$ part dominates. 
Together, this yields a spectral dimension which looks like the purely scaling part with an extra local maximum superposed at scale $\tau\approx  1/\omega$. 
This is exactly the qualitative behaviour found in Fig.~\ref{fig:specdim1_Lambda}.
In this way, the simplified example gives an explanation of the mechanism  underlying the effect of a cosmological constant on the spectral dimension of spin foams.

\section{Discussion}\label{sec:Discussion}

\subsection{Renormalization and the spectral dimension}

From the perspective of classical general relativity, return probability and spectral dimension are reasonable observables. 
The return probability is coordinate independent since it is a trace of the heat kernel over the entire space-time. Equivalently, it is given by the exponentiated integral of the spectrum of the Laplace operator. 
Therefore it should be a suitable observable in the context of quantum gravity.

Diffeomorphism invariance is typically broken 
in spin foam models, yet we expect its restoration to be tied to (a notion of) discretization independence \cite{Dittrich:2008pw,Bahr:2009ku,Bahr:2009qc,Bahr:2011uj,Dittrich:2013xwa,Bahr:2015gxa,Bahr:2016hwc,Steinhaus:2020lgb,Asante:2022dnj}. 
This is vital if we interpret the spin foam 2-complex as a fiducial object, providing a regularization of the theory. 
Then, predictions of the theory must be consistent for different regulator choices, i.e. the regulator can be removed and the observables remain well defined in a suitable refinement limit. In principle this logic should apply to the spectral dimension as well, yet it is more subtle. 

To point out the particularities of renormalization in a background independent setting, it is helpful to first consider the spectral dimension on two different discretizations of the same manifold, where one is the refinement of the other. For diffusion times much larger than the typical length scale of the triangulation, their spectral dimension will agree with the continuum result. By definition, for the coarser triangulation this scale is larger, therefore its spectral dimension will deviate from the continuum result for larger diffusion compared to the finer triangulation. This is perfectly expected as the coarser triangulation is ignorant to dynamics below its discretization scale. In the context of spin foams however, this question is more intricate.

In a spin foam setting, consider two 2-complexes, where the coarser one arises from coarse graining the fine one. Let us additionally assume that we can compute the coarse graining flow of spin foam amplitudes, such that we can assign a theory to the fine 2-complex and its coarse grained version to the coarse one. Crucially, in a background independent theory, the discretization scale is not a parameter but part of the variables we are summing over. 
Thus, instead of the scale, the difference between coarse and fine theory is that the fine theory features more variables and can thus capture more configurations than the coarse one. Some of these fine configurations will correspond to representations of coarse configurations on the finer discretization, between which one could relate with embedding maps, yet some configurations cannot be captured in the coarse case. 
Therefore, one would expect differences to occur when the spectral dimension probes these fine configurations, assuming one is using the same definition of Laplace operator. 
This might be avoided by also coarse graining the Laplace operator such that the coarse version effectively reflects how the scalar field probes the fine version configurations leading to a modified spectrum of the renormalized operator.\footnote{The interpretation is that the coarse theory arises as an effective dynamics from the fine one after integrating out fine degrees of freedom. This would apply to the scalar field even though it is unphysical.} However, this procedure is ambitious and goes beyond the scope of this article.

Another method to employ a refinement limit is to study the same observable on finer and finer 2-complexes, with the same theory assigned to each complex. This is closer to the setup of this article, but strictly speaking not a renormalization procedure. The idea is to send the number of building blocks to infinity with the goal to identify whether the observable
converges under refinement; and further whether there are indications for a phase transition, if it is indeed an order parameter for this transition. In this way one might determine the set of parameters for which the original theory approximates a potential fixed point. 
Close to such a fixed point discretization independence would be approximately satisfied.\footnote{Conceptually, the ideas outlined here can be straightforwardly applied for 2-complexes without a boundary. For 2-complexes with boundary, one still needs to relate boundary states in different Hilbert spaces to describe the same transition.}

Beyond the technical challenges, one must first define how to systematically refine the 2-complex in order to define such a limit (if it indeed exists). In our setup of $\mathcal{N}$-periodic spin foams on hypercubic lattices, this is straightforwardly the limit $\mathcal{N} \rightarrow \infty$, which simultaneously removes the assumption of periodicity. 
To distinguish this limit from the refinement limit as determined by a renormalization procedure\footnote{It is not straightforward to implement $\mathcal{N}$-periodicity in a coarse graining procedure: imagine coarse graining an $\mathcal{N}$-periodic spin foam such that the number of degrees of freedom are halved. Unfortunately, without additional restrictions the resulting spin foam will not be an $\mathcal{N}/2$-periodic spin foam.
}, 
we call the limit $\mathcal{N} \rightarrow \infty$ together with $j_\text{max} \rightarrow \infty$ the thermodynamic limit. 
We discuss the implications of our analytical results for the thermodynamic limit in the next section.

\subsection{The thermodynamic limit}\label{subsec:The thermodynamic limit}

Several assumptions underlie the numerical results of Sec.~\ref{sec:results}, the strongest of which is clearly that of $\mathcal{N}$-periodicity of the geometric configurations. Furthermore, we truncated the number of dressed vertex amplitudes to $\mathcal{N}$ and introduced an upper cut-off $j_{\mathrm{max}}$ for a feasible numerical implementation. 
Both the cut-off and the periodicity need to be removed for physically viable results in a limit $j_{\mathrm{max}}\rightarrow\infty$ and $\mathcal{N}\rightarrow\infty$, respectively. In the following, we discuss these limits and the resulting interpretation of the spectral dimension.

As shown, the cut-offs $j_{\mathrm{min}}$ and $j_{\mathrm{max}}$ mark the boundaries for the scale $\tau$ between which an intermediate spectral dimension is possible; outside these values it flows to the value zero and four, respectively. In the limit $j_{\mathrm{max}}\rightarrow\infty$, we therefore expect an indefinite continuation of the intermediate regime to large scales. 
Due to the restrictive value of $j_{\mathrm{max}} = 201$ in Sec.~\ref{subsec:2-periodic spectral dimension} and the intricate form of $\mathcal{C}_v$ for $\mathcal{N}>1$ in Sec.~\ref{subsec:Analytical estimate of the spectral dimension}, inferring the explicit form of this continued intermediate regime for $\mathcal{N}\geq 2$ remains an open challenge. 
Further, the simplified integrable model of Sec.~\ref{sec:integrable model} has shown that oscillations can result in a flow to the topological dimension around the scale given by the frequency of the oscillations independent of an upper cut-off. Whether this generalizes to the intricate oscillatory dependence of spin foams is an intriguing direction for future research.

The plateau of the spectral dimension for purely scaling amplitudes, given in~\eqref{eq:intermediate D}, depends on the number of configuration variables $n$ as well as the number of vertices $V$. Within the approximations we have imposed, $n$ and $V$ are related to the periodicity $\mathcal{N}$ via $n = 2\mathcal{N}$ and $V = \mathcal{N}^4$, respectively. Then, the interval $[\alpha_{\mathrm{min}},\alpha_{\mathrm{max}}]$, for which an intermediate dimension $0 < \Ds^\alpha < 4$ exists can be re-expressed in terms of $\mathcal{N}$ 
\begin{equation}
\alpha_{\mathrm{min}} = \frac{1}{12}\left(9-\frac{2\mathcal{N}+2}{\mathcal{N}^4}\right), \qquad \alpha_{\mathrm{max}} = \frac{1}{12}\left(9-\frac{2\mathcal{N}}{\mathcal{N}^4}\right).
\end{equation}
Clearly, in the limit $\mathcal{N}\rightarrow\infty$, the interval shrinks to a single point $\alpha_*$, corresponding to the value at which the amplitudes are scale invariant. This is due to the fact that we are considering higher and higher powers of the amplitudes. Following~\cite{Steinhaus:2018aav}, in the context of cuboids, this point marks a phase transition since the spectral dimension changes discontinuously from $0$ to $4$ at $\alpha_*$. Since the degrees of freedom as well as the combinatorial length $L = \mathcal{N}$ are taken to infinity while keeping their ratio fixed, $\mathcal{N}\rightarrow\infty$ corresponds to a thermodynamic limit. Numerical results for $1$-periodic frusta, presented in Sec.~\ref{subsec:1-periodic spectral dimension}, are in alignment with the analytical formula in~\eqref{eq:intermediate D}. Also, the interval $[\alpha_{\mathrm{min}},\alpha_{\mathrm{max}}]$ of admissible $\alpha$-values to observe such an intermediate regime are supported by our results. 

Generalizing to $2$-periodic or higher configurations, the parameter space of the theory extends as the amplitudes now depend on $(\alpha, G,\bi,\Lambda)$. Qualitatively, for fixed $(G,\bi,\Lambda)$, we still observe that $\alpha$ controls the value of the intermediate dimension and that the interval between $\alpha_{\mathrm{min}}$ and $\alpha_{\mathrm{max}}$ is smaller in comparison to the $1$-periodic case. This suggests that a similar argument as in~\cite{Steinhaus:2018aav} applies: In the limit $\mathcal{N}\rightarrow\infty$, the spectral dimension exhibits a non-trivial flow only when $\alpha$ is tuned to some fixed value $\alpha_*$. Notice that, in comparison to cuboids, frusta amplitudes are strictly speaking \emph{never} scale-invariant. Consequently, $\alpha_*$ is in general not given by $\alpha_* = \frac{3}{4}$. Crucially, we observe that $(G,\bi,\Lambda)$ impact the spectral dimension only in the intermediate regime, which is controlled by $\alpha$. Conversely, outside the relevant $\alpha$ interval, we did not observe these parameters inducing a change in the spectral dimension.

If we vary $(G,\bi,\Lambda)$, the value $\alpha_*$ might change in principle and become a function of these parameters. Geometrically, this would imply in general a non-trivial surface $\Sigma_*$ of co-dimension one in $4$-dimensional parameter space $(\alpha,G,\bi,\Lambda)$ which marks the phase transition in the limit $\mathcal{N}\rightarrow\infty$. 
As demonstrated in Sec.~\ref{subsec:2-periodic spectral dimension}, the values of $\alpha_{\mathrm{min}}$ and $\alpha_{\mathrm{max}}$ are indeed dependent on $G$ and $\Lambda$.\footnote{We fixed $\bi$ to render the simulations feasible. However, if not fixed by other arguments such as from black hole entropy matching~\cite{Perez:2017cmj}, we expect that $\bi$ plays a similar role as $G$ and $\Lambda$, influencing the value of $\alpha_*$. We therefore assume in the following that $\alpha_*$ depends also on $\bi$.} 
In the limit $\mathcal{N}\rightarrow\infty$, the interval of $\alpha$'s will shrink to a point, the value of which depends on $G,\bi$ and $\Lambda$. Consequently, $\alpha_*(G,\bi,\Lambda)$ defines a non-trivial embedded surface $\Sigma_*$ which, tentatively speaking, marks the critical surface of a phase transition. 

\begin{figure}
    \centering
    \includegraphics[width=0.6\textwidth]{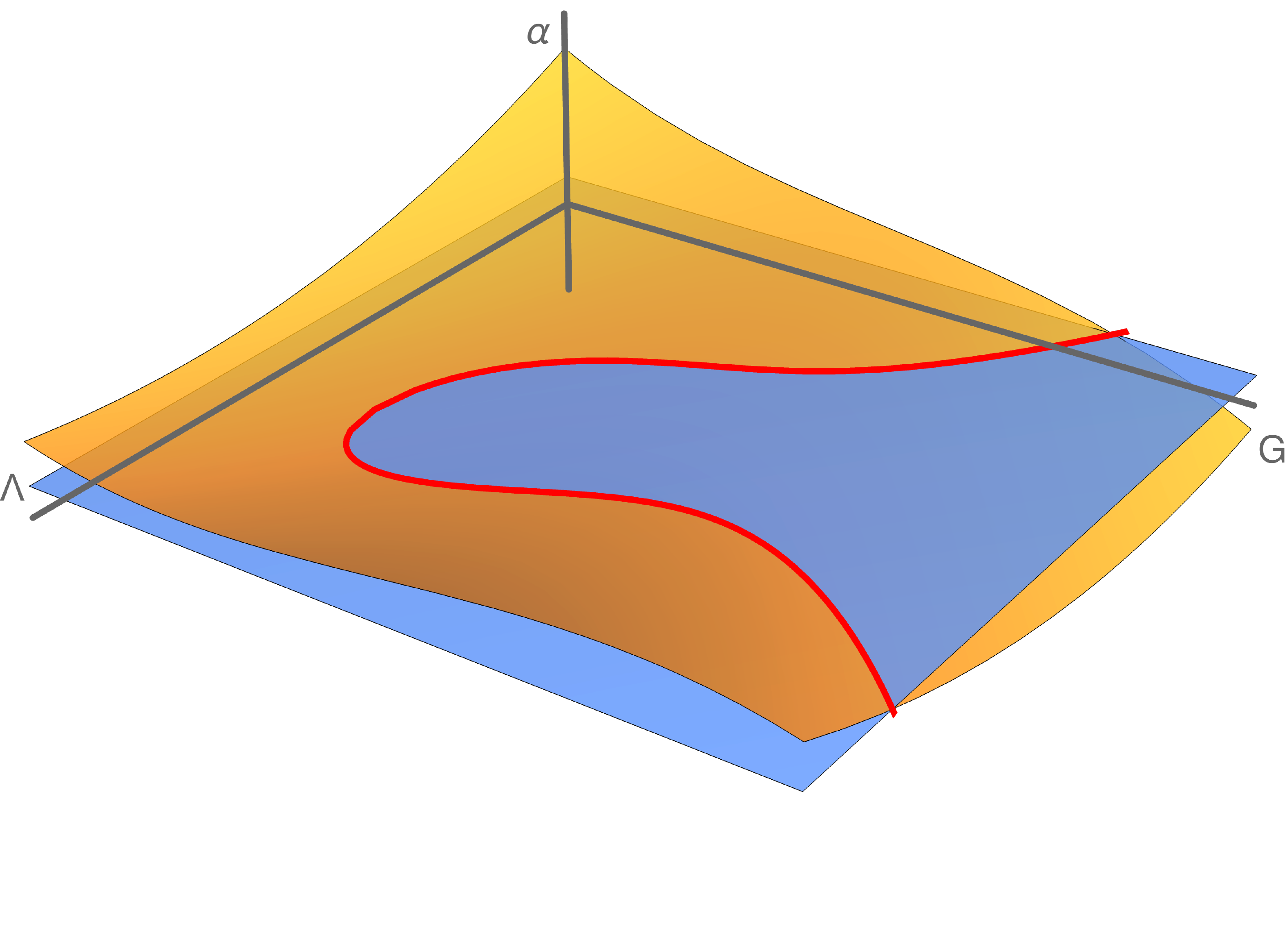}
    \caption{Sketch of a critical surface $\Sigma_*$ in the parameter space of theory with the $\bi$-direction suppressed. Given that $\alpha$ is fixed (blue plane), the intersection with $\Sigma_*$, indicated by the red curve, marks the critical values of $G$ and $\Lambda$.}
    \label{fig:critical_surface}
\end{figure}

Taking the perspective that $\bi$ and $\alpha$ are fixed and non-flowing parameters%
\footnote{Matching the black hole entropy from LQG to the Bekenstein-Hawking formula requires $\bi$ to take a specific value~\cite{Perez:2017cmj}. We remind the reader that $\alpha$ parametrizes an ambiguity in the face amplitudes. Choosing a specific model with a given face amplitude therefore corresponds to a fixed non-flowing $\alpha$.
}, 
the intersection of the $\alpha$ and $\bi$ hypersurfaces together with $\Sigma_*$ yield the lines of $(G,\Lambda)$, which we interpret as critical lines. A graphical intuition for that is presented in Fig.~\ref{fig:critical_surface} with the $\bi$-direction suppressed.
This intersection might also be empty though, depending on the values of $\alpha$ and $\bi$. 

In light of the previous discussion on renormalization, we emphasize that the large-$\mathcal{N}$ limit is not to be understood as a renormalization flow but rather as a removal of the $\mathcal{N}$-periodicity assumption. Defining a spin foam renormalization flow requires a refinement of the combinatorial structure as well as a relation between coarse and fine geometric variables. In contrast, the large-$\mathcal{N}$ limit simply considers the addition of building blocks and does not affect the variables or the states. Consequently, the RG-fixed points are in general not related to the ``critical values'' $(\alpha_*,G_*,\bi_*,\Lambda_*)$ of the limit $\mathcal{N}\rightarrow\infty$, but might provide indications for such a fixed point.

\subsection{Information about $G$, $\Lambda$ and $\gamma_{\mathrm{BI}}$ from the spectral dimension}\label{subsec:Interpretation of the results with a cosmological constant}

As demonstrated in Sec.~\ref{subsec:Cosmological constant} and Sec.~\ref{subsec:2-periodic spectral dimension}, Newton's constant $G$ and the cosmological constant $\Lambda$ have an immediate effect on the flow of the spectral dimension. It is furthermore to be expected that $\bi$ plays a similar role for frusta of $\mathcal{N}>2$. In principle, this could open the possibility to extract the values of $G,\bi$ and $\Lambda$ in a given regime from determining the effective, spectral dimension. In this section we discuss this possibility and its limitations.

Consider the $1$-periodic case discussed in Sec.~\ref{subsec:Cosmological constant} first, where the Regge curvature vanishes and only the ratio $\Lambda/G$ becomes relevant. Given that an intermediate regime exists, the spectral dimension will locally flow to a larger value in the vicinity of the scale at which the amplitude completes the first oscillation. This scale is directly related to the value of $\Lambda/G$. In principle, a given $1$-periodic spectral dimension flow therefore provides insight to the value of $\Lambda/G$. 

This picture changes drastically when going to higher periodicities where $S_R\neq 0$. In these cases the flow of the spectral dimension is highly sensitive to the three parameters $(G,\bi,\Lambda)$, where the effects do not seem to be localized around a certain scale. Again, a given measurement of the spectral dimension could, in principle, be used to attain partial knowledge on the values of these parameters. 
Various caveats and limitations follow these suggestions, the most important ones of which we discuss in the following.

First, for computing the $2$-periodic spectral dimension we have employed semi-classical amplitudes. 
Therefore, quantum effects have been neglected on small scales $\tau\sim j_{\mathrm{min}}$. 
Such effects have presented themselves in two ways. As discussed in Sec.~\ref{subsec:1-periodic spectral dimension}, quantum amplitudes show a non-constant modified scaling behaviour. Following~\cite{Allen:2022unb}, oscillating quantum amplitudes also show a phase shift with respect to the semi-classical ones on low scales. Clearly, both of these quantum effects need to be taken into account when considering the spectral dimension as an observable quantity.

Second, the results of Sec.~\ref{subsec:2-periodic spectral dimension} suggest that the map between the parameters $(G,\bi,\Lambda)$ and the corresponding spectral dimension is only surjective, such that one cannot extract a unique triple $(G,\bi,\Lambda)$ from a measured flow of $\Ds$. Moreover, for $\mathcal{N}>1$, the amplitudes exhibit an intricate oscillatory behaviour which leads to a spectral dimension that is highly sensitive to the values of $(G,\bi,\Lambda)$. 
{Indeed, we expect that it is necessary to measure several observables to determine these parameters accurately. Nevertheless, knowing $\Ds$ for all scales contains a lot, albeit coarse grained, information of quantum space-time that should allow us to constraint the range of admissible parameters, e.g. to distinguish whether $G$ is large or small.} 

Third, the triple ($G$, $\bi$, $\Lambda$) entering the semi-classical amplitudes in~\eqref{eq:Spin4Vamp} is considered as the \emph{bare} parameters.\footnote{This is discussed in more detail in the previous two sections. For discussion on the flow of $\bi$, we refer the interested reader to~\cite{Benedetti:2011yb}.} 
Thus, under a renormalization flow and under the assumption that we project back onto the original theory space e.g. as in \cite{Bahr:2018gwf}, it is in general to be expected that ($G,\bi,\Lambda)$ flow as well. As stressed above, for physical viability, the spectral dimension must be considered at a fixed point of the renormalization group flow to ensure discretization independence of the result. The values of $(G,\bi,\Lambda)$ at this fixed point are then the quantities that can be in principle observed. 

An observable effect of the parameters on the spectral dimension requires the existence of an intermediate regime. As discussed in Sec.~\ref{subsec:The thermodynamic limit}, in the thermodynamic limit $j_{\mathrm{max}}\rightarrow\infty$ and $\mathcal{N}\rightarrow\infty$, the values of $(\alpha, G, \bi, \Lambda)$ for which such a regime exists shrink to a point. At this transition point, the spectral dimension is expected to show a non-trivial behaviour. However, with the assumptions of finite upper cut-off $j_{\mathrm{max}}$ and periodicity $\mathcal{N}$, necessary for the computations in Sec.~\ref{sec:results}, determining the spectral dimension at the transition point is currently out of reach. 

As a last caveat, we remind the reader that the model we consider here presupposes a Euclidean space-time signature. The kinematics of physical scalar fields is governed by the d'Alembert operator and not the four-dimensional Laplace operator. Hence, deviations between predictions based on Euclidean models and results from measurements are in general to be expected. In addition to Lorentzian effects, we recall that the Laplace operator was defined via its action on scalar test fields. Physical fields used for actual measurements couple non-trivially to the geometry of space-time and lead to back-reactions on the gravitational field, see e.g. \cite{Ali:2022vhn}. As a result, the spectrum of the Laplace operator and therefore the return probability as well as the scaling behaviour of spin foam and matter amplitudes are modified.

\section{Conclusions} \label{Sec:Conclusion}

In this work, we have studied the effects of quantum amplitudes and oscillations on the spectral dimension within the EPRL-FK model, restricted to $\mathcal{N}$-periodic frusta geometries. This marks a significant expansion of previous work \cite{Steinhaus:2018aav} on flat, non-oscillatory and purely semi-classical cuboid geometries \cite{Bahr:2015gxa}. 
To by-pass the steep numerical costs for computing quantum frusta amplitudes already at small spins \cite{Allen:2022unb}, we have presented a method to extrapolate quantum amplitudes for $\mathcal{N}=1$ to large spins in Sec.~\ref{sec:Quantum amplitudes from extrapolation}, serving as an improved approximation compared to semi-classical amplitude, in particular at low spins. This marks the first result of our work. 

Computing the spectral dimension with respect to extrapolated amplitudes, we find additive corrections at low scales compared to the semi-classical results. Supported by analytical computations, these quantum corrections can be traced back to a modified effective scaling behaviour of the extrapolated amplitudes, constituting our second result. 

As a third result we have found that curvature induced by a cosmological constant~$\Lambda$ yields additive corrections of the $1$-periodic spectral dimension at scales $\tau\sim1/\sqrt{\Lambda}$. 
We have shown that such effects of $\Lambda$ can also be understood qualitatively by considering the effective scaling of the amplitudes. 
Furthermore, we have given an explanation of the mechanism underlying these results in terms of a simplified integrable model with such oscillating measure.

Finally, we found that $2$-periodic amplitudes with an intricate oscillatory behaviour lead to a flow of the spectral dimension which depends on the full set of parameters $(\alpha, G,\bi,\Lambda)$. 
Summarizing, our numerical and analytical results show that curvature is an essential factor for the spectral dimension that requires further study. 

\

In an overarching perspective, the results of the present work are an intermediate step towards understanding the spectral dimension of more general quantum geometries. 
Spin foam frusta with their inherent high degree of symmetry present a strong restriction of the quantum geometry compared to the general case. Retaining hypercubic combinatorics, a feasible scenario would be to construct a more general restricted model, which however quickly becomes cumbersome and the numerical challenges in the quantum regime remain. Furthermore, there is evidence that the EPRL-FK model for higher valent vertices differs from the one defined on triangulations in the implementation of simplicity constraints~\cite{Bahr:2017ajs,Assanioussi:2020fml}, and geometric critical points with torsion and non-metricity exist~\cite{Bahr:2015gxa,Dona:2020yao}. Therefore, if all restrictions on the geometry are lifted, it is advantageous to directly work on triangulations, where the semi-classical amplitudes are well studied~\cite{Conrady:2008mk,Barrett:2009ci,Barrett:2009mw,Kaminski:2017eew,Liu:2018gfc,Simao:2021qno,Han:2021bln} and more numerical methods are available and in development~\cite{Dona:2019dkf,Gozzini:2021kbt,Asante:2020qpa,Han:2021kll,Asante:2022lnp}. 

In the following, we briefly discuss the challenges one faces when defining the spectral dimension on unrestricted spin foams defined on a triangulation. 

\begin{itemize}
    \item {\textit{$\mathcal{N}$-periodicity on a triangulation:}}
    The introduction of $\mathcal{N}$-periodicity~\cite{Steinhaus:2018aav} is highly beneficial for reducing the computational effort. In particular, formulating the spectrum of the Laplace operator on momentum space is far more efficient than directly diagonalizing the Laplace operator and can be straightforwardly generalized to larger 2-complexes. The notion of $\mathcal{N}$-periodicity is tied to combinatorial ``directions'', which are naturally defined on a hypercubic lattice. On 4-dimensional triangulations
    these intuitions are not applicable and $\mathcal{N}$-periodicity is not straightforwardly defined.
    \item \textit{Vector geometries in the semi-classical limit:} The semi-classical amplitudes for a 4-simplex exhibit different types of critical points, depending on the boundary data~\cite{Barrett:2009ci}. While Regge geometries present one class of solutions, so-called vector geometries~\cite{Barrett:2009ci} contribute with the same degree of polynomial decay. Since such configurations do not correspond to geometric 4-simplices, their $4$-volume is not defined, and it is not obvious how to generalize the definition of a Laplace operator.
    \item \textit{Non-matching simplices / complex critical points:} The restriction to cuboid or frusta geometries is special, as each 4d building block is evaluated on a critical point and glued along matching 3d polyhedra. In recent years, there has been growing evidence that configurations beyond real critical points must be taken into account for large, but finite spins. In effective spin foams \cite{Asante:2020qpa,Asante:2021zzh} and the hybrid algorithm representation \cite{Asante:2022lnp} these are parametrized as geometric but non-matching simplices, i.e. the shared 3d building blocks have different shapes as seen from their 4d building blocks. Alternatively, such configurations are called complex critical points \cite{Han:2021kll,Han:2023cen}: such configurations can be interpreted as exhibiting non-vanishing torsion and play an important role in understanding and circumventing the flatness problem \cite{Bonzom:2009hw,Hellmann:2013gva,Oliveira:2017osu,Engle:2020ffj,Asante:2020qpa,Han:2023cen}. Following our definition, the Laplace operator for such a non-matching configuration might not be symmetric any more.
    \item \textit{Pre-geometric configurations:} Beyond a modified scaling behaviour, the deep quantum regime of spin foams additionally features pre-geometric configurations 
    which are not peaked (as coherent states) on the shape of semi-classical polyhedra. It is an intriguing question how a scalar test field can probe such a quantum geometry.
    \item \textit{Numerical challenges in and beyond the quantum regime:} Although the computation of quantum amplitudes is more feasible in the general simplicial case, the mere number of ten spin configurations per vertex presents serious numerical challenges.
    \item \textit{Lorentzian signature:} The studies of the spectral dimension presented here essentially examine a diffusion process on Euclidean quantum space-time. It will be interesting to see whether these concepts can be generalized to Lorentzian test fields probing a Lorentzian quantum space-time, see e.g.~\cite{Eichhorn:2013ova} for an implementation in causal set theory. For the gravitational side, choosing the Lorentzian EPRL-FK model might be beneficial compared to the Euclidean one: semi-classical Lorentzian 4-simplices feature oscillations proportional to $\bi$.
\end{itemize}

 We expect that several of the features mentioned above will have an impact on the spectral dimension, which we currently cannot estimate. In particular, it will be interesting to see whether the effective scaling of the amplitudes can still explain the behaviour of the spectral dimension if pre-geometric configurations are considered. Therefore, it is imperative to lift the restrictions of the frusta model and work towards exploring the spectral dimension on unrestricted spin foam quantum geometries.

A possibility to define $\mathcal{N}$-periodicity in a simplicial context could be to triangulate $\mathcal{N}$-periodic cubical lattices. There exist several inequivalent options to realize that, two examples of which are given in~\cite{ScottMara1976} and~\cite{Dittrich:2022yoo}, respectively. Geometrically, these configurations correspond to different triangulations of the torus. Explicitly setting up $\mathcal{N}$-periodic triangulations and computing the spectral dimension thereof is left for future work.

Within the context of spin foam frusta, a conceivable method to compute the spectral dimension for higher periodicities and larger cut-offs $j_{\mathrm{max}}$ is to restrict to configurations with small dihedral angles, e.g. via a Gaussian. Previous work~\cite{Steinhaus:2018aav} supports the conjecture that this restriction still captures the relevant geometric information. If valid, this reduction of configurations would greatly simplify the numerical effort required to compute $\Ds$, enabling exploration of regimes currently out of reach. Also for less restricted geometries on a triangulation, a linearization around flat configurations could be advantageous as this simultaneously simplifies the form of the amplitudes and the Laplace operator.

Going beyond small periodicities is numerically challenging regardless of the specific model at hand. 
This is even more true for triangulations without geometric restrictions, where we must consider vastly more variables compared to the symmetry restricted frusta cases. To reliably compute expectation values of observables, it is therefore imperative to use a numerical method that does not exponentially scale with the number of variables of the system. In many areas of physics, Monte Carlo methods serve this role, yet cannot be readily applied in spin foams. Due to the oscillatory nature of spin foam amplitudes, the spin foam partition function might not be naively usable as a probability distribution to sample results with. A way around this might be to define Markov Chain Monte Carlo on Lefschetz thimbles of the spin foam partition function \cite{Han:2020npv}, where the integration contour is changed such that the imaginary part of the system is constant and thus non-oscillatory. Alternatively, a potential strategy might be to propose a new probability distribution to sample configurations with. Recently, random sampling of bulk quantum numbers was applied to approximate spin foam amplitudes with many bulk faces in \cite{Dona:2023myv}.

To conclude, the spectral dimension remains an intriguing observable in spin foam models that is far from understood. Our results show that its value at small scales depends on all the parameters $\alpha, G,\bi,\Lambda$, though the specific relations are sensitive to the specific restricted model. Many effects such as pre-geometric configurations might alter its behaviour. Hence, our work shows how the spectral dimension of spin foam geometry can provide us with a deeper, coarse grained understanding of quantum space-time while at the same time allowing us to connect to other approaches of quantum gravity as well as continuum physics.

\section*{Acknowledgements}

AFJ would like to thank Florian Atteneder for frequent support with the numerical implementation. The authors would like to thank the anonymous referee for their insightful and constructive comments.

AFJ and SSt gratefully acknowledge support by the Deutsche Forschungsgemeinschaft (DFG, German Research Foundation) project number 422809950. 
AFJ kindly acknowledges support from the MCQST via the seed funding Aost 862983-4 and from the Deutsche Forschungsgemeinschaft (DFG) under Grant No 406116891 within the Research Training Group RTG 2522/1.
JT's research was funded by DFG grant number 418838388 and Germany's Excellence Strategy EXC 2044--390685587, Mathematics M\"unster: Dynamics - Geometry - Structure. 



\bibliographystyle{JHEP}
\bibliography{main}

\end{document}